\documentclass{aastex631}

\usepackage{graphicx}
\usepackage{mwe}
\usepackage{booktabs}
\usepackage{amsmath}
\usepackage[utf8]{inputenc}    
\usepackage[T1]{fontenc}       
\usepackage{lmodern}           

\begin{document}

\title{Testing the Young FRB Progenitor Hypothesis: A Crossmatch of Catalog-1 CHIME Bursts with Historic Local Universe Supernovae}

\correspondingauthor{Wanqing Liu \& Mohit Bhardwaj}
\email{wanqing2@andrew.cmu.edu, mohitb@andrew.cmu.edu}

  \author[0009-0008-3199-2627]{Wanqing Liu}
\affiliation{Department of Physics, Carnegie Mellon University, 5000 Forbes Avenue, Pittsburgh, 15213, PA, USA}

  \author[0000-0002-3615-3514]{Mohit Bhardwaj}
\affiliation{Department of Physics, Carnegie Mellon University, 5000 Forbes Avenue, Pittsburgh, 15213, PA, USA}

\author[0000-0001-8405-2649]{Ben Margalit}
  \affiliation{School of Physics and Astronomy, University of Minnesota, Minneapolis, MN 55455, USA}

\begin{abstract}

Fast radio bursts (FRBs) are among the most energetic and enigmatic transients in the radio sky, with mounting evidence suggesting newborn, highly magnetized neutron stars formed in core‐collapse supernovae (CCSNe) as their progenitors. A definitive spatial association between an FRB and a historic CCSN would confirm this link and tightly constrain young neutron star source models. Here we report on the first systematic cross‐matching of 886 spectroscopically classified CCSNe in the local Universe ($z\leq0.043$) against 241 CHIME/FRB Catalog 1 events, applying rigorous spatial, dispersion measure (DM), and scattering time ($\tau$) criteria. We identify four positional overlaps, all consistent with chance alignment; however, one pair, FRB 20190412B–SN 2009gi, also satisfies independent host‐DM and $\tau$ constraints, making it a promising candidate for targeted follow‐up. Next, we search for compact (persistent or transient) radio emission at all matched supernova sites using multi‐epoch VLASS data and detect none. Treating every CCSN sight line as a non‐detection, we derive Poisson upper limits on the FRB burst rate at these locations, which lie well below the rates observed for the most active repeaters unless their activity is heavily suppressed by beaming, intermittency, or residual free‐free absorption. We then develop a galaxy‐integrated FRB‐rate model that incorporates an intrinsic spectral index, secular magnetar‐activity decay, and frequency‐dependent free‐free opacity. Applying this formalism to existing FRB data shows that reproducing the observed CHIME/CRAFT all‐sky rate ratio requires a steep decline in magnetar burst rates with age. Finally, our work underscores the necessity of sub‐arcsecond localizations and multiwavelength follow‐up to definitively test the young neutron star source hypothesis.

\end{abstract}


\section{Introduction} \label{sec:intro}

Fast radio bursts (FRBs) are energetic, millisecond‐duration pulses of coherent radio emission detected at cosmological distances \citep{Lorimer2007,Thornton2013}. Since the first discovery in 2007, over 1000 FRBs have been reported.\footnote{For an up-to-date catalog, see the TNS: \url{https://www.wis-tns.org/} \citep{2020TNSAN..70....1Y}.} Despite extensive observational progress, their origin remains an unsolved mystery, although models invoking young, highly magnetized neutron stars (“magnetars”) are increasingly favored \citep[see e.g.,][ for a recent review]{2023RvMP...95c5005Z}. Moreover, the all‐sky FRB rate of $\sim10^3\,\mathrm{day^{-1}}$ \citep{2023ApJS..264...53C} implies a volumetric occurrence comparable to or exceeding that of core‐collapse supernovae (CCSNe), suggesting that most FRB sources likely repeat even though only $\sim5\%$ of events are observed to do so \citep{2019raviNat,2021ApJ...919L..24B,2023PASA...40...57J,2023ApJ...947...83C}. This young‐neutron‐star origin hypothesis was bolstered by the detection of FRB‐like burstS from the Galactic magnetar SGR\,1935$+$2154, confirming that magnetars can produce radio bursts with energetics and durations analogous to low‐luminosity extragalactic FRBs \citep{2020SGR,Bochenek2020}.

FRB host demographic studies further reinforce the young‐magnetar scenario: well‐localized FRBs predominantly occur in star‐forming galaxies, consistent with massive‐star progenitors \citep{2021ApJ...907L..31B,2022AJ....163...69B,Gordon+23,2024ApJ...971L..51B,2024Natur.635...61S,Loudas+25}. Additionally, several hyperactive repeaters coincide with compact, persistent radio emission interpreted as magnetar wind nebulae of age $\lesssim100\,$yr \citep{2017ApJ...841...14M,2017Natur.541...58C,2017ApJ...834L...8M,2018ApJ...868L...4M,2022Natur.606..873N,2024Natur.632.1014B,2024arXiv241213121B,2024arXiv241219358B,2025A&A...695L..12B,2025arXiv250401125M}. Taken together, the Galactic magnetar burst detection, FRB volumetric rate, and host‐environment evidence imply that CCSNe constitute the primary formation channel for FRB sources, with a minority of events in quiescent or globular‐cluster environments hinting at a secondary, delayed channel \citep{2019ApJ...886..110M,2021ApJ...910L..18B,2022Natur.602..585K,2023ApJ...950..175S,2023ApJ...950..134M,2025ApJ...979L..21S,2025ApJ...979L..22E,HorowiczMargalit2025}.

A direct spatial association between an FRB and a historic supernova would provide an unambiguous link to a progenitor stellar‐death event, fixing both the source age and its natal environment. Theoretical models predict that the FRB burst rate and energetics peak within $\sim10^2\,$yr of neutron‐star birth, when rotational and magnetic energy reservoirs are largest \citep{2017ApJ...841...14M,2017ApJ...843...84N,2020ApJ...896..142B}. Thus, a confirmed FRB-SN match would calibrate the delay between core collapse and FRB activity, anchor magnetar spin down and nebular evolution models, and would serve as a reference to interpret the growing FRB sample.

CCSNe form a heterogeneous class of explosions from massive stars ($M\gtrsim8\,M_{\odot}$). They are classified by spectral and light‐curve characteristics into subtypes—dominant ones are Type IIP, IIL, IIb, IIn, Ib, and Ic—reflecting differences in progenitor structure and envelope composition \citep{1997ARA&A..35..309F,2007AIPC..937..187T,2017MNRAS.469.2672P}. Progenitor mass‐loss history, metallicity, and binary interactions determine whether the star retains its hydrogen envelope (Type II) or loses it prior to collapse (Type Ib/c) \citep{2003ApJ...591..288H,2016ApJ...821...38S}. Identifying which CCSN subtypes produce magnetars capable of powering FRBs remains an outstanding observational challenge.

The Canadian Hydrogen Intensity Mapping Experiment (CHIME) has transformed the discovery of FRB by continuously surveying the northern sky in the 400–800\,MHz band with an instantaneous field of view of $\sim200\,\mathrm{deg}^2$ \citep{2018ApJ...863...48C}. Its first catalog (Catalog-1 hereafter) comprises 536 bursts detected between July 2018 and July 2019—including 62 from 18 repeating sources—and represents the first large, uniformly selected FRB sample, albeit with localization uncertainties of tens of arcminutes \citep{2021ApJS..257...59C}. In this study, we cross-match Catalog-1 events with historic CCSNe from the Sternberg Astronomical Institute catalog (6545 SNe up to October 17, 2014; \citealt{2004AstL...30..729T}\footnote{Available at \url{http://www.sai.msu.su/sn/sncat/download.html}.}), focusing on explosions within 200 Mpc. By enforcing different criteria to minimize false positives, we search for FRB–SN associations that can directly test the young-progenitor hypothesis.

The remainder of this paper is organized as follows. 
In \S\ref{sec:sample_selection}, we describe the construction of our supernova (\S\ref{sec:sample_sn}) and FRB samples (\S\ref{sec:sample_frb}). 
In \S\ref{sec:cross-match}, we present the cross‐matching methodology and identify candidate FRB–SN associations. 
\S\ref{sec:frb_prs} details our search for compact (\(\lesssim 2\,\text{kpc}\)) radio emission at the CCSN sites coincident with CHIME FRBs. 
In \S\ref{sec:discussion}, we first examine the impact of free–free opacity on FRB detectability (\S\ref{sec:free-free_constraint}), then compare our non‐detections to the activity levels of known repeaters (\S\ref{sec:burst_limits}), and finally assess prospects for definitive FRB–SN associations with future observations (\S\ref{sec:frb_sn_prospects}). 
In \S\ref{sec:magnetar_constraints}, we explore the implications of our results for young‐magnetar FRB models. 
Finally, we summarize our conclusions in \S\ref{sec:conclusion}.

\noindent We adopt the \cite{2020A&A...641A...6P} cosmology with $H_{0}=67.7\,\mathrm{km\,s^{-1}\,Mpc^{-1}}$.



\section{Sample Selection}\label{sec:sample_selection}

Here we summarize the construction of the historic supernova and
FRB samples used in the cross-matching analysis. \S\ref{sec:sample_sn} describes the CCSN sample, while \S\ref{sec:sample_frb} outlines the selection applied to CHIME/FRB Catalog-1. 

\subsection{SAI Supernovae Catalog}
\label{sec:sample_sn}

We draw supernova data from the SAI catalog \citep{2004AstL...30..729T}, including SN name, type, coordinates, redshift, and host‐galaxy position for each event. We apply the following criteria to select SNe for this study:

\begin{enumerate}
\item \textit{Event type:}  Only spectroscopically classified Type II 
and Type Ib/c (in the catalog, they are reported as: Ib, Ic, IPec, Ipec:, Ib/c, Ib/c:, Ib/cPec, Ib/cPec:, Ib:, IbPec, IbPec:, Ic:, IcPec, IcPec:) 
events are considered.
\item \textit{CHIME visibility:}  Only supernovae falling within the non-zero exposure region of the CHIME/FRB visibility map are retained \citep{2021ApJS..257...59C}.
\item \textit{Distance limit:}  The host galaxy must appear in the Heraklion Extragalactic CATaloguE \citep[HECATE;][]{2021MNRAS.506.1896K} with a luminosity distance \(D_{\mathrm{L}}\le 200\) Mpc ( or \(z\le 0.043\)). This limit ensures volumetric completeness and matches the maximum dispersion measure (DM) estimated distance cut imposed on the FRB sample (\S\ref{sec:sample_frb}) to reduce false positive associations. This is in line with the hypothesis‐testing framework described by \citep{2024ApJ...971L..51B}.

\end{enumerate}

\begin{figure}
\gridline{
  \fig{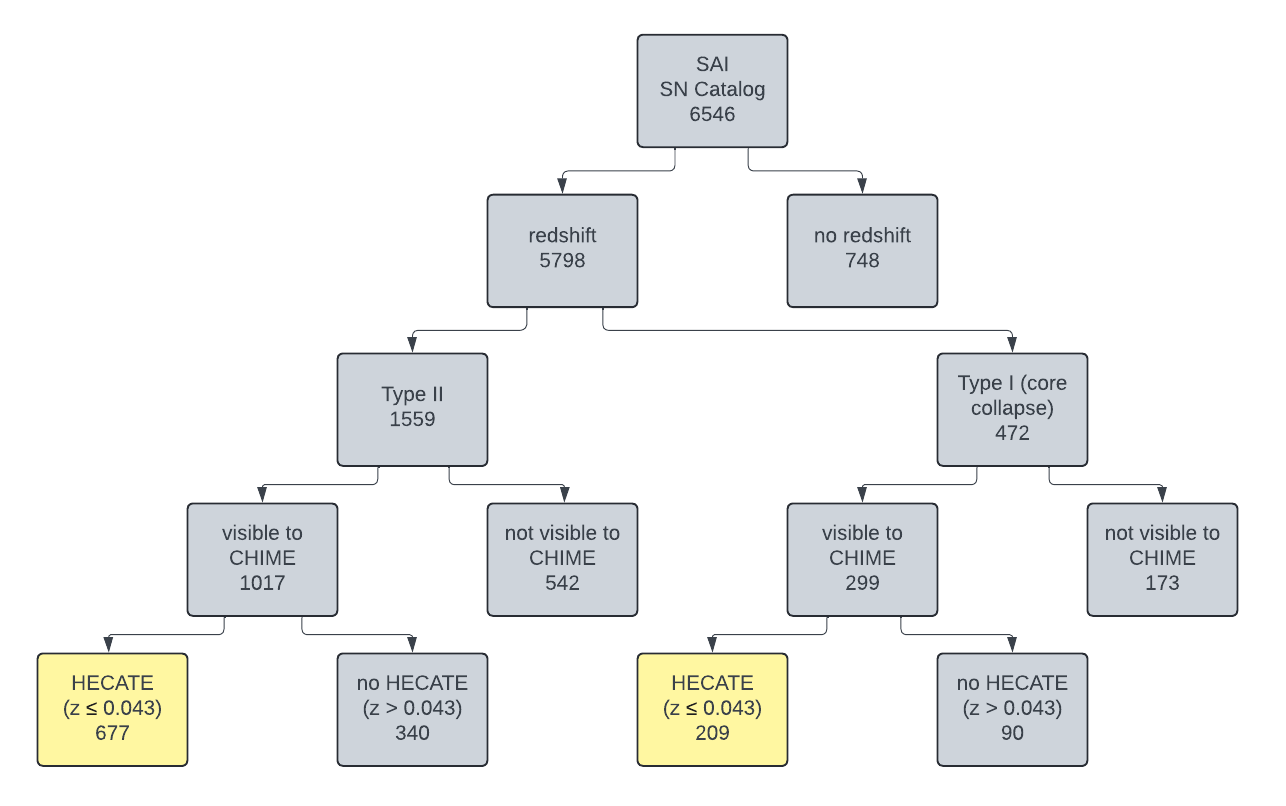}{0.55\textwidth}{(a)}
  \fig{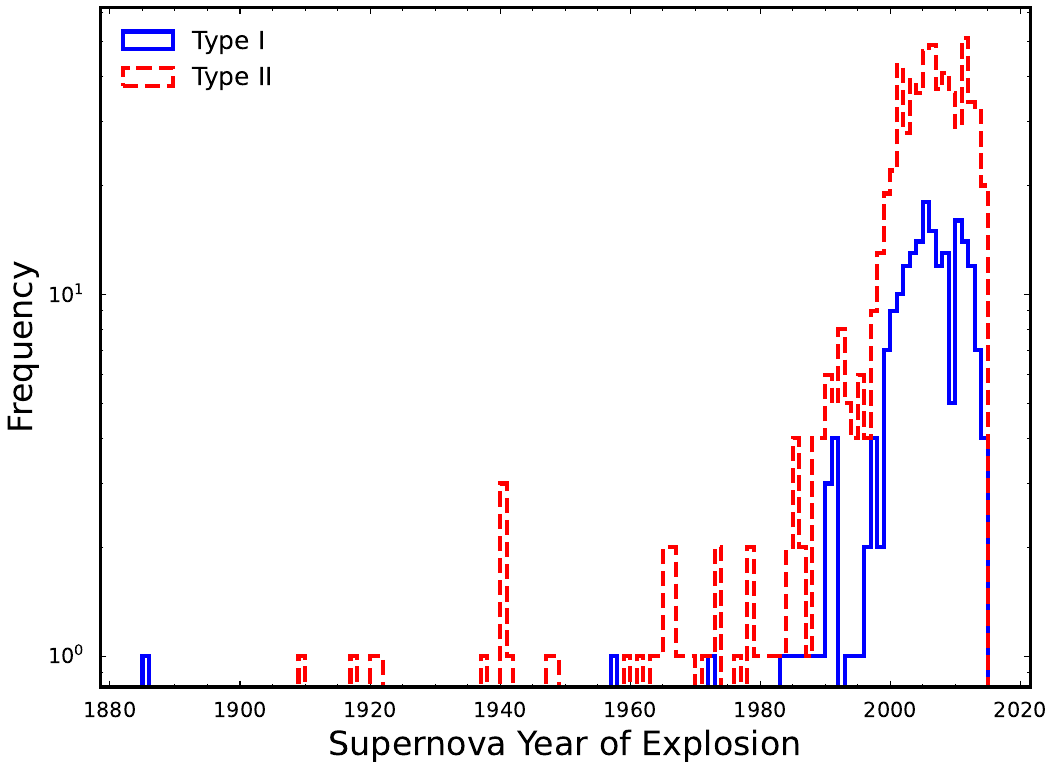}{0.44\textwidth}{(b)}
}
\vspace{0.5em}
\gridline{
  \fig{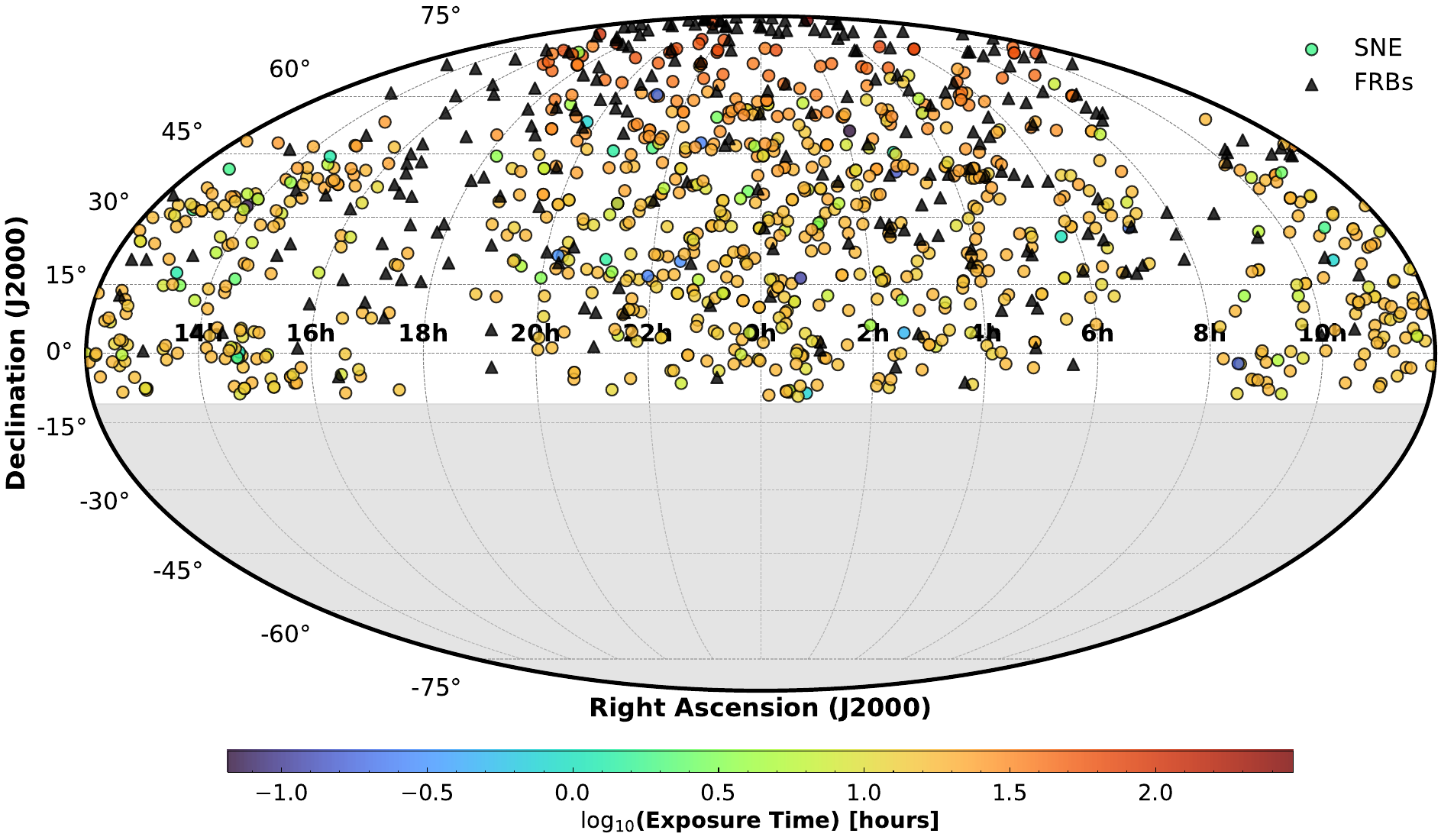}{\textwidth}{(c)}
}
\caption{
Selection overview and basic properties of the selected SN samples.
\textbf{(a)} Flow chart illustrating the successive cuts applied
to the Sternberg supernova catalogue and the resulting sample sizes.
\textbf{(b)} Discovery-year histogram for the 886 retained
core-collapse supernovae, with Type II events shown by the
dashed red line and Type Ib/c by the solid blue line
(1885–2014).
\textbf{(c)} Mollweide projection of the 886 supernovae
(coloured circles), where the colour scale indicates the mean CHIME/FRB
exposure from Catalog-1. The 241 CHIME bursts that satisfy the DM and
localisation cuts are plotted as green triangles. The light-grey band
marks declinations south of \(-11^{\circ}\), which lie outside the
CHIME field of view \citep{2018ApJ...863...48C}.
}
\label{fig:sne_frb_summary}
\end{figure}

The final SN sample comprises 886 events—677 Type II and 209 Type Ib/c. The sequence of selection cuts is illustrated by the flow chart in Figure \ref{fig:sne_frb_summary}(a), while the discovery‐year histogram in Figure \ref{fig:sne_frb_summary}(b) and the sky distribution in Figure \ref{fig:sne_frb_summary}(c) summarize the temporal and spatial characteristics of the sample.

\subsection{CHIME/FRB Catalog-1}\label{sec:sample_frb}

CHIME/FRB Catalog-1 contains 536 bursts detected between 25 July 2018 and 1 July 2019 at 400–800 MHz \citep{2021ApJS..257...59C}.  For each event, the catalog lists various parameters, such as sky coordinates, dispersion measure (DM), and scattering time.  62 of the bursts originate from 18 previously reported repeaters.  Using the published more precise localizations for these 18 sources \citep{CHIMEFRB2019,2020ApJ...891L...6F,2020Natur.577..190M,2021ApJ...919L..24B,2023ApJ...950..134M,2024ApJ...971L..51B,2024ApJ...961...99I}, we confirm that none coincide with the 886 selected supernovae; repeating events are therefore excluded, leaving 474 apparently non-repeating bursts.

Given that most Catalog-1 localizations cover areas of
\(\gtrsim 0.1~\mathrm{deg}^{2}\), we restrict the FRB sample to the same
volume adopted for the supernovae, namely \(z \le 0.043\)
(\(D_{\mathrm{L}} \le 200\) Mpc).  This distance cut
corresponds to an DM-excess requirement
\[
\mathrm{DM}_{\mathrm{excess}}
      = \mathrm{DM}_{\mathrm{FRB}} - \mathrm{DM}_{\mathrm{ISM}}
      \le 500~\mathrm{pc\,cm^{-3}},
\]
where \(\mathrm{DM}_{\mathrm{ISM}}\) is the Milky-Way disk contribution
estimated with the NE2001 and YMW16 electron–density models
\citep{Cordes2002,Yao2017}.  The derivation of the 500 pc cm\(^{-3}\)
threshold is provided in Appendix~\ref{app:max-dmexcess}.  Applying this
criterion reduces the non-repeating sample from 474 to 241 bursts.

For each of the 241 bursts we adopt the most precise localization
available: (i) 70 events have baseband positions from the first
CHIME/FRB baseband catalog \citep{2024ApJ...969..145C}; these
footprints are at least a factor of 500 smaller in area than the
header regions provided in Catalog-1, and we therefore employ them in our
cross-matching analysis. (ii) Of the remaining 171 bursts, two are classified as side-lobe detections.  We replace their header regions with the refined intensity localizations provided by \citet{2024ApJ...975...75L}, which improve the
area by a factor of \(\gtrsim 50\). (iii) For the other 169 events, we employ the header localizations reported in Catalog-1. In all four cases, we use the 90\% confidence regions in the
cross-matching procedure described in \S\ref{sec:cross-match}.
The sky distribution of the selected FRB sample is shown in
Figure \ref{fig:sne_frb_summary}(c).

\section{Cross-matching analysis}\label{sec:cross-match}

\begin{figure}
\gridline{
  \fig{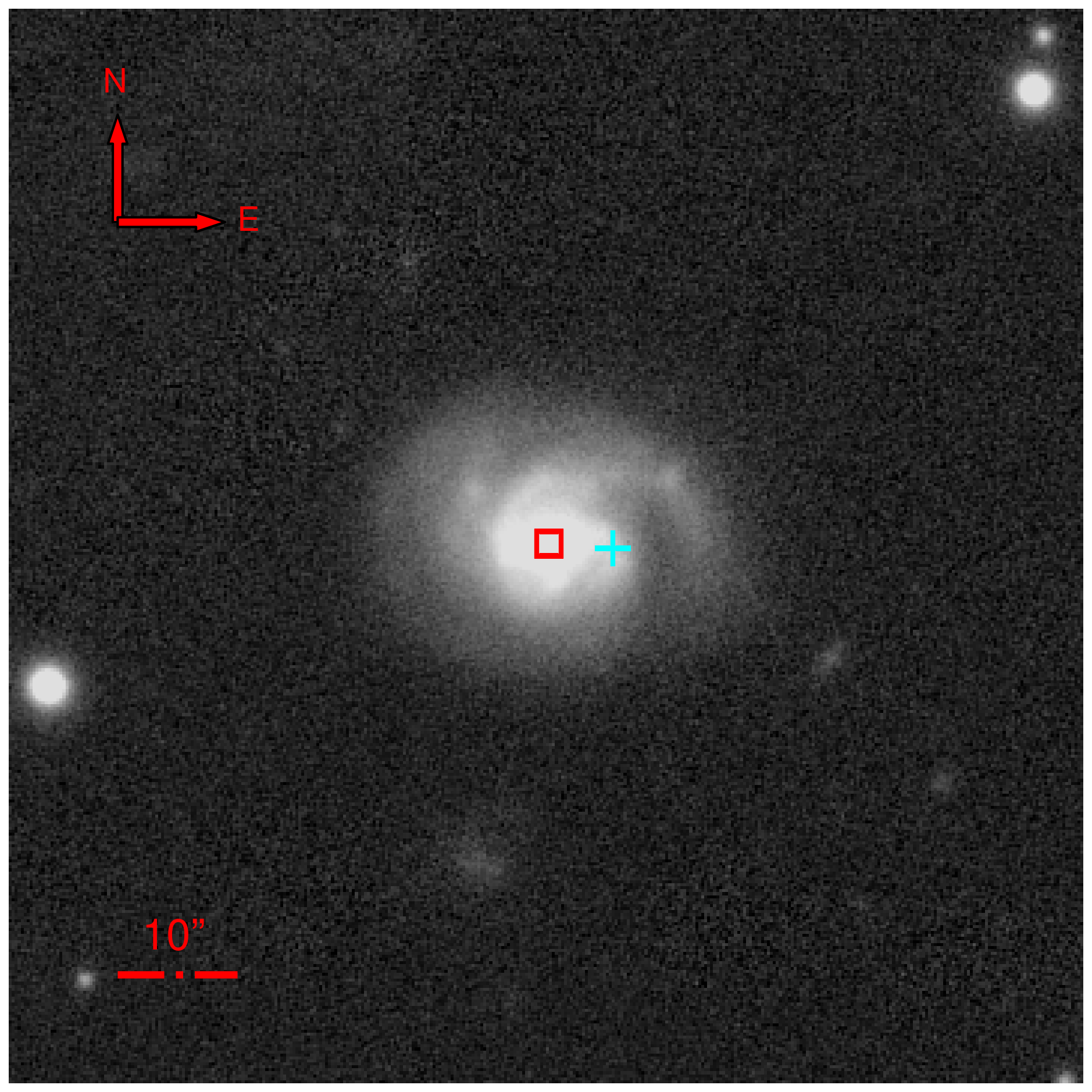}{0.22\textwidth}{(a) SN 2003la}
  \fig{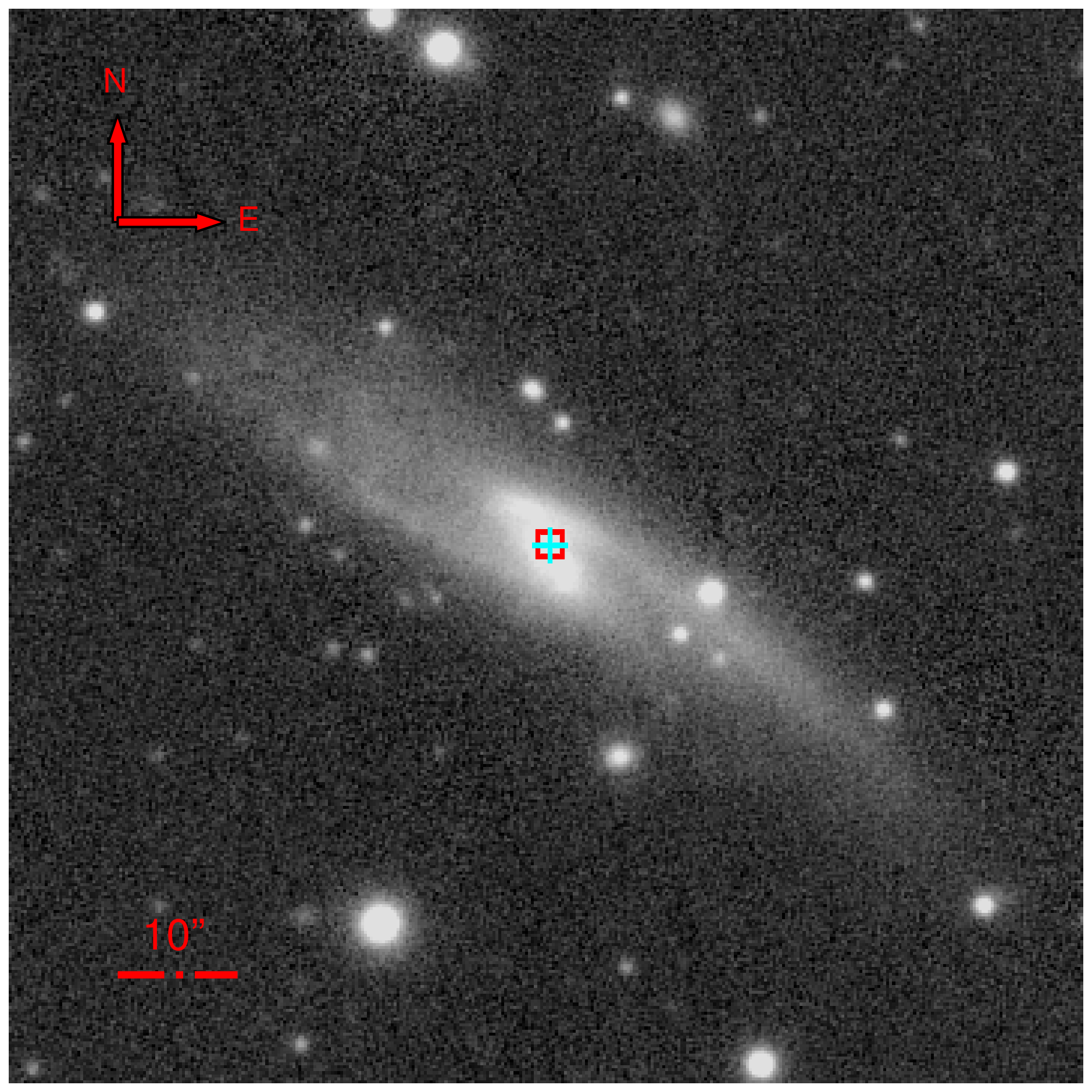}{0.22\textwidth}{(b) SN 2014ay}
  \fig{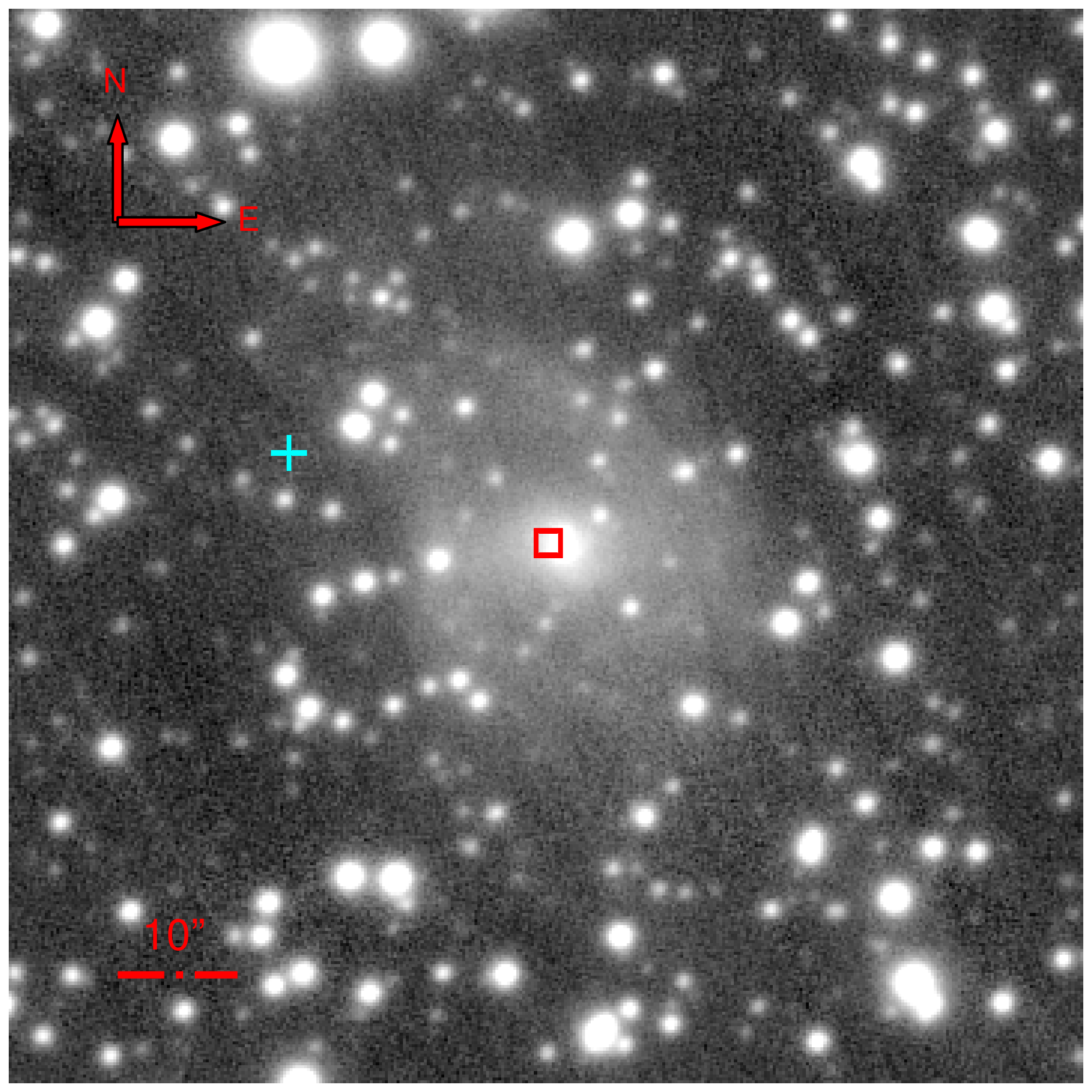}{0.22\textwidth}{(c) SN 2009gi}
  \fig{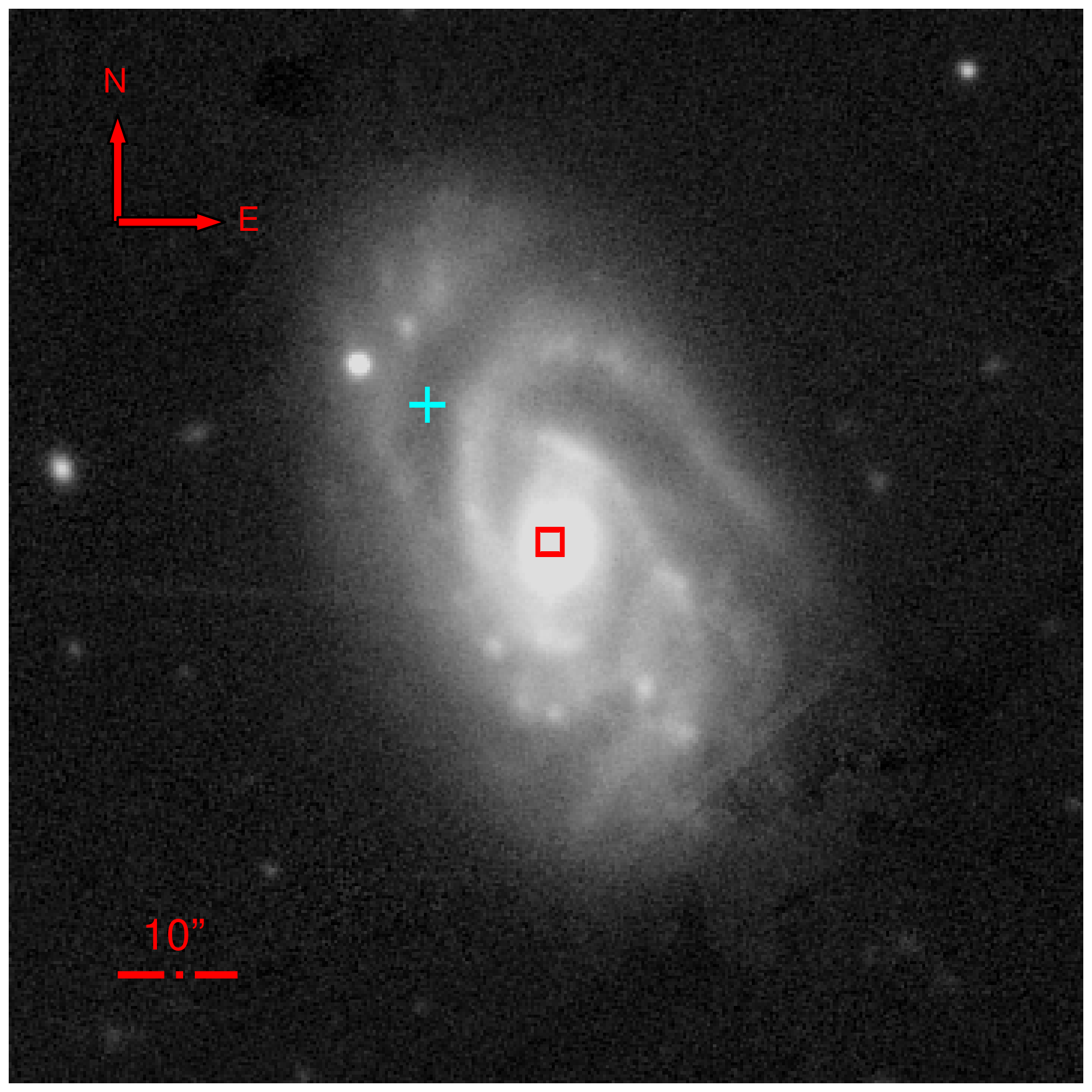}{0.22\textwidth}{(d) SN 2001ab}
}
\caption{Pan-STARRS $r$-band images of the galaxies hosting the four CCSNe (see Table~\ref{tab:sn_properties}) that lie within CHIME/FRB localization regions, labeled (a)--(d). The cyan cross marks each supernova’s position, while the red box indicates the galaxy’s photometric center.}
\label{fig:sn_images}
\end{figure}

\begin{figure*}
    \centering
    \renewcommand{\arraystretch}{0}
    
    \begin{tabular}{c c}
        \includegraphics[width=0.5\linewidth]{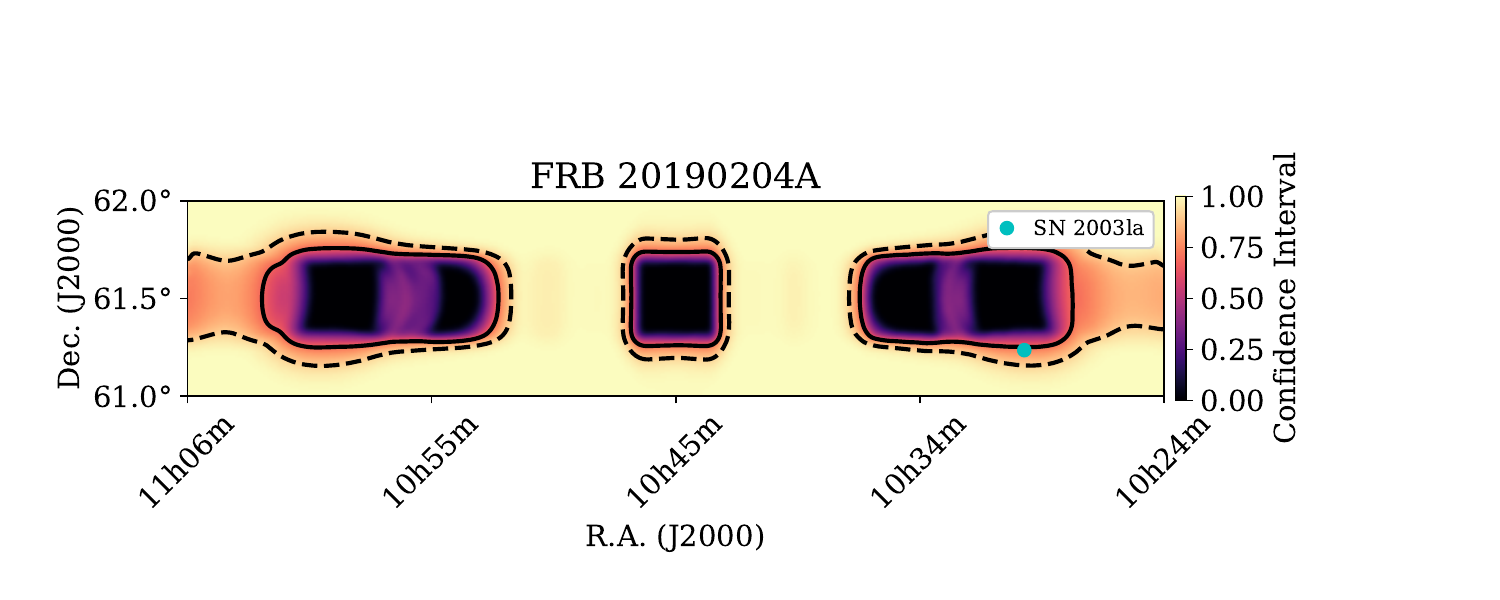} & 
        \includegraphics[width=0.5\linewidth]{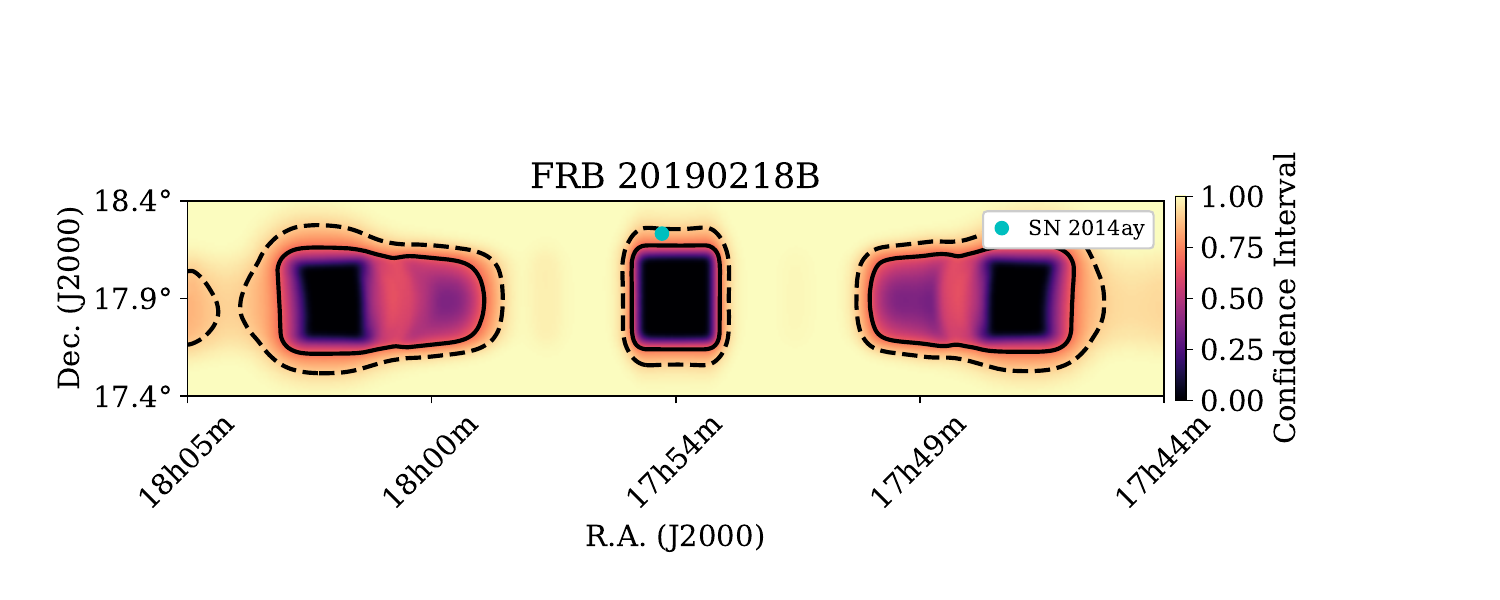} \\

        \includegraphics[width=0.5\linewidth]{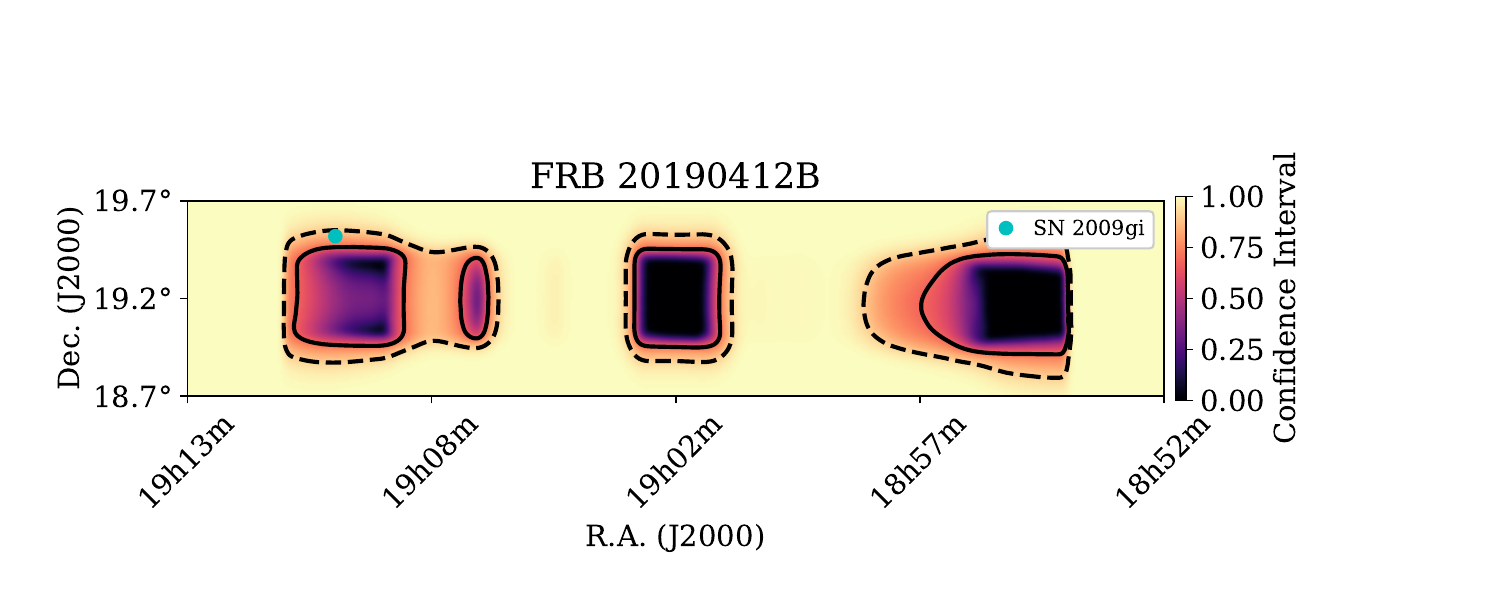} & 
        \includegraphics[width=0.5\linewidth]{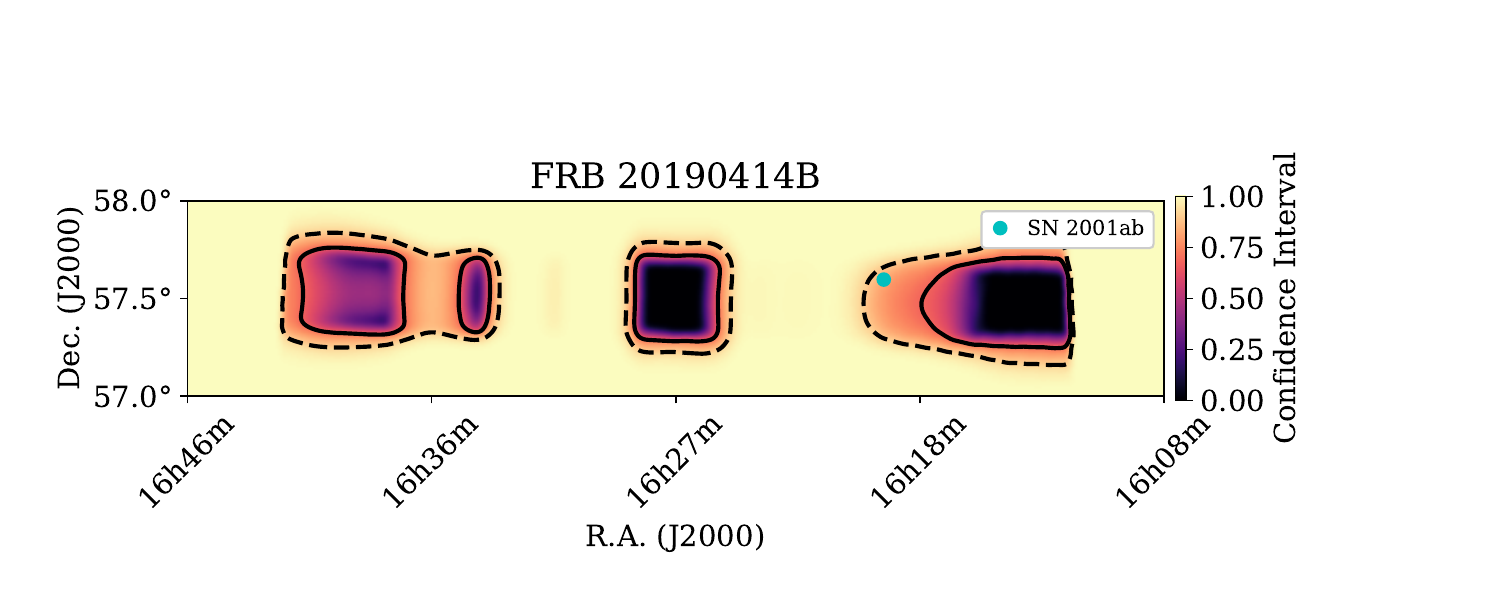} \\
    \end{tabular}

    \caption{Header localization confidence regions for the four CHIME/FRB sources that coincide with a supernova.  Solid and dashed curves mark the 68\% and 90\% confidence contours, respectively. The matched supernova position is indicated by the cyan dot.  Each panel is centered on the beam of maximum detection and spans 5$^{\circ}$ in right ascension (scaled by cos$\delta$) and 1$^{\circ}$ in declination ($\delta$).}
\label{fig:frb_loc}
\end{figure*}

As outlined in \S\ref{sec:sample_frb}, we adopt the 90\% confidence localization for each FRB (baseband, refined-intensity, or header, as detailed in \S\ref{sec:sample_frb}) to identify potential FRB-SN associations. A supernova is deemed coincident with an FRB when its position falls inside the corresponding localization footprint. Moreover, astrometric uncertainties for supernovae ($\lesssim0.3''$) are more than two orders of magnitude smaller than the smallest FRB regions and are therefore ignored in this analysis. Applying the spatial–overlap criterion to the full sample of 886 CCSNe and 241 CHIME bursts yields four FRB–SN positional matches. The key parameters of the four CCSNe and FRBs are listed in Tables~\ref{tab:sn_properties} and \ref{tab:frb_properties}, respectively. Figure~\ref{fig:sn_images} presents the Panoramic Survey Telescope \& Rapid Response System \citep[Pan-STARRS;][]{chambers2016pan} $r$-band cutouts of the SN host galaxies while Figure~\ref{fig:frb_loc} shows the FRB localization contours with the supernova positions overlaid. Finally, stellar-population metrics from HECATE and morphology from Pan-STARRS cutouts suggest that the four SN hosts are star-forming spiral galaxies.

Next, we evaluate the likelihood of obtaining four or more matches by chance in two independent ways.  
(i)~\emph{Monte-Carlo sampling:} Keeping the 241 FRB positions fixed, we draw 886 artificial supernovae uniformly within the CHIME field of view ($\delta>-11^{\circ}$) and repeat the cross-match 10\,000 times. At least four coincidences occur in 96.6 \% of the trials, yielding $P_{\mathrm{MC}}=0.966$. (ii)~\emph{Analytic estimate:}  Treating the 886 CCSNe as a uniform surface density over the CHIME field of view
(\(A_{0}=20,600\ \mathrm{deg^{2}}\)), and noting that the combined
90\% localization footprints of the 241 bursts subtend
\(A_{1}=197\ \mathrm{deg^{2}}\), the expected number of coincidental
FRB–SN overlaps is $\lambda = N A_{1}/A_{0} =(886)(197)/(20,600)\simeq8.47.$ Assuming Poisson statistics, the chance of observing four or more coincidences is $P(k\ge4)=1-\sum_{k=0}^{3}e^{-\lambda}\lambda^{k}/k!\approx0.97,$ consistent with the Monte-Carlo result. This is consistent with the Monte-Carlo sampling result. We therefore conclude that the four positional overlaps are fully compatible with random expectation. Hence, no statistically significant FRB–SN association is found.

Although the global statistics argue against a true association, individual pairs can be examined for physical consistency.  For each of the four bursts we used the Catalog-1 scattering time $\tau$ to place an upper limit on the host-galaxy DM, $\mathrm{DM}_{\mathrm{host},\tau}$, following the formalism by \cite{2022ApJ...931...88C} as discussed in Appendix~\ref{app:host_dm_scattering}. Combining this with the priors on the Milky-Way halo (50 pc cm$^{-3}$) and the IGM contribution at the supernova redshift ($\le40$ pc cm$^{-3}$ as delineated in Appendix~\ref{app:max-dmexcess}) yields the criterion,
\[
\mathrm{DM}_{\mathrm{excess}}
      \;\lesssim\;
      \mathrm{DM}_{\mathrm{MW,halo}}
      +\mathrm{DM}_{\mathrm{IGM}}
      +\mathrm{DM}_{\mathrm{host},\tau,\max}.
\]
Only FRB 20190412B satisfies this inequality if SN 2009gi is assumed to be associated with the FRB. Moreover, the FRB’s low $\mathrm{DM}_{\mathrm{excess}}$ is consistent with the supernova’s large projected offset of 6.3 kpc ($\simeq2.9\,r_e$) which imply a relatively less dense local environment based on the MW disk model \citep{Yao2017}.  
The other three pairs do not satisfy the aforementioned criterion and are therefore unlikely to be physically related. Hence, while four FRB–SN positional coincidences are expected by chance and show no global significance, FRB 20190412B–SN 2009gi survives the scattering time and DM-budget consistency checks. This pair thus remains the sole viable candidate and merits targeted follow-up to establish or refute a physical connection.

\begin{table}
\centering
\caption{Key properties of the four CCSNe spatially coincident with the selected CHIME/FRB bursts.}
\begin{tabular}{cccccccccc}
\toprule
\shortstack{Pair \\ Number} &
\shortstack{SN \\ $ $} &
\shortstack{Host Galaxy \\ $ $} &
\shortstack{$z$ \\ $ $} &
\shortstack{Type \\ $ $} &
\shortstack{RA \\ (deg)} &
\shortstack{Dec \\ (deg)} &
\shortstack{CHIME Exposure$^{a}$ \\ T$_{\mathrm{exp}}$ (hrs)} &
\shortstack{Host $r_{e}$ \\ (kpc)} &
\shortstack{Proj. SN host offset \\ (kpc)} \\
\midrule
1 & 2003la & MCG+10-15-089 & 0.031 & II & 157.5842 & 61.2635 & 21.5 & 4.4 & 3.3 \\
2 & 2014ay & UGC11037 & 0.011 & II & 268.7726 & 18.2574 & 15.6 & 2.5 & 0\\
3 & 2009gi & PGC1596871 & 0.013 & IIb & 287.5796 & 19.5596 & 14.4 & 2.2 & 6.3 \\
4 & 2001ab & NGC6130 & 0.017 & II & 244.8951 & 57.6182 & 29.6 & 6.3 & 5.2 \\
\bottomrule
\end{tabular}
\smallskip
\footnotesize%
$^{a}$ Total CHIME/FRB exposure at each source’s upper transit during the Catalog-1 period.
\label{tab:sn_properties}
\end{table}

\begin{table}[ht]
\centering
\caption{Key parameters of the four CHIME/FRB bursts spatially coincident with historic CCSNe.}
\makebox[\textwidth][c]{
\resizebox{\textwidth}{!}{
\begin{tabular}{cccccccccc}
\toprule
\shortstack{Pair \\ Number} &
\shortstack{FRB \\ $ $ } &
\shortstack{RA \\ (deg)} &
\shortstack{Dec \\ (deg)} &
\shortstack{DM \\ (pc cm$^{-3}$)} &
\shortstack{DM$_{\mathrm{excess,NE2001}}$ \\ (pc cm$^{-3}$)} &
\shortstack{DM$_{\mathrm{excess,YMW16}}$ \\ (pc cm$^{-3}$)} &
\shortstack{DM$_{\mathrm{Host,\tau,max}}^{\,a}$ \\ (pc cm$^{-3}$)} &
\shortstack{$\tau$ \\ (s)} &
\shortstack{Temp.\ offset$^{\,b}$ \\ (months)} \\
\midrule
1 & 20190204A & 161.3 & 61.53 & 449.6 & 414 & 423 &  49 & 0.0008(2) & 183 \\
2 & 20190218B & 268.7 & 17.93 & 547.9 & 466 & 483 & 200 & 0.014(2) &  56 \\
3 & 20190412B & 285.7 & 19.25 & 375.8 & 111 & 110 & 196 & 0.015(3) & 116 \\
4 & 20190414B & 246.9 & 57.53 & 506.5 & 469 & 476 & 127 & $<0.0058$ & 155 \\
\bottomrule
\end{tabular}}}

\smallskip
\footnotesize%
$^{a}$\,95\% credible upper limit on $\mathrm{DM}_{\rm host}$ inferred from the scattering time-scale (Appendix \ref{app:host_dm_scattering}).\;
$^{b}$\,Temporal offset between the FRB detection date and the reported supernova explosion date.
\label{tab:frb_properties}
\end{table}

\subsection{Search for compact radio emission at nearby SN sites}\label{sec:frb_prs}

Several hyperactive repeaters—most notably FRB 20121102A, FRB 20190520B and FRB 20201124A—are co-located with compact, persistent radio sources interpreted as $\lesssim100$ yr magnetar wind nebulae \citep{2024arXiv241219358B,2025arXiv250401125M}.  Such nebulae provide a direct probe of young FRB engines.  We therefore inspected the positions of all 886 CCSNe in our CHIME-visible sample in the multi-epoch mosaics of the VLA Sky Survey \citep[VLASS;][]{2020PASP..132c5001L}, which reach a median rms of $\simeq120\;\mu$Jy beam$^{-1}$ at 2–4 GHz. Note that VLASS provides the highest spatial resolution of $\approx 2.6$ arcsec (or $\approx 2.3$ kpc at $z = 0.043$) among all existing wide-sky radio surveys to date.

No compact VLASS source, persistent or transient, is detected at the positions of the four CCSNe that overlap CHIME localizations (\S\ref{sec:cross-match}).  For the most promising candidate, FRB 20190412B–SN 2009gi (at z=0.013), the 3 GHz map provides a \(3\sigma\) flux-density limit of \(\approx0.35\) mJy, which corresponds to \(L_{\nu}<2\times10^{27}\ \mathrm{erg\,s^{-1}\,Hz^{-1}}\).  This limit rules out a compact radio source, as seen in the case of several FRBs \citep{2017Natur.541...58C,2022Natur.606..873N, 2024Natur.632.1014B,2024arXiv241213121B, 2025A&A...695L..12B}, at the SN location.  

We also identified about a dozen other CCSNe, with no spatial overlap with our FRB sample, that show compact radio emission in the VLASS. A comprehensive study of these radio‐bright supernovae, including their spectral properties and variability, lies beyond the scope of this paper and will be presented elsewhere.

\section{Discussion} \label{sec:discussion}

\subsection{Free-free optical depth constraints for the detection of FRBs in the CHIME band} \label{sec:free-free_constraint}

The dense SN ejecta in which the putative FRB source is embedded can have a significant impact on the observability of FRBs \citep[e.g.,][]{Piro2016,Piro&Gaensler18,Margalit+18_CLOUDY}. In particular, low-frequency radio waves will suffer free-free absorption while propagating through this medium. As the SN ejecta expands, the density drops and the impact of free-free absorption becomes less significant. We can define a free-free transparency timescale $t_{\rm ff}$ as the time at which the free-free optical depth drops to $\tau_{\rm ff} = 1$. At earlier times, radio waves will be absorbed by the ejecta ($\tau_{\rm ff} \gg 1$) while at later times, radio emission can escape the ejecta uninhibited by free-free absorption.
The free-free transparency timescale can be estimated as
\begin{equation}
\label{eq:t_ff}
    t_{\rm ff} \approx 77.6\,{\rm yr}\,\left(\frac{\nu}{600\,{\rm MHz}}\right)^{-0.42} \left(\frac{E}{10^{51}\,{\rm erg}}\right)^{-1/2} \left(\frac{M_{\rm ej}}{10M_\odot}\right)^{9/10} \left(\frac{T_e}{10^4\,{\rm K}}\right)^{-0.26} \left(\frac{f_{\rm ion}}{0.1}\right)^{2/5} ,
\end{equation}
where $\nu$ is the frequency of interest (chosen here to be the center of the CHIME band), $E$ is the SN explosion energy, $M_{\rm ej}$ is the SN ejecta mass, and $T_e$ and $f_{\rm ion}$ are the appropriately averaged SN ejecta temperature and ionization fraction, respectively \citep[e.g.,][]{Margalit+18_CLOUDY}.
This simplified estimate assumes a homogenous ejecta of density $n \propto M_{\rm ej} / (v_{\rm ej} t )^3$ expanding at a constant velocity $v_{\rm ej} \propto \sqrt{E/M_{\rm ej}}$. 
For stripped envelope SNe such as Type Ib and Ic the characteristic ejecta mass is $M_{\rm ej} \sim 3 M_\odot$, which leads to an expected free-free transparency timescale of $t_{\rm ff}({\rm Ibc}) \sim 26\,{\rm yr}$. On the other hand, common subtypes of hydrogen-rich SNe such as Type IIP are thought to have characteristic ejecta masses of $M_{\rm ej} \sim 12 M_\odot$. This implies a free-free transparency timescale of $t_{\rm ff}({\rm IIP}) \sim 91\,{\rm yr}$.

However, the estimate of $t_{\rm ff}$ should be regarded as indicative rather than exact, since substantial uncertainties can alter its value. In particular, Rayleigh–Taylor instabilities and other inhomogeneities in the ejecta can significantly modify the transparency timescale. For example, in a toy-model scenario where the ejecta fragments into a large number of clumps and the density of each clump is enhanced by a factor $\delta > 1$ compared to the volume-averaged density $n$, the free-free optical depth will---on average---be {\it enhanced} by a similar factor such that $\tau_{\rm ff} \propto \delta$.\footnote{
The free-free optical depth scales as $\tau_{\rm ff} \propto \delta$ even though $\tau_{\rm ff}$ depends on the density squared because the path-length that intersects clumps is $< R$. If $N_{\rm clump}$ is the number of clumps and $\delta = n_{\rm clump} / n$ is the density enhancement of each clump, then the characteristic size of a clump is $\ell_{\rm clump} \sim R N_{\rm clump}^{-1/3} \delta^{-1/3}$ and the number of clumps intersected along a typical sight-line is $N_{\rm int} \propto \ell_{\rm clump}^2 (N_{\rm clump}/R^3) R \sim N_{\rm clump}^{1/3} \delta^{-2/3}$. The free-free optical depth is then $\tau_{\rm ff} \propto \int n^2 dr \sim N_{\rm int} n_{\rm clump}^2 \ell_{\rm clump} \sim n^2 R \delta \propto \delta$
(where we have assumed a fixed temperature $T_e$).
} 
The free-free transparency timescale will therefore increase as $t_{\rm ff} \propto \delta^{1/5}$ for a typical sight-line. Moreover, sight‐line‐to‐sight‐line variation is expected unless the number of clumps is extremely large. As a result, $t_{\rm ff}$ for an individual supernova may differ markedly from the homogeneous‐ejecta value given in Equation \ref{eq:t_ff}.

Furthermore, both the ejecta temperature $T_e$ and the ionization fraction $f_{\rm ion}$ are poorly constrained and evolve over time. \cite{Margalit+18_CLOUDY} performed detailed radiative‐transfer calculations—including a central ionizing source such as a millisecond magnetar—to model these quantities as functions of radius. While their fiducial values agree approximately with Equation \ref{eq:t_ff}, actual supernova conditions (e.g., composition and shock evolution) can differ. Similarly, \cite{Piro2016} and \cite{Piro&Gaensler18} estimated $T_e$ and $f_{\rm ion}$ by modeling forward and reverse shocks in the ejecta and circumstellar medium, without any central engine. Local supernova remnants such as SN 1986J \citep{Bietenholz&Bartel17} also exhibit a range of behaviors, some consistent with Equation \ref{eq:t_ff} and others indicating faster or slower clearing.

Given these uncertainties—including clumping, temperature evolution, ionization state, shock physics, and composition—Equation \ref{eq:t_ff} may differ from the true free–free transparency timescale by factors of several. Consequently, although the temporal offsets in Table \ref{tab:frb_properties} are substantially shorter than the nominal $t_{\rm ff}$ values, the candidate SN–FRB associations identified here—particularly FRB 20190412B–SN 2009gi—remain plausible and warrant follow‐up observations to confirm or refute a physical connection. 

More importantly, if FRBs are detected from historical SNe on timescales as short as $\sim$several years post-explosion, this would have implications for the environment and/or explosion properties of the progenitor star. Such short free-free transparency times would imply that the FRB propagates along an especially dilute path out of the SN ejecta. This could be explained either by small-scale inhomogeneities in the ejecta (e.g., due to clumping as discussed above) or by a large-scale anisotropy in the explosion or the surrounding circumstellar material. Such anisotropies are observed in many local-Universe SN-remnants, which often show a bipolar or equatorial structure. Alternatively, a short free-free transparency timescale may potentially be explained by a particularly luminous central source of ionizing radiation if such ionizing radiation can Compton-heat the surrounding SN ejecta to very high temperatures $\gg 10^4\,{\rm K}$. Of course, rare types of core-collapse explosions may also potentially lead to short $t_{\rm ff}$. 
For example, ultra-stripped SNe SNe where the ejecta mass is very low ($\lesssim1\,M_\odot$) would be associated with short free-free transparency timescales.
Additionally, Broad-lined Type Ic SNe and superluminous SNe are associated with large explosion energies ($E_{\rm ej} \sim 10^{52}\,{\rm erg}$) which would shorten $t_{\rm ff}$. Furthermore, if such events are accompanied by relativistic jets that successfully burrow out of the surrounding SN ejecta, these jets could evacuate the polar regions of the ejecta and potentially allow FRBs to escape in these directions uninhibited by free-free absorption. Any of the potential scenarios mentioned above would be associated with testable predictions that would have to be verified in the event that FRBs were observed to occur shortly ($\lesssim$ several years) after a CCSN.


\subsection{Burst-rate limits and comparison with active repeaters}\label{sec:burst_limits}

If the four FRB–SN overlaps in \S\ref{sec:cross-match} are chance
alignments, each of the 886 CCSNe in our volume-limited sample
represents a line of sight along which no FRB activity was detected
during CHIME/FRB Catalog-1 duration.  For every supernova we extracted the cumulative CHIME exposure time, \(T_{\mathrm{exp}}\), accumulated in
the more sensitive upper transit between 25 July 2018 and 1 July 2019
(Table \ref{tab:sn_properties}; Figure~\ref{fig:sne_frb_summary}).
Using the mean Catalog-1 completeness limit of
\(F_{\mathrm{th}}=5\) Jy ms at 600 MHz \citep{2021ApJS..257...59C},
the 1 $\sigma$ Poisson upper limit on the burst rate is $R_{\mathrm{th}}
    =1.84/T_{\mathrm{exp}}\ \mathrm{hr^{-1}}$ \citep{1986ApJ...303..336G}.

\begin{figure}[ht]
\begin{center}
\includegraphics[width=0.6\linewidth]
{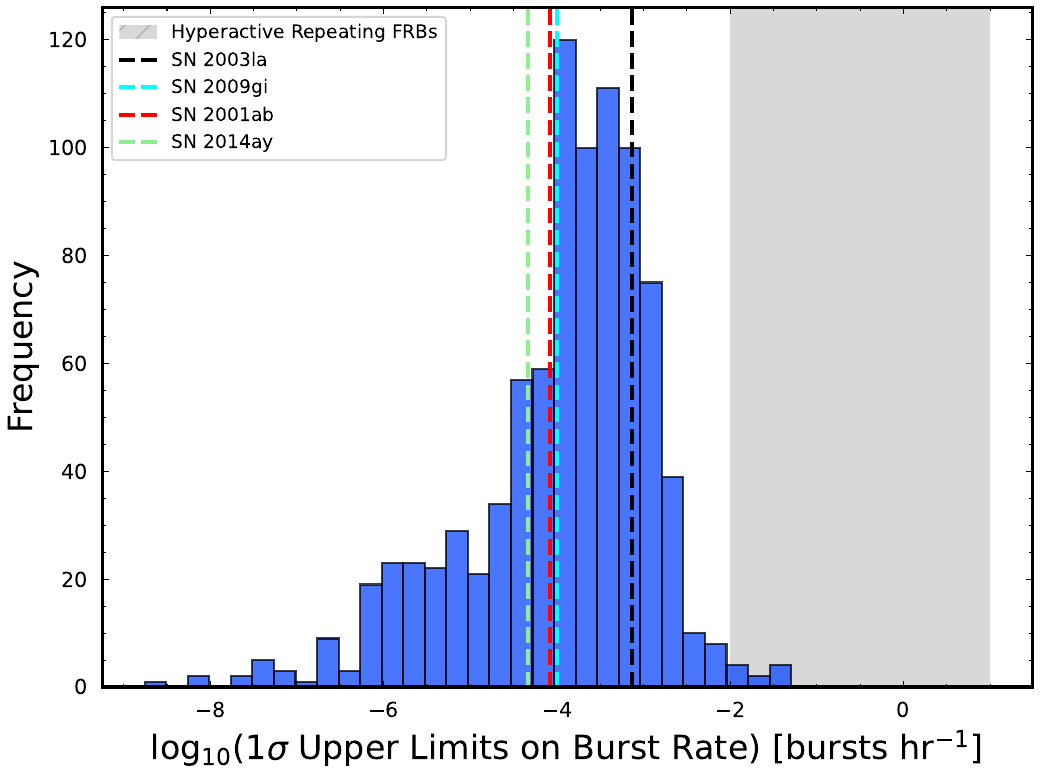}
\end{center}
\caption{Histogram of 1$\sigma$ Poisson upper limits on the burst rate for the 886 CCSN sites, scaled to a fiducial isotropic energy of \(10^{39}\) erg (see \S\ref{sec:burst_limits}).  Limits are
computed from the CHIME/FRB exposure stated in Table \ref{tab:sn_properties} at each position and converted using a cumulative energy distribution \(N(>E)\propto E^{-1.5}\).
The vertical grey band marks the range 0.01–10 h\(^{-1}\), characteristic of the most active repeaters (see \S\ref{sec:burst_limits}); all supernova sites lie below this level. The four colored, vertical dashed lines mark the SNe that coincide spatially with CHIME/FRB localizations.}
\label{fig:burst_rate}
\end{figure}

Because a single fluence threshold corresponds to different isotropic
energies at different distances, we rescale each limit to the fiducial
energy \(10^{39}\) erg using the redshifts of 886 SNe in our sample.  Adopting the cumulative energy distribution
\(N(>E)\propto E^{-1.5}\)
\citep{2023ApJ...947...83C,2024ApJ...971L..51B}, we obtain
\[
R_{39}=R_{\mathrm{th}}
        \left(\frac{E_{\mathrm{th}}}{10^{39}\,\mathrm{erg}}\right)^{1.5},
\]
where \(E_{\mathrm{th}}=4\pi D_{L}^{2}F_{\mathrm{th}}\Delta\nu\) with $\Delta\nu$ = 400 MHz.
The resulting distribution peaks near
\(10^{-3.5}\ \mathrm{h^{-1}}\) (Figure~\ref{fig:burst_rate}). We note that our rate limits assume the FRB burst‐energy distribution extends without an intrinsic low-energy cutoff; introducing such a cutoff would change the scaled limits and alter the comparative rates among the supernova sample.

Hyperactive CHIME repeaters—FRB 20180916B, FRB 20201124A,
FRB 20220912A, and FRB 20240114A
\citep{2020Natur.582..351C,2022ApJ...927...59L,2022ATel15679....1M,2025arXiv250513297S}—reach
episodic burst rates of \(10^{-2}\)–\(10\ \mathrm{h^{-1}}\) at the same
fiducial energy and are thought to be magnetars younger than
\(\sim100\) yr
\citep{2024ApJ...967L..44L,2024arXiv241106996X,2024arXiv241219358B}.
If equally active sources existed at any CCSN site and their 600 MHz
emission was not absorbed, CHIME should have recorded multiple bursts.
The non-detections therefore imply that such hyperactive repeaters are
rare among ordinary CCSN remnants.

Three caveats moderate this conclusion.  
(i) FRB activity is intermittent; long quiescent intervals can lower
the time-averaged detection probability.  
(ii) FRB emission may be strongly beamed; a solid angle
\(\Omega\lesssim0.5\) sr, combined with a $\sim$20 percent magnetar
fraction among CCSNe \citep{2016ApJ...821...38S,2019MNRAS.487.1426B,2021MNRAS.508..516N,2021Natur.589...29B,2025arXiv250311875S}, could hide many active FRB sources from our sight-line.
(iii) Dense or clumpy ejecta may still absorb 600 MHz radiation even
decades after explosion. 

Longer monitoring with CHIME/FRB Catalog 2, targeted observations at
higher radio frequencies, and improved beaming statistics will
quantify these effects and tighten burst rate limits at CCSN sites.

\subsection{Prospects for associating FRBs with historic supernovae}\label{sec:frb_sn_prospects}

The present study tests for purely two–dimensional overlaps between CHIME/FRB Localizations and historic core-collapse supernovae.  With localization radii of tens of arcminutes the chance-coincidence probability, $P_{\mathrm{cc}}$, is correspondingly high: for the Catalog-1 burst sample a circular error of $30'$ yields $P_{\mathrm{cc}}\simeq0.3$ per burst (Appendix~\ref{app:pcc}).  Reducing the radius to $\sim10\arcsec$ lowers $P_{\mathrm{cc}}$ to the percent level, yet even at that scale misleading associations can arise.  A cautionary case is FRB 20250316A, initially thought to be possibly linked to SNe 2008X and 2009E in NGC 4141 \citep{2025ATel17081....1N,Boles2008,Madison2008,Pastorello2011}, but a subsequent $\gtrsim200$ pc very-long-baseline (VLBI) position from the CHIME/FRB Outriggers excluded both events as progenitors \citep{2025ATel17086....1L}.  Robust FRB–SN associations therefore require sub-arcsecond accuracy for both transients. Moreover, such precision is becoming routine.  Most cataloged SNe already have $\lesssim0\farcs3$ astrometry and spectroscopic redshifts, while next-generation arrays $-$ ASKAP/CRACO, BURSTT, CHIME/FRB Outriggers, CHORD, DSA-2000, HIREX, and SKA \citep{2025PASA...42....5W,2022PASP..134i4106L,2025arXiv250405192F,2019clrp.2020...28V,2019BAAS...51g.255H,2016SPIE.9906E..5XN,2015aska.confE..55M} $-$ are expected to deliver comparable accuracy for hundreds of FRBs per year.  Once sub-arcsecond positions are available, spatial coincidence can be strengthened by demanding redshift concordance or by weighting offsets with the host’s supernova surface-density profile.  Archival incompleteness, however, remains a limitation: optical surveys miss many core-collapse explosions older than $\sim$30 yr, so the absence of a recorded SN near an FRB is not definitive.  Expanding the search to radio or X-ray supernova-remnant catalogs \citep[e.g.][]{Bozzetto2017,Green2019} can extend the look-back window to $10^{3}$–$10^{4}$ yr, albeit at the cost of coarser astrometry.

A qualitatively different regime arises when VLBI places an FRB within $\lesssim10$ pc of a recently characterized SN.  In that case, the two-dimensional probability of chance falls below $10^{-6}$, effectively demanding a physical link.  At face value, such youth conflicts with models in which supernova ejecta remain opaque to $\lesssim1$ GHz radiation for $\gtrsim50$ yr \citep{Piro2016,Piro&Gaensler18,Margalit+18_CLOUDY}.  Several scenarios can relieve the tension: low-ejecta-mass Type Ib/c explosions, bipolar SNe that carve polar channels, or rapid clearing by a magnetar-driven wind (see \S\ref{sec:free-free_constraint}).  Distinguishing among them requires combining VLBI co-location with broadband follow-up and late-time light-curve modeling.  Only when these diagnostics are satisfied should an FRB be considered firmly associated with a particular historical supernova.

\section{Constraints on Young Neutron Star Source Models from the FRB Rates}
\label{sec:magnetar_constraints}

In the young‐magnetar framework, FRBs are powered by highly magnetized neutron stars born in the aftermath of CCSNe \citep[e.g.,][]{2014MNRAS.442L...9L,2017ApJ...843L..26B,2017MNRAS.468.2726K,Metzger2019}.  As a magnetar spins down and its magnetospheric activity wanes, its burst rate is expected to decay roughly as \(r(t)\propto t^{-\beta}\) \citep[e.g.,][]{Margalit+20_SGR1935}.  At the same time, the expanding supernova ejecta impose a frequency‐dependent “fog” of free–free opacity: bursts at meter wavelengths remain trapped until the ejecta become transparent \citep[e.g.,][]{Piro2016,Margalit+18_CLOUDY,2024ApJ...977..122B} on a timescale \(t_{\rm ff}\propto\nu^{-0.42}\) from Equation~\ref{eq:t_ff}, so that low‐frequency surveys preferentially miss the youngest, most active sources (see \S\ref{sec:free-free_constraint}).  This interplay between intrinsic aging and propagation $-$ secular decline versus spectral clearing $-$ predicts a measurable difference in the all‐sky FRB rates reported by low‐ and high‐frequency instruments. CHIME/FRB and CRAFT surveys thus provide precisely such complementary datasets.  CHIME reports an all‐sky rate of $[525\pm30\ (\mathrm{stat}\;1\sigma)]\,^{+142}_{-131}\ (\mathrm{sys})\;\mathrm{sky^{-1}\,day^{-1}}$
above 5 Jy ms at 600 MHz \citep{2023ApJS..264...53C}, while ASKAP measures $37\pm8\;\mathrm{sky^{-1}\,day^{-1}}$
above 26 Jy ms at 1.4 GHz \citep{2018Natur.562..386S}.  Scaling the CHIME rate to the common 26 Jy ms threshold using the observed fluence index \(\alpha=-1.4\) \citep{2023ApJS..264...53C}, we find $\frac{R(1.4\,\mathrm{GHz})}{R(0.6\,\mathrm{GHz})}
   =\bigl[0.71\pm0.16\ (\mathrm{stat}\;1\sigma)\bigr]\,^{+0.24}_{-0.15}\ (\mathrm{sys})\,.$

Moreover, it has been suggested that the intrinsic burst fluence itself decreases with frequency  characterize by fitting a power law of the form F$_{\nu} \propto \nu^{-\Gamma}$, with $\Gamma$ in the range [1,3]. \citep{2018Natur.562..386S,2019ApJ...872L..19M,2025ApJ...982..158C}. However, we note that individual bursts from several repeating sources are known to show stochastic intrinsic spectral features which varies at a range of timescales \citep{2016Natur.531..202S}.
To translate the rate‐ratio measurement reported above into constraints on the magnetar parameters, we numerically evaluate the frequency‐dependent, galaxy‐integrated FRB rate ratio Equation \ref{equation:ratio} derived in Appendix~\ref{sec:rate_ratio_derivation}.  Figure~\ref{fig:ratio} shows, as a shaded band, the region in the \((\Gamma,\beta)\) plane satisfying the above ratio for an assumed active lifetime \(T_{\rm active}=10^5\) yr. Overlaid are the two \(\Gamma\) measurements: \citet{2023ApJ...944..105S} find \(\Gamma=1.39^{+1.19}_{-0.86}\), which is too broad to yield a meaningful bound on \(\beta\), whereas the tighter measurement \(\Gamma=2.29\pm0.29\) from \citet{2025ApJ...982..158C} intersects our band only for $1.5 \lesssim \beta \lesssim 8.5$. 
Such steep decay indices imply a rapid secular decline in burst rate, which also includes a simple dipole‐spin‐down expectation (\(\beta\sim2\)) and may point to evolving magnetospheric processes or to an age‐dependent burst–energy distribution suggesting more active FRB sources are likely younger ones compare to non-repeating ones after accounting for various observational biases. However, flat Future refinement of the spectral‐index \(\Gamma\) after properly accounting for propagation effects, and extension of this comparison to still higher frequencies, will further elucidate the physical mechanism of young FRB sources.  

\begin{figure}[ht]
\centering
\includegraphics[width=0.8\columnwidth]{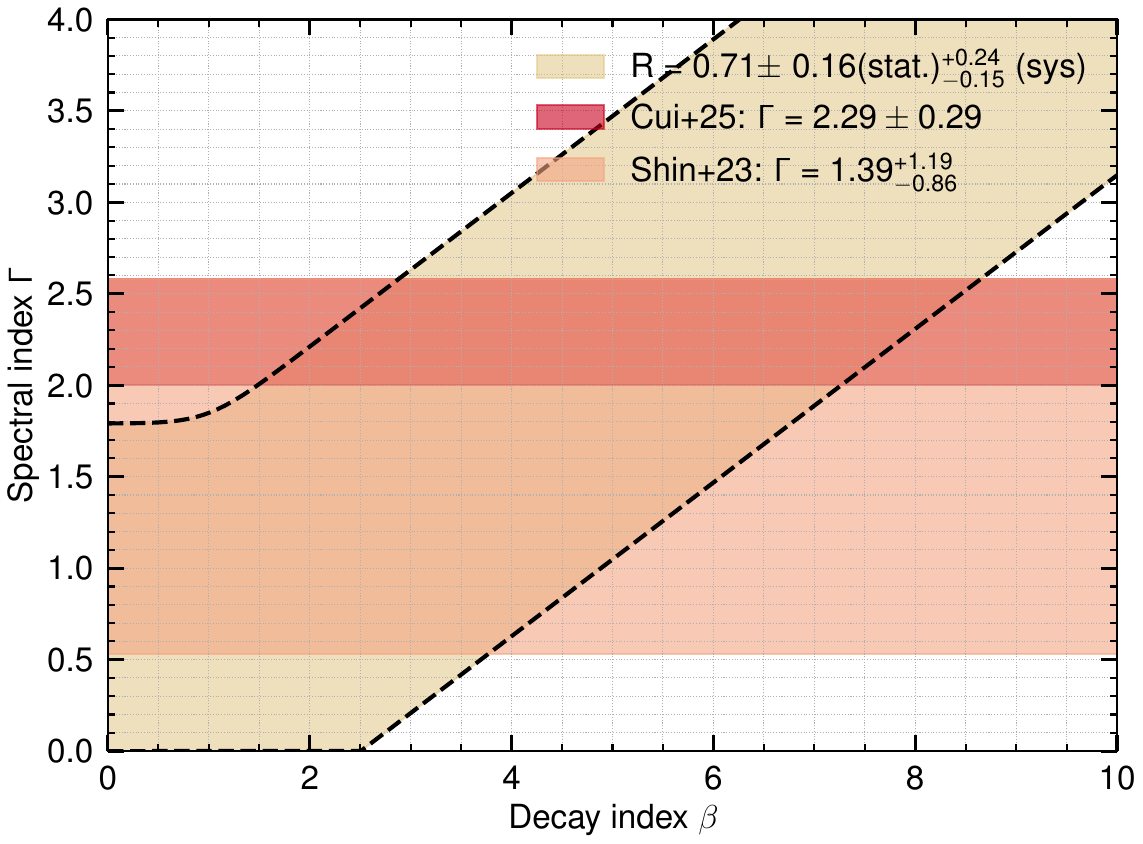}
\caption{Constraints on the intrinsic spectral index $\Gamma$ and burst‐rate decay index $\beta$ of young magnetars, derived by matching the observed all‐sky FRB rate ratio \(R(1.4\,\mathrm{GHz})/R(0.6\,\mathrm{GHz})=0.71\pm0.16\,({\rm stat})\,^{+0.24}_{-0.15}\,({\rm sys})\).  The shaded band marks the region of \((\Gamma,\beta)\) that reproduces this ratio for an assumed active lifetime \(T_{\rm active}=10^5\) yr. Vertical bands show independent \(\Gamma\) measurements: \(\Gamma=1.39^{+1.19}_{-0.86}\) from \cite{2023ApJ...944..105S} (blue) and \(\Gamma=2.29\pm0.29\) from \cite{2025ApJ...982..158C} (red). The intersection of the red band with the shaded region yields \(1.5\lesssim\beta\lesssim8.6\).}
\label{fig:ratio}
\end{figure}


\section{Conclusion and Future Work} 
\label{sec:conclusion}

We have performed the first systematic search for spatial coincidences between CCSNe within 200\,Mpc and CHIME/FRB Catalog-1 events, applying rigorous distance, DM, and localization cuts to minimize false positives. Out of 886 CCSNe and 241 apparently non‐repeating bursts, four positional overlaps are found—exactly the number expected by chance (\(P\gtrsim0.96\) by Monte Carlo and analytic estimates). Only one of these, FRB 20190412B and SN 2009gi, passes a self‐consistency test comparing its DM budget and scattering time; this pair remains the only candidate for a genuine FRB–SN association. Treating all 886 SN sight‐lines as non‐detections, we set 1\(\sigma\) Poisson upper limits on the burst rate of putative young FRB sources of \(\lesssim10^{-3.5}\,\mathrm{h^{-1}}\) (scaled to \(E\ge10^{39}\)\,erg). These limits lie well below the episodic rates of hyper‐active repeaters (\(10^{-2}\)–10\,h\(^{-1}\)), indicating that such extreme activity is uncommon among ordinary CCSN remnants unless suppressed by beaming, intermittency, or local absorption.

In parallel, we have developed a general framework that can provide constraint on the posposed young magnetar engine using the all‐sky FRB rate as a function of frequency, accounting for an intrinsic spectral slope \(\Gamma\), secular decay \(r\propto t^{-\beta}\), and free–free opacity \(\tau_{\rm ff}\propto\nu^{-2.1}t^{-5}\).  Comparing the observed rate ratio \(R(1.4\,\mathrm{GHz})/R(0.6\,\mathrm{GHz})=0.71\pm0.16_{\rm stat}{}^{+0.24}_{-0.15}{}_{\rm sys}\) yields meaningful constraints only when adopting tight spectral‐index measurements.  In particular, using \(\Gamma=2.29\pm0.29\) \citep{2025ApJ...982..158C} requires a steep decay index \(1.5\lesssim\beta\lesssim8.6\), suggestive of rapid magnetospheric evolution or energy‐cutoff effects in young magnetars.

Looking forward, decisive tests of the CCSN–magnetar FRB hypothesis will hinge on three advances:

1. \textit{Sub‐arcsecond localizations.}  VLBI follow‐up of CHIME/FRB Outrigger detections and next‐generation interferometers such as DSA-2000, CHORD, and SKA will routinely deliver \(\lesssim0.1''\) positions, driving chance‐coincidence probabilities below \(10^{-6}\) and enabling unambiguous FRB–SN matches.

2. \textit{Broadband and temporal monitoring.}  Multi‐frequency observations—from hundreds of MHz to several GHz—will chart the free–free transparency evolution of young SNRs, directly measuring \(t_{\rm ff}(\nu)\) and refining models of ejecta clumping and ionization.  Simultaneously, long‐baseline monitoring with CHIME/FRB and ASKAP’s CRACO, complemented by high‐frequency surveys, will improve burst‐rate statistics and constrain intermittency.

3. \textit{Expanded historic SN samples.}  Continued mining of archival optical, radio, and X-ray data (e.g. Rubin LSST, Roman Space Telescope, SKA1-Mid, ngVLA) will extend the catalog of well‐localized CCSNe and their remnants out to \(\sim200\)–400 Mpc and ages \(\gtrsim50\) yr.  This broader SN census, combined with precise FRB localizations, will power far more sensitive cross‐matching analyses. 

By integrating these capabilities with the analytical framework provided here, the coming decade promises either the first secure identification of an FRB with its natal supernova or stringent limits that will reshape our understanding of young magnetars and their radio‐burst activity.

\begin{acknowledgments}
We acknowledge that CHIME is located on the traditional, ancestral, and unceded territory of the Syilx/Okanagan people. CHIME is funded by a grant from the Canada Foundation for Innovation (CFI) 2012 Leading Edge Fund (Project 31170) and by contributions from the provinces of British Columbia, Qu\'ebec and Ontario. The CHIME/FRB Project is funded by a grant from the CFI 2015 Innovation Fund (Project 33213) and by contributions from the provinces of British Columbia and Qu\'ebec, and by the Dunlap Institute for Astronomy and Astrophysics at the University of Toronto. Additional support was provided by the Canadian Institute for Advanced Research (CIFAR), McGill University and the McGill Space Institute via the Trottier Family Foundation, and the University of British Columbia. The Pan-STARRS1 Surveys (PS1) and the PS1 public science archive have been made possible through contributions by the Institute for Astronomy, the University of Hawaii, the Pan-STARRS Project Office, the Max-Planck Society and its participating institutes, the Max Planck Institute for Astronomy, Heidelberg and the Max Planck Institute for Extraterrestrial Physics, Garching, The Johns Hopkins University, Durham University, the University of Edinburgh, the Queen's University Belfast, the Harvard-Smithsonian Center for Astrophysics, the Las Cumbres Observatory Global Telescope Network Incorporated, the National Central University of Taiwan, the Space Telescope Science Institute, the National Aeronautics and Space Administration under Grant No. NNX08AR22G issued through the Planetary Science Division of the NASA Science Mission Directorate, the National Science Foundation Grant No. AST-1238877, the University of Maryland, Eotvos Lorand University (ELTE), the Los Alamos National Laboratory, and the Gordon and Betty Moore Foundation. This work makes use of data from the VLA Sky Survey (VLASS), conducted with the Karl G. Jansky Very Large Array (VLA), which is operated by the National Radio Astronomy Observatory (NRAO). The NRAO is a facility of the National Science Foundation and is operated under a cooperative agreement by Associated Universities, Inc. 
M.B. is a McWilliams fellow and an International Astronomical Union Gruber fellow. M.B. also receives support from several McWilliams seed grants. 
\end{acknowledgments}

\vspace{5mm}
\software{
    Astropy \citep{2022ApJ...935..167A},
    Matplotlib \citep{2007CSE.....9...90H}, 
    Numpy \citep{harris2020array},
    Healpy \citep{zonca2020healpy},
    H5py \citep{collette2023python},
    Scienceplots \citep{scienceplots}
}

\facilities{CHIME\citep{2018ApJ...863...48C}, VLA \citep{1980ApJS...44..151T}, Pan-STARRS\citep{chambers2016pan}}

\bibliographystyle{aasjournal}
\bibliography{SNe_FRB.bib}

\begin{thebibliography}{}
\expandafter\ifx\csname natexlab\endcsname\relax\def\natexlab#1{#1}\fi
\providecommand{\url}[1]{\href{#1}{#1}}
\providecommand{\dodoi}[1]{doi:~\href{http://doi.org/#1}{\nolinkurl{#1}}}
\providecommand{\doeprint}[1]{\href{http://ascl.net/#1}{\nolinkurl{http://ascl.net/#1}}}
\providecommand{\doarXiv}[1]{\href{https://arxiv.org/abs/#1}{\nolinkurl{https://arxiv.org/abs/#1}}}

\bibitem[{{ CHIME/FRB Collaboration} {et~al.}(2025){ CHIME/FRB Collaboration}, {Amiri}, {Andersen}, {Andrew}, {Bandura}, {Bhardwaj}, {Bhopi}, {Bidula}, {Boyle}, {Brar}, {Carlson}, {Cassanelli}, {Cassity}, {Chatterjee}, {Cliche}, {Curtin}, {Darlinger}, {DeBoer}, {Dobbs}, {Dong}, {Eadie}, {Fonseca}, {Gaensler}, {Gusinskaia}, {Halpern}, {Hendricksen}, {Hessels}, {Joseph}, {Kaczmarek}, {Kaspi}, {Khairy}, {Landecker}, {Lanman}, {Kit Lau}, {Lazda}, {Leung}, {Main}, {Masui}, {Mckinven}, {Mena-Parra}, {Meyers}, {Michilli}, {Milutinovic}, {Nimmo}, {Noble}, {Pandhi}, {Pearlman}, {Peterson}, {Petroff}, {Pleunis}, {Pollak}, {Rafiei-Ravandi}, {Renard}, {Sammons}, {Sand}, {Sanghavi}, {Scholz}, {Shah}, {Shin}, {Siegel}, {Siemion}, {Sievers}, {Smith}, {Spear}, {Stairs}, {Vanderlinde}, {Wang}, {Willis}, \& {Zegmott}}]{2025arXiv250405192F}
{ CHIME/FRB Collaboration}, {Amiri}, M., {Andersen}, B.~C., {et~al.} 2025, arXiv e-prints, arXiv:2504.05192, \dodoi{10.48550/arXiv.2504.05192}

\bibitem[{{Akeson} {et~al.}(2019){Akeson}, {Armus}, {Bachelet}, {Bailey}, {Bartusek}, {Bellini}, {Benford}, {Bennett}, {Bhattacharya}, {Bohlin}, {Boyer}, {Bozza}, {Bryden}, {Calchi Novati}, {Carpenter}, {Casertano}, {Choi}, {Content}, {Dayal}, {Dressler}, {Dor{\'e}}, {Fall}, {Fan}, {Fang}, {Filippenko}, {Finkelstein}, {Foley}, {Furlanetto}, {Kalirai}, {Gaudi}, {Gilbert}, {Girard}, {Grady}, {Greene}, {Guhathakurta}, {Heinrich}, {Hemmati}, {Hendel}, {Henderson}, {Henning}, {Hirata}, {Ho}, {Huff}, {Hutter}, {Jansen}, {Jha}, {Johnson}, {Jones}, {Kasdin}, {Kelly}, {Kirshner}, {Koekemoer}, {Kruk}, {Lewis}, {Macintosh}, {Madau}, {Malhotra}, {Mandel}, {Massara}, {Masters}, {McEnery}, {McQuinn}, {Melchior}, {Melton}, {Mennesson}, {Peeples}, {Penny}, {Perlmutter}, {Pisani}, {Plazas}, {Poleski}, {Postman}, {Ranc}, {Rauscher}, {Rest}, {Roberge}, {Robertson}, {Rodney}, {Rhoads}, {Rhodes}, {Ryan}, {Sahu}, {Sand}, {Scolnic}, {Seth}, {Shvartzvald}, {Siellez}, {Smith}, {Spergel}, {Stassun}, {Street}, {Strolger}, {Szalay},
  {Trauger}, {Troxel}, {Turnbull}, {van der Marel}, {von der Linden}, {Wang}, {Weinberg}, {Williams}, {Windhorst}, {Wollack}, {Wu}, {Yee}, \& {Zimmerman}}]{2019arXiv190205569A}
{Akeson}, R., {Armus}, L., {Bachelet}, E., {et~al.} 2019, arXiv e-prints, arXiv:1902.05569, \dodoi{10.48550/arXiv.1902.05569}

\bibitem[{{Arcodia} {et~al.}(2024){Arcodia}, {Bauer}, {Cenko}, {Dage}, {Haggard}, {Ho}, {Kara}, {Koss}, {Liu}, {Mallick}, {Negro}, {Pradhan}, {Quirola-V{\'a}squez}, {Reynolds}, {Ricci}, {Rothschild}, {Sridhar}, {Troja}, \& {Yao}}]{2024Univ...10..316A}
{Arcodia}, R., {Bauer}, F.~E., {Cenko}, S.~B., {et~al.} 2024, Universe, 10, 316, \dodoi{10.3390/universe10080316}

\bibitem[{{Astropy Collaboration} {et~al.}(2022){Astropy Collaboration}, {Price-Whelan}, {Lim}, {Earl}, {Starkman}, {Bradley}, {Shupe}, {Patil}, {Corrales}, {Brasseur}, {N{\"o}the}, {Donath}, {Tollerud}, {Morris}, {Ginsburg}, {Vaher}, {Weaver}, {Tocknell}, {Jamieson}, {van Kerkwijk}, {Robitaille}, {Merry}, {Bachetti}, {G{\"u}nther}, {Aldcroft}, {Alvarado-Montes}, {Archibald}, {B{\'o}di}, {Bapat}, {Barentsen}, {Baz{\'a}n}, {Biswas}, {Boquien}, {Burke}, {Cara}, {Cara}, {Conroy}, {Conseil}, {Craig}, {Cross}, {Cruz}, {D'Eugenio}, {Dencheva}, {Devillepoix}, {Dietrich}, {Eigenbrot}, {Erben}, {Ferreira}, {Foreman-Mackey}, {Fox}, {Freij}, {Garg}, {Geda}, {Glattly}, {Gondhalekar}, {Gordon}, {Grant}, {Greenfield}, {Groener}, {Guest}, {Gurovich}, {Handberg}, {Hart}, {Hatfield-Dodds}, {Homeier}, {Hosseinzadeh}, {Jenness}, {Jones}, {Joseph}, {Kalmbach}, {Karamehmetoglu}, {Ka{\l}uszy{\'n}ski}, {Kelley}, {Kern}, {Kerzendorf}, {Koch}, {Kulumani}, {Lee}, {Ly}, {Ma}, {MacBride}, {Maljaars}, {Muna}, {Murphy}, {Norman},
  {O'Steen}, {Oman}, {Pacifici}, {Pascual}, {Pascual-Granado}, {Patil}, {Perren}, {Pickering}, {Rastogi}, {Roulston}, {Ryan}, {Rykoff}, {Sabater}, {Sakurikar}, {Salgado}, {Sanghi}, {Saunders}, {Savchenko}, {Schwardt}, {Seifert-Eckert}, {Shih}, {Jain}, {Shukla}, {Sick}, {Simpson}, {Singanamalla}, {Singer}, {Singhal}, {Sinha}, {Sip{\H{o}}cz}, {Spitler}, {Stansby}, {Streicher}, {{\v{S}}umak}, {Swinbank}, {Taranu}, {Tewary}, {Tremblay}, {de Val-Borro}, {Van Kooten}, {Vasovi{\'c}}, {Verma}, {de Miranda Cardoso}, {Williams}, {Wilson}, {Winkel}, {Wood-Vasey}, {Xue}, {Yoachim}, {Zhang}, {Zonca}, \& {Astropy Project Contributors}}]{2022ApJ...935..167A}
{Astropy Collaboration}, {Price-Whelan}, A.~M., {Lim}, P.~L., {et~al.} 2022, \apj, 935, 167, \dodoi{10.3847/1538-4357/ac7c74}

\bibitem[{{Beloborodov}(2017)}]{2017ApJ...843L..26B}
{Beloborodov}, A.~M. 2017, \apjl, 843, L26, \dodoi{10.3847/2041-8213/aa78f3}

\bibitem[{{Beloborodov}(2020)}]{2020ApJ...896..142B}
---. 2020, \apj, 896, 142, \dodoi{10.3847/1538-4357/ab83eb}

\bibitem[{{Beniamini} {et~al.}(2019){Beniamini}, {Hotokezaka}, {van der Horst}, \& {Kouveliotou}}]{2019MNRAS.487.1426B}
{Beniamini}, P., {Hotokezaka}, K., {van der Horst}, A., \& {Kouveliotou}, C. 2019, \mnras, 487, 1426, \dodoi{10.1093/mnras/stz1391}

\bibitem[{{Bhandari} {et~al.}(2022){Bhandari}, {Heintz}, {Aggarwal}, {Marnoch}, {Day}, {Sydnor}, {Burke-Spolaor}, {Law}, {Xavier Prochaska}, {Tejos}, {Bannister}, {Butler}, {Deller}, {Ekers}, {Flynn}, {Fong}, {James}, {Lazio}, {Luo}, {Mahony}, {Ryder}, {Sadler}, {Shannon}, {Han}, {Lee}, \& {Zhang}}]{2022AJ....163...69B}
{Bhandari}, S., {Heintz}, K.~E., {Aggarwal}, K., {et~al.} 2022, \aj, 163, 69, \dodoi{10.3847/1538-3881/ac3aec}

\bibitem[{{Bhardwaj} {et~al.}(2024{\natexlab{a}}){Bhardwaj}, {Lee}, \& {Ji}}]{2024Natur.634.1065B}
{Bhardwaj}, M., {Lee}, J., \& {Ji}, K. 2024{\natexlab{a}}, \nat, 634, 1065, \dodoi{10.1038/s41586-024-08065-w}

\bibitem[{{Bhardwaj} {et~al.}(2024{\natexlab{b}}){Bhardwaj}, {Palmese}, {Maga{\~n}a Hernandez}, {D'Emilio}, \& {Morisaki}}]{2024ApJ...977..122B}
{Bhardwaj}, M., {Palmese}, A., {Maga{\~n}a Hernandez}, I., {D'Emilio}, V., \& {Morisaki}, S. 2024{\natexlab{b}}, \apj, 977, 122, \dodoi{10.3847/1538-4357/ad9023}

\bibitem[{{Bhardwaj} {et~al.}(2021{\natexlab{a}}){Bhardwaj}, {Kirichenko}, {Michilli}, {Mayya}, {Kaspi}, {Gaensler}, {Rahman}, {Tendulkar}, {Fonseca}, {Josephy}, {Leung}, {Merryfield}, {Petroff}, {Pleunis}, {Sanghavi}, {Scholz}, {Shin}, {Smith}, \& {Stairs}}]{2021ApJ...919L..24B}
{Bhardwaj}, M., {Kirichenko}, A.~Y., {Michilli}, D., {et~al.} 2021{\natexlab{a}}, \apjl, 919, L24, \dodoi{10.3847/2041-8213/ac223b}

\bibitem[{{Bhardwaj} {et~al.}(2021{\natexlab{b}}){Bhardwaj}, {Gaensler}, {Kaspi}, {Landecker}, {Mckinven}, {Michilli}, {Pleunis}, {Tendulkar}, {Andersen}, {Boyle}, {Cassanelli}, {Chawla}, {Cook}, {Dobbs}, {Fonseca}, {Kaczmarek}, {Leung}, {Masui}, {Mnchmeyer}, {Ng}, {Rafiei-Ravandi}, {Scholz}, {Shin}, {Smith}, {Stairs}, \& {Zwaniga}}]{2021ApJ...910L..18B}
{Bhardwaj}, M., {Gaensler}, B.~M., {Kaspi}, V.~M., {et~al.} 2021{\natexlab{b}}, \apjl, 910, L18, \dodoi{10.3847/2041-8213/abeaa6}

\bibitem[{{Bhardwaj} {et~al.}(2024{\natexlab{c}}){Bhardwaj}, {Michilli}, {Kirichenko}, {Modilim}, {Shin}, {Kaspi}, {Andersen}, {Cassanelli}, {Brar}, {Chatterjee}, {Cook}, {Dong}, {Fonseca}, {Gaensler}, {Ibik}, {Kaczmarek}, {Lanman}, {Leung}, {Masui}, {Pandhi}, {Pearlman}, {Petroff}, {Pleunis}, {Prochaska}, {Rafiei-Ravandi}, {Sand}, {Scholz}, \& {Smith}}]{2024ApJ...971L..51B}
{Bhardwaj}, M., {Michilli}, D., {Kirichenko}, A.~Y., {et~al.} 2024{\natexlab{c}}, \apjl, 971, L51, \dodoi{10.3847/2041-8213/ad64d1}

\bibitem[{{Bhattacharya} {et~al.}(2024){Bhattacharya}, {Murase}, \& {Kashiyama}}]{2024arXiv241219358B}
{Bhattacharya}, M., {Murase}, K., \& {Kashiyama}, K. 2024, arXiv e-prints, arXiv:2412.19358, \dodoi{10.48550/arXiv.2412.19358}

\bibitem[{{Bhusare} {et~al.}(2024){Bhusare}, {Maan}, \& {Kumar}}]{2024arXiv241213121B}
{Bhusare}, Y., {Maan}, Y., \& {Kumar}, A. 2024, arXiv e-prints, arXiv:2412.13121, \dodoi{10.48550/arXiv.2412.13121}

\bibitem[{{Bietenholz} \& {Bartel}(2017)}]{Bietenholz&Bartel17}
{Bietenholz}, M.~F., \& {Bartel}, N. 2017, \apj, 851, 124, \dodoi{10.3847/1538-4357/aa98d9}

\bibitem[{{Bochenek} {et~al.}(2020){Bochenek}, {Ravi}, {Belov}, {Hallinan}, {Kocz}, {Kulkarni}, \& {McKenna}}]{Bochenek2020}
{Bochenek}, C.~D., {Ravi}, V., {Belov}, K.~V., {et~al.} 2020, \nat, 587, 59, \dodoi{10.1038/s41586-020-2872-x}

\bibitem[{{Bochenek} {et~al.}(2021){Bochenek}, {Ravi}, \& {Dong}}]{2021ApJ...907L..31B}
{Bochenek}, C.~D., {Ravi}, V., \& {Dong}, D. 2021, \apjl, 907, L31, \dodoi{10.3847/2041-8213/abd634}

\bibitem[{{Boles}(2008)}]{Boles2008}
{Boles}, T. 2008, Central Bureau Electronic Telegrams, 1239, 1

\bibitem[{{Bozzetto} {et~al.}(2017){Bozzetto}, {Filipovi{\'c}}, {Vukoti{\'c}}, {Pavlovi{\'c}}, {Uro{\v{s}}evi{\'c}}, {Kavanagh}, {Arbutina}, {Maggi}, {Sasaki}, {Haberl}, {Crawford}, {Roper}, {Grieve}, \& {Points}}]{Bozzetto2017}
{Bozzetto}, L.~M., {Filipovi{\'c}}, M.~D., {Vukoti{\'c}}, B., {et~al.} 2017, \apjs, 230, 2, \dodoi{10.3847/1538-4365/aa653c}

\bibitem[{{Bruni} {et~al.}(2024){Bruni}, {Piro}, {Yang}, {Quai}, {Zhang}, {Palazzi}, {Nicastro}, {Feruglio}, {Tripodi}, {O'Connor}, {Gardini}, {Savaglio}, {Rossi}, {Nicuesa Guelbenzu}, \& {Paladino}}]{2024Natur.632.1014B}
{Bruni}, G., {Piro}, L., {Yang}, Y.-P., {et~al.} 2024, \nat, 632, 1014, \dodoi{10.1038/s41586-024-07782-6}

\bibitem[{{Bruni} {et~al.}(2025){Bruni}, {Piro}, {Yang}, {Palazzi}, {Nicastro}, {Rossi}, {Savaglio}, {Maiorano}, \& {Zhang}}]{2025A&A...695L..12B}
{Bruni}, G., {Piro}, L., {Yang}, Y.~P., {et~al.} 2025, \aap, 695, L12, \dodoi{10.1051/0004-6361/202453233}

\bibitem[{{Burrows} \& {Vartanyan}(2021)}]{2021Natur.589...29B}
{Burrows}, A., \& {Vartanyan}, D. 2021, \nat, 589, 29, \dodoi{10.1038/s41586-020-03059-w}

\bibitem[{{Chambers} {et~al.}(2016){Chambers}, {Magnier}, {Metcalfe}, {Flewelling}, {Huber}, {Waters}, {Denneau}, {Draper}, {Farrow}, {Finkbeiner}, {Holmberg}, {Koppenhoefer}, {Price}, {Rest}, {Saglia}, {Schlafly}, {Smartt}, {Sweeney}, {Wainscoat}, {Burgett}, {Chastel}, {Grav}, {Heasley}, {Hodapp}, {Jedicke}, {Kaiser}, {Kudritzki}, {Luppino}, {Lupton}, {Monet}, {Morgan}, {Onaka}, {Shiao}, {Stubbs}, {Tonry}, {White}, {Ba{\~n}ados}, {Bell}, {Bender}, {Bernard}, {Boegner}, {Boffi}, {Botticella}, {Calamida}, {Casertano}, {Chen}, {Chen}, {Cole}, {Deacon}, {Frenk}, {Fitzsimmons}, {Gezari}, {Gibbs}, {Goessl}, {Goggia}, {Gourgue}, {Goldman}, {Grant}, {Grebel}, {Hambly}, {Hasinger}, {Heavens}, {Heckman}, {Henderson}, {Henning}, {Holman}, {Hopp}, {Ip}, {Isani}, {Jackson}, {Keyes}, {Koekemoer}, {Kotak}, {Le}, {Liska}, {Long}, {Lucey}, {Liu}, {Martin}, {Masci}, {McLean}, {Mindel}, {Misra}, {Morganson}, {Murphy}, {Obaika}, {Narayan}, {Nieto-Santisteban}, {Norberg}, {Peacock}, {Pier}, {Postman}, {Primak}, {Rae}, {Rai},
  {Riess}, {Riffeser}, {Rix}, {R{\"o}ser}, {Russel}, {Rutz}, {Schilbach}, {Schultz}, {Scolnic}, {Strolger}, {Szalay}, {Seitz}, {Small}, {Smith}, {Soderblom}, {Taylor}, {Thomson}, {Taylor}, {Thakar}, {Thiel}, {Thilker}, {Unger}, {Urata}, {Valenti}, {Wagner}, {Walder}, {Walter}, {Watters}, {Werner}, {Wood-Vasey}, \& {Wyse}}]{chambers2016pan}
{Chambers}, K.~C., {Magnier}, E.~A., {Metcalfe}, N., {et~al.} 2016, arXiv e-prints, arXiv:1612.05560, \dodoi{10.48550/arXiv.1612.05560}

\bibitem[{{Chatterjee} {et~al.}(2017){Chatterjee}, {Law}, {Wharton}, {Burke-Spolaor}, {Hessels}, {Bower}, {Cordes}, {Tendulkar}, {Bassa}, {Demorest}, {Butler}, {Seymour}, {Scholz}, {Abruzzo}, {Bogdanov}, {Kaspi}, {Keimpema}, {Lazio}, {Marcote}, {McLaughlin}, {Paragi}, {Ransom}, {Rupen}, {Spitler}, \& {van Langevelde}}]{2017Natur.541...58C}
{Chatterjee}, S., {Law}, C.~J., {Wharton}, R.~S., {et~al.} 2017, \nat, 541, 58, \dodoi{10.1038/nature20797}

\bibitem[{{CHIME/FRB Collaboration} {et~al.}(2018){CHIME/FRB Collaboration}, {Amiri}, {Bandura}, {Berger}, {Bhardwaj}, {Boyce}, {Boyle}, {Brar}, {Burhanpurkar}, {Chawla}, {Chowdhury}, {Cliche}, {Cranmer}, {Cubranic}, {Deng}, {Denman}, {Dobbs}, {Fandino}, {Fonseca}, {Gaensler}, {Giri}, {Gilbert}, {Good}, {Guliani}, {Halpern}, {Hinshaw}, {H{\"o}fer}, {Josephy}, {Kaspi}, {Landecker}, {Lang}, {Liao}, {Masui}, {Mena-Parra}, {Naidu}, {Newburgh}, {Ng}, {Patel}, {Pen}, {Pinsonneault-Marotte}, {Pleunis}, {Rafiei Ravandi}, {Ransom}, {Renard}, {Scholz}, {Sigurdson}, {Siegel}, {Smith}, {Stairs}, {Tendulkar}, {Vanderlinde}, \& {Wiebe}}]{2018ApJ...863...48C}
{CHIME/FRB Collaboration}, {Amiri}, M., {Bandura}, K., {et~al.} 2018, \apj, 863, 48, \dodoi{10.3847/1538-4357/aad188}

\bibitem[{{CHIME/FRB Collaboration} {et~al.}(2019){CHIME/FRB Collaboration}, {Andersen}, {Bandura}, {Bhardwaj}, {Boubel}, {Boyce}, {Boyle}, {Brar}, {Cassanelli}, {Chawla}, {Cubranic}, {Deng}, {Dobbs}, {Fandino}, {Fonseca}, {Gaensler}, {Gilbert}, {Giri}, {Good}, {Halpern}, {Hill}, {Hinshaw}, {H{\"o}fer}, {Josephy}, {Kaspi}, {Kothes}, {Landecker}, {Lang}, {Li}, {Lin}, {Masui}, {Mena-Parra}, {Merryfield}, {Mckinven}, {Michilli}, {Milutinovic}, {Naidu}, {Newburgh}, {Ng}, {Patel}, {Pen}, {Pinsonneault-Marotte}, {Pleunis}, {Rafiei-Ravandi}, {Rahman}, {Ransom}, {Renard}, {Scholz}, {Siegel}, {Singh}, {Smith}, {Stairs}, {Tendulkar}, {Tretyakov}, {Vanderlinde}, {Yadav}, \& {Zwaniga}}]{CHIMEFRB2019}
{CHIME/FRB Collaboration}, {Andersen}, B.~C., {Bandura}, K., {et~al.} 2019, \apjl, 885, L24, \dodoi{10.3847/2041-8213/ab4a80}

\bibitem[{{CHIME/FRB Collaboration} {et~al.}(2020{\natexlab{a}}){CHIME/FRB Collaboration}, {:}, {Andersen}, {Band ura}, {Bhardwaj}, {Bij}, {Boyce}, {Boyle}, {Brar}, {Cassanelli}, {Chawla}, {Chen}, {Cliche}, {Cook}, {Cubranic}, {Curtin}, {Denman}, {Dobbs}, {Dong}, {Fandino}, {Fonseca}, {Gaensler}, {Giri}, {Good}, {Halpern}, {Hill}, {Hinshaw}, {H{\"o}fer}, {Josephy}, {Kania}, {Kaspi}, {Landecker}, {Leung}, {Li}, {Lin}, {Masui}, {Mckinven}, {Mena-Parra}, {Merryfield}, {Meyers}, {Michilli}, {Milutinovic}, {Mirhosseini}, {M{\"u}nchmeyer}, {Naidu}, {Newburgh}, {Ng}, {Patel}, {Pen}, {Pinsonneault-Marotte}, {Pleunis}, {Quine}, {Rafiei-Ravandi}, {Rahman}, {Ransom}, {Renard}, {Sanghavi}, {Scholz}, {Shaw}, {Shin}, {Siegel}, {Singh}, {Smegal}, {Smith}, {Stairs}, {Tan}, {Tendulkar}, {Tretyakov}, {Vanderlinde}, {Wang}, {Wulf}, \& {Zwaniga}}]{2020SGR}
{CHIME/FRB Collaboration}, {:}, {Andersen}, B.~C., {et~al.} 2020{\natexlab{a}}, arXiv e-prints, arXiv:2005.10324.
\newblock \doarXiv{2005.10324}

\bibitem[{{CHIME/FRB Collaboration} {et~al.}(2020{\natexlab{b}}){CHIME/FRB Collaboration}, {Amiri}, {Andersen}, {Bandura}, {Bhardwaj}, {Boyle}, {Brar}, {Chawla}, {Chen}, {Cliche}, {Cubranic}, {Deng}, {Denman}, {Dobbs}, {Dong}, {Fandino}, {Fonseca}, {Gaensler}, {Giri}, {Good}, {Halpern}, {Hessels}, {Hill}, {H{\"o}fer}, {Josephy}, {Kania}, {Karuppusamy}, {Kaspi}, {Keimpema}, {Kirsten}, {Landecker}, {Lang}, {Leung}, {Li}, {Lin}, {Marcote}, {Masui}, {McKinven}, {Mena-Parra}, {Merryfield}, {Michilli}, {Milutinovic}, {Mirhosseini}, {Naidu}, {Newburgh}, {Ng}, {Nimmo}, {Paragi}, {Patel}, {Pen}, {Pinsonneault-Marotte}, {Pleunis}, {Rafiei-Ravandi}, {Rahman}, {Ransom}, {Renard}, {Sanghavi}, {Scholz}, {Shaw}, {Shin}, {Siegel}, {Singh}, {Smegal}, {Smith}, {Stairs}, {Tendulkar}, {Tretyakov}, {Vanderlinde}, {Wang}, {Wang}, {Wulf}, {Yadav}, \& {Zwaniga}}]{2020Natur.582..351C}
{CHIME/FRB Collaboration}, {Amiri}, M., {Andersen}, B.~C., {et~al.} 2020{\natexlab{b}}, \nat, 582, 351, \dodoi{10.1038/s41586-020-2398-2}

\bibitem[{{CHIME/FRB Collaboration} {et~al.}(2021){CHIME/FRB Collaboration}, {Amiri}, {Andersen}, {Bandura}, {Berger}, {Bhardwaj}, {Boyce}, {Boyle}, {Brar}, {Breitman}, {Cassanelli}, {Chawla}, {Chen}, {Cliche}, {Cook}, {Cubranic}, {Curtin}, {Deng}, {Dobbs}, {Dong}, {Eadie}, {Fandino}, {Fonseca}, {Gaensler}, {Giri}, {Good}, {Halpern}, {Hill}, {Hinshaw}, {Josephy}, {Kaczmarek}, {Kader}, {Kania}, {Kaspi}, {Landecker}, {Lang}, {Leung}, {Li}, {Lin}, {Masui}, {McKinven}, {Mena-Parra}, {Merryfield}, {Meyers}, {Michilli}, {Milutinovic}, {Mirhosseini}, {M{\"u}nchmeyer}, {Naidu}, {Newburgh}, {Ng}, {Patel}, {Pen}, {Petroff}, {Pinsonneault-Marotte}, {Pleunis}, {Rafiei-Ravandi}, {Rahman}, {Ransom}, {Renard}, {Sanghavi}, {Scholz}, {Shaw}, {Shin}, {Siegel}, {Sikora}, {Singh}, {Smith}, {Stairs}, {Tan}, {Tendulkar}, {Vanderlinde}, {Wang}, {Wulf}, \& {Zwaniga}}]{2021ApJS..257...59C}
---. 2021, \apjs, 257, 59, \dodoi{10.3847/1538-4365/ac33ab}

\bibitem[{{CHIME/FRB Collaboration} {et~al.}(2023{\natexlab{a}}){CHIME/FRB Collaboration}, {Amiri}, {Andersen}, {Bandura}, {Berger}, {Bhardwaj}, {Boyce}, {Boyle}, {Brar}, {Breitman}, {Cassanelli}, {Chawla}, {Chen}, {Cliche}, {Cook}, {Cubranic}, {Curtin}, {Deng}, {Dobbs}, {Dong}, {Eadie}, {Fandino}, {Fonseca}, {Gaensler}, {Giri}, {Good}, {Halpern}, {Hill}, {Hinshaw}, {Josephy}, {Kaczmarek}, {Kader}, {Kania}, {Kaspi}, {Landecker}, {Lang}, {Leung}, {Li}, {Lin}, {Masui}, {McKinven}, {Mena-Parra}, {Merryfield}, {Meyers}, {Michilli}, {Milutinovic}, {Mirhosseini}, {M{\"u}nchmeyer}, {Naidu}, {Newburgh}, {Ng}, {Patel}, {Pen}, {Petroff}, {Pinsonneault-Marotte}, {Pleunis}, {Rafiei-Ravandi}, {Rahman}, {Ransom}, {Renard}, {Sanghavi}, {Scholz}, {Shaw}, {Shin}, {Siegel}, {Sikora}, {Singh}, {Smith}, {Stairs}, {Tan}, {Tendulkar}, {Vanderlinde}, {Wang}, {Wulf}, \& {Zwaniga}}]{2023ApJS..264...53C}
---. 2023{\natexlab{a}}, \apjs, 264, 53, \dodoi{10.3847/1538-4365/acb54c}

\bibitem[{{CHIME/FRB Collaboration} {et~al.}(2023{\natexlab{b}}){CHIME/FRB Collaboration}, {Andersen}, {Bandura}, {Bhardwaj}, {Boyle}, {Brar}, {Cassanelli}, {Chatterjee}, {Chawla}, {Cook}, {Curtin}, {Dobbs}, {Dong}, {Faber}, {Fandino}, {Fonseca}, {Gaensler}, {Giri}, {Herrera-Martin}, {Hill}, {Ibik}, {Josephy}, {Kaczmarek}, {Kader}, {Kaspi}, {Landecker}, {Lanman}, {Lazda}, {Leung}, {Lin}, {Masui}, {McKinven}, {Mena-Parra}, {Meyers}, {Michilli}, {Ng}, {Pandhi}, {Pearlman}, {Pen}, {Petroff}, {Pleunis}, {Rafiei-Ravandi}, {Rahman}, {Ransom}, {Renard}, {Sand}, {Sanghavi}, {Scholz}, {Shah}, {Shin}, {Siegel}, {Smith}, {Stairs}, {Su}, {Tendulkar}, {Vanderlinde}, {Wang}, {Wulf}, \& {Zwaniga}}]{2023ApJ...947...83C}
{CHIME/FRB Collaboration}, {Andersen}, B.~C., {Bandura}, K., {et~al.} 2023{\natexlab{b}}, \apj, 947, 83, \dodoi{10.3847/1538-4357/acc6c1}

\bibitem[{{CHIME/FRB Collaboration} {et~al.}(2024){CHIME/FRB Collaboration}, {Amiri}, {Andersen}, {Andrew}, {Bandura}, {Bhardwaj}, {Boyle}, {Brar}, {Breitman}, {Cassanelli}, {Chawla}, {Cook}, {Curtin}, {Dobbs}, {Dong}, {Eadie}, {Fonseca}, {Gaensler}, {Giri}, {Herrera-Martin}, {Hopkins}, {Ibik}, {Joseph}, {Kaczmarek}, {Kader}, {Kaspi}, {Lanman}, {Lazda}, {Leung}, {Liu}, {Masui}, {McKinven}, {Mena-Parra}, {Merryfield}, {Michilli}, {Ng}, {Nimmo}, {Noble}, {Pandhi}, {Patel}, {Pearlman}, {Pen}, {Petroff}, {Pleunis}, {Rafiei-Ravandi}, {Rahman}, {Ransom}, {Sand}, {Scholz}, {Shah}, {Shin}, {Shpunarska}, {Siegel}, {Smith}, {Stairs}, {Stenning}, {Vanderlinde}, {Wang}, {White}, \& {Wulf}}]{2024ApJ...969..145C}
{CHIME/FRB Collaboration}, {Amiri}, M., {Andersen}, B.~C., {et~al.} 2024, \apj, 969, 145, \dodoi{10.3847/1538-4357/ad464b}

\bibitem[{{Collette} {et~al.}(2023){Collette}, {Kluyver}, {Caswell}, {Tocknell}, {Kieffer}, {Jelenak}, {Scopatz}, {Dale}, {Chen}, {VINCENT}, {Einhorn}, {Payno}, {Juliagarriga}, {Sciarelli}, {Valls}, {Ghosh}, {Kofoed Pedersen}, {Kittisopikul}, {Jakirkham}, {Raspaud}, {Danilevski}, {Abbasi}, {Readey}, {M{\"u}hlbauer}, {Paramonov}, {Chan}, {De Schepper}, {Sol{\'e}}, {Jialin}, \& {Hay Guest}}]{collette2023python}
{Collette}, A., {Kluyver}, T., {Caswell}, T.~A., {et~al.} 2023, {h5py/h5py: 3.8.0}, 3.8.0,  Zenodo, \dodoi{10.5281/zenodo.7560547}

\bibitem[{{Cook} {et~al.}(2023){Cook}, {Bhardwaj}, {Gaensler}, {Scholz}, {Eadie}, {Hill}, {Kaspi}, {Masui}, {Curtin}, {Dong}, {Fonseca}, {Herrera-Martin}, {Kaczmarek}, {Lanman}, {Lazda}, {Leung}, {Meyers}, {Michilli}, {Pandhi}, {Pearlman}, {Pleunis}, {Ransom}, {Rahman}, {Sand}, {Shin}, {Smith}, {Stairs}, \& {Stenning}}]{2023ApJ...946...58C}
{Cook}, A.~M., {Bhardwaj}, M., {Gaensler}, B.~M., {et~al.} 2023, \apj, 946, 58, \dodoi{10.3847/1538-4357/acbbd0}

\bibitem[{Cordes \& Lazio(2002)}]{Cordes2002}
Cordes, J.~M., \& Lazio, T. J.~W. 2002, arXiv e-prints.
\newblock \doarXiv{astro-ph/0207156}

\bibitem[{{Cordes} {et~al.}(2022){Cordes}, {Ocker}, \& {Chatterjee}}]{2022ApJ...931...88C}
{Cordes}, J.~M., {Ocker}, S.~K., \& {Chatterjee}, S. 2022, \apj, 931, 88, \dodoi{10.3847/1538-4357/ac6873}

\bibitem[{{Cruise} {et~al.}(2025){Cruise}, {Guainazzi}, {Aird}, {Carrera}, {Costantini}, {Corrales}, {Dauser}, {Eckert}, {Gastaldello}, {Matsumoto}, {Osten}, {Petrucci}, {Porquet}, {Pratt}, {Rea}, {Reiprich}, {Simionescu}, {Spiga}, \& {Troja}}]{2025NatAs...9...36C}
{Cruise}, M., {Guainazzi}, M., {Aird}, J., {et~al.} 2025, Nature Astronomy, 9, 36, \dodoi{10.1038/s41550-024-02416-3}

\bibitem[{{Cui} {et~al.}(2025){Cui}, {James}, {Li}, \& {Zhang}}]{2025ApJ...982..158C}
{Cui}, X.-h., {James}, C.~W., {Li}, D., \& {Zhang}, C.-m. 2025, \apj, 982, 158, \dodoi{10.3847/1538-4357/adbbcb}

\bibitem[{{Dall'Osso} \& {Stella}(2022)}]{2022ASSL..465..245D}
{Dall'Osso}, S., \& {Stella}, L. 2022, in Astrophysics and Space Science Library, Vol. 465, Astrophysics and Space Science Library, ed. S.~{Bhattacharyya}, A.~{Papitto}, \& D.~{Bhattacharya}, 245--280, \dodoi{10.1007/978-3-030-85198-9_8}

\bibitem[{{Dewdney} {et~al.}(2009){Dewdney}, {Hall}, {Schilizzi}, \& {Lazio}}]{2009IEEEP..97.1482D}
{Dewdney}, P.~E., {Hall}, P.~J., {Schilizzi}, R.~T., \& {Lazio}, T.~J.~L.~W. 2009, IEEE Proceedings, 97, 1482, \dodoi{10.1109/JPROC.2009.2021005}

\bibitem[{{Di Francesco} {et~al.}(2019){Di Francesco}, {Chalmers}, {Denman}, {Fissel}, {Friesen}, {Gaensler}, {Hlavacek-Larrondo}, {Kirk}, {Matthews}, {O'Dea}, {Robishaw}, {Rosolowsky}, {Rupen}, {Sadavoy}, {Sa-Harb}, {Sivakoff}, {Tahani}, {van der Marel}, {White}, \& {Wilson}}]{2019clrp.2020...32D}
{Di Francesco}, J., {Chalmers}, D., {Denman}, N., {et~al.} 2019, in Canadian Long Range Plan for Astronomy and Astrophysics White Papers, Vol. 2020, 32, \dodoi{10.5281/zenodo.3765763}

\bibitem[{{Dubner} \& {Giacani}(2015)}]{2015A&ARv..23....3D}
{Dubner}, G., \& {Giacani}, E. 2015, \aapr, 23, 3, \dodoi{10.1007/s00159-015-0083-5}

\bibitem[{{Eftekhari} {et~al.}(2025){Eftekhari}, {Dong}, {Fong}, {Shah}, {Simha}, {Andersen}, {Andrew}, {Bhardwaj}, {Cassanelli}, {Chatterjee}, {Coulter}, {Fonseca}, {Gaensler}, {Gordon}, {Hessels}, {Ibik}, {Joseph}, {Kahinga}, {Kaspi}, {Kharel}, {Kilpatrick}, {Lanman}, {Lazda}, {Leung}, {Liu}, {Mas-Ribas}, {Masui}, {Mckinven}, {Mena-Parra}, {Miller}, {Nimmo}, {Pandhi}, {Patil}, {Pearlman}, {Pleunis}, {Prochaska}, {Rafiei-Ravandi}, {Sammons}, {Scholz}, {Shin}, {Smith}, \& {Stairs}}]{2025ApJ...979L..22E}
{Eftekhari}, T., {Dong}, Y., {Fong}, W., {et~al.} 2025, \apjl, 979, L22, \dodoi{10.3847/2041-8213/ad9de2}

\bibitem[{{Filippenko}(1997)}]{1997ARA&A..35..309F}
{Filippenko}, A.~V. 1997, \araa, 35, 309, \dodoi{10.1146/annurev.astro.35.1.309}

\bibitem[{{Fonseca} {et~al.}(2020){Fonseca}, {Andersen}, {Bhardwaj}, {Chawla}, {Good}, {Josephy}, {Kaspi}, {Masui}, {Mckinven}, {Michilli}, {Pleunis}, {Shin}, {Tendulkar}, {Bandura}, {Boyle}, {Brar}, {Cassanelli}, {Cubranic}, {Dobbs}, {Dong}, {Gaensler}, {Hinshaw}, {Landecker}, {Leung}, {Li}, {Lin}, {Mena-Parra}, {Merryfield}, {Naidu}, {Ng}, {Patel}, {Pen}, {Rafiei-Ravandi}, {Rahman}, {Ransom}, {Scholz}, {Smith}, {Stairs}, {Vanderlinde}, {Yadav}, \& {Zwaniga}}]{2020ApJ...891L...6F}
{Fonseca}, E., {Andersen}, B.~C., {Bhardwaj}, M., {et~al.} 2020, \apjl, 891, L6, \dodoi{10.3847/2041-8213/ab7208}

\bibitem[{Garrett(2021)}]{scienceplots}
Garrett, J.~D. 2021, {garrettj403/SciencePlots}, 1.0.9,  Zenodo, \dodoi{10.5281/zenodo.4106649}

\bibitem[{{Gehrels}(1986)}]{1986ApJ...303..336G}
{Gehrels}, N. 1986, \apj, 303, 336, \dodoi{10.1086/164079}

\bibitem[{{Gordon} {et~al.}(2023){Gordon}, {Fong}, {Kilpatrick}, {Eftekhari}, {Leja}, {Prochaska}, {Nugent}, {Bhandari}, {Blanchard}, {Caleb}, {Day}, {Deller}, {Dong}, {Glowacki}, {Gourdji}, {Mannings}, {Mahoney}, {Marnoch}, {Miller}, {Paterson}, {Rastinejad}, {Ryder}, {Sadler}, {Scott}, {Sears}, {Shannon}, {Simha}, {Stappers}, \& {Tejos}}]{Gordon+23}
{Gordon}, A.~C., {Fong}, W.-f., {Kilpatrick}, C.~D., {et~al.} 2023, \apj, 954, 80, \dodoi{10.3847/1538-4357/ace5aa}

\bibitem[{{Green}(2019)}]{Green2019}
{Green}, D.~A. 2019, Journal of Astrophysics and Astronomy, 40, 36, \dodoi{10.1007/s12036-019-9601-6}

\bibitem[{{Hallinan} {et~al.}(2019){Hallinan}, {Ravi}, {Weinreb}, {Kocz}, {Huang}, {Woody}, {Lamb}, {D'Addario}, {Catha}, {Law}, {Kulkarni}, {Phinney}, {Eastwood}, {Bouman}, {McLaughlin}, {Ransom}, {Siemens}, {Cordes}, {Lynch}, {Kaplan}, {Brazier}, {Bhatnagar}, {Myers}, {Walter}, \& {Gaensler}}]{2019BAAS...51g.255H}
{Hallinan}, G., {Ravi}, V., {Weinreb}, S., {et~al.} 2019, in Bulletin of the American Astronomical Society, Vol.~51, 255, \dodoi{10.48550/arXiv.1907.07648}

\bibitem[{{Hambleton} {et~al.}(2023){Hambleton}, {Bianco}, {Street}, {Bell}, {Buckley}, {Graham}, {Hernitschek}, {Lund}, {Mason}, {Pepper}, {Pr{\v{s}}a}, {Rabus}, {Raiteri}, {Szab{\'o}}, {Szkody}, {Andreoni}, {Antoniucci}, {Balmaverde}, {Bellm}, {Bonito}, {Bono}, {Botticella}, {Brocato}, {Bu{\v{c}}ar Bricman}, {Cappellaro}, {Carnerero}, {Chornock}, {Clarke}, {Cowperthwaite}, {Cucchiara}, {D'Ammando}, {Dage}, {Dall'Ora}, {Davenport}, {de Martino}, {de Somma}, {Di Criscienzo}, {Di Stefano}, {Drout}, {Fabrizio}, {Fiorentino}, {Gandhi}, {Garofalo}, {Giannini}, {Gomboc}, {Greggio}, {Hartigan}, {Hundertmark}, {Johnson}, {Johnson}, {Jurkic}, {Khakpash}, {Leccia}, {Li}, {Magurno}, {Malanchev}, {Marconi}, {Margutti}, {Marinoni}, {Mauron}, {Molinaro}, {M{\"o}ller}, {Moniez}, {Muraveva}, {Musella}, {Ngeow}, {Pastorello}, {Petrecca}, {Piranomonte}, {Ragosta}, {Reguitti}, {Righi}, {Ripepi}, {Rivera Sandoval}, {Stassun}, {Stroh}, {Terreran}, {Trimble}, {Tsapras}, {van Velzen}, {Venuti}, \& {Vink}}]{2023PASP..135j5002H}
{Hambleton}, K.~M., {Bianco}, F.~B., {Street}, R., {et~al.} 2023, \pasp, 135, 105002, \dodoi{10.1088/1538-3873/acdb9a}

\bibitem[{{Harris} {et~al.}(2020){Harris}, {Millman}, {van der Walt}, {Gommers}, {Virtanen}, {Cournapeau}, {Wieser}, {Taylor}, {Berg}, {Smith}, {Kern}, {Picus}, {Hoyer}, {van Kerkwijk}, {Brett}, {Haldane}, {del R{\'\i}o}, {Wiebe}, {Peterson}, {G{\'e}rard-Marchant}, {Sheppard}, {Reddy}, {Weckesser}, {Abbasi}, {Gohlke}, \& {Oliphant}}]{harris2020array}
{Harris}, C.~R., {Millman}, K.~J., {van der Walt}, S.~J., {et~al.} 2020, \nat, 585, 357, \dodoi{10.1038/s41586-020-2649-2}

\bibitem[{{Heger} {et~al.}(2003){Heger}, {Fryer}, {Woosley}, {Langer}, \& {Hartmann}}]{2003ApJ...591..288H}
{Heger}, A., {Fryer}, C.~L., {Woosley}, S.~E., {Langer}, N., \& {Hartmann}, D.~H. 2003, \apj, 591, 288, \dodoi{10.1086/375341}

\bibitem[{{Horowicz} \& {Margalit}(2025)}]{HorowiczMargalit2025}
{Horowicz}, A., \& {Margalit}, B. 2025, arXiv e-prints, arXiv:2504.08038, \dodoi{10.48550/arXiv.2504.08038}

\bibitem[{{Hunter}(2007)}]{2007CSE.....9...90H}
{Hunter}, J.~D. 2007, Computing in Science and Engineering, 9, 90, \dodoi{10.1109/MCSE.2007.55}

\bibitem[{{Ibik} {et~al.}(2024){Ibik}, {Drout}, {Gaensler}, {Scholz}, {Michilli}, {Bhardwaj}, {Kaspi}, {Pleunis}, {Cassanelli}, {Cook}, {Dong}, {Kaczmarek}, {Leung}, {Lu}, {Masui}, {Pearlman}, {Rafiei-Ravandi}, {Sand}, {Shin}, {Smith}, \& {Stairs}}]{2024ApJ...961...99I}
{Ibik}, A.~L., {Drout}, M.~R., {Gaensler}, B.~M., {et~al.} 2024, \apj, 961, 99, \dodoi{10.3847/1538-4357/ad0893}

\bibitem[{{Ivezi{\'c}} {et~al.}(2019){Ivezi{\'c}}, {Kahn}, {Tyson}, {Abel}, {Acosta}, {Allsman}, {Alonso}, {AlSayyad}, {Anderson}, {Andrew}, {Angel}, {Angeli}, {Ansari}, {Antilogus}, {Araujo}, {Armstrong}, {Arndt}, {Astier}, {Aubourg}, {Auza}, {Axelrod}, {Bard}, {Barr}, {Barrau}, {Bartlett}, {Bauer}, {Bauman}, {Baumont}, {Bechtol}, {Bechtol}, {Becker}, {Becla}, {Beldica}, {Bellavia}, {Bianco}, {Biswas}, {Blanc}, {Blazek}, {Blandford}, {Bloom}, {Bogart}, {Bond}, {Booth}, {Borgland}, {Borne}, {Bosch}, {Boutigny}, {Brackett}, {Bradshaw}, {Brandt}, {Brown}, {Bullock}, {Burchat}, {Burke}, {Cagnoli}, {Calabrese}, {Callahan}, {Callen}, {Carlin}, {Carlson}, {Chandrasekharan}, {Charles-Emerson}, {Chesley}, {Cheu}, {Chiang}, {Chiang}, {Chirino}, {Chow}, {Ciardi}, {Claver}, {Cohen-Tanugi}, {Cockrum}, {Coles}, {Connolly}, {Cook}, {Cooray}, {Covey}, {Cribbs}, {Cui}, {Cutri}, {Daly}, {Daniel}, {Daruich}, {Daubard}, {Daues}, {Dawson}, {Delgado}, {Dellapenna}, {de Peyster}, {de Val-Borro}, {Digel}, {Doherty}, {Dubois},
  {Dubois-Felsmann}, {Durech}, {Economou}, {Eifler}, {Eracleous}, {Emmons}, {Fausti Neto}, {Ferguson}, {Figueroa}, {Fisher-Levine}, {Focke}, {Foss}, {Frank}, {Freemon}, {Gangler}, {Gawiser}, {Geary}, {Gee}, {Geha}, {Gessner}, {Gibson}, {Gilmore}, {Glanzman}, {Glick}, {Goldina}, {Goldstein}, {Goodenow}, {Graham}, {Gressler}, {Gris}, {Guy}, {Guyonnet}, {Haller}, {Harris}, {Hascall}, {Haupt}, {Hernandez}, {Herrmann}, {Hileman}, {Hoblitt}, {Hodgson}, {Hogan}, {Howard}, {Huang}, {Huffer}, {Ingraham}, {Innes}, {Jacoby}, {Jain}, {Jammes}, {Jee}, {Jenness}, {Jernigan}, {Jevremovi{\'c}}, {Johns}, {Johnson}, {Johnson}, {Jones}, {Juramy-Gilles}, {Juri{\'c}}, {Kalirai}, {Kallivayalil}, {Kalmbach}, {Kantor}, {Karst}, {Kasliwal}, {Kelly}, {Kessler}, {Kinnison}, {Kirkby}, {Knox}, {Kotov}, {Krabbendam}, {Krughoff}, {Kub{\'a}nek}, {Kuczewski}, {Kulkarni}, {Ku}, {Kurita}, {Lage}, {Lambert}, {Lange}, {Langton}, {Le Guillou}, {Levine}, {Liang}, {Lim}, {Lintott}, {Long}, {Lopez}, {Lotz}, {Lupton}, {Lust}, {MacArthur}, {Mahabal},
  {Mandelbaum}, {Markiewicz}, {Marsh}, {Marshall}, {Marshall}, {May}, {McKercher}, {McQueen}, {Meyers}, {Migliore}, {Miller}, \& {Mills}}]{2019ApJ...873..111I}
{Ivezi{\'c}}, {\v{Z}}., {Kahn}, S.~M., {Tyson}, J.~A., {et~al.} 2019, \apj, 873, 111, \dodoi{10.3847/1538-4357/ab042c}

\bibitem[{{James}(2023)}]{2023PASA...40...57J}
{James}, C.~W. 2023, \pasa, 40, e057, \dodoi{10.1017/pasa.2023.51}

\bibitem[{{Kirsten} {et~al.}(2022){Kirsten}, {Marcote}, {Nimmo}, {Hessels}, {Bhardwaj}, {Tendulkar}, {Keimpema}, {Yang}, {Snelders}, {Scholz}, {Pearlman}, {Law}, {Peters}, {Giroletti}, {Paragi}, {Bassa}, {Hewitt}, {Bach}, {Bezrukovs}, {Burgay}, {Buttaccio}, {Conway}, {Corongiu}, {Feiler}, {Forss{\'e}n}, {Gawro{\'n}ski}, {Karuppusamy}, {Kharinov}, {Lindqvist}, {Maccaferri}, {Melnikov}, {Ould-Boukattine}, {Possenti}, {Surcis}, {Wang}, {Yuan}, {Aggarwal}, {Anna-Thomas}, {Bower}, {Blaauw}, {Burke-Spolaor}, {Cassanelli}, {Clarke}, {Fonseca}, {Gaensler}, {Gopinath}, {Kaspi}, {Kassim}, {Lazio}, {Leung}, {Li}, {Lin}, {Masui}, {Mckinven}, {Michilli}, {Mikhailov}, {Ng}, {Orbidans}, {Pen}, {Petroff}, {Rahman}, {Ransom}, {Shin}, {Smith}, {Stairs}, \& {Vlemmings}}]{2022Natur.602..585K}
{Kirsten}, F., {Marcote}, B., {Nimmo}, K., {et~al.} 2022, \nat, 602, 585, \dodoi{10.1038/s41586-021-04354-w}

\bibitem[{Kovacs {et~al.}(2024)Kovacs, Mao, Basu, Ma, Pakmor, Spitler, \& Walker}]{Kovacs2024}
Kovacs, T.~O., Mao, S.~A., Basu, A., {et~al.} 2024, Astronomy \& Astrophysics, 690, A47, \dodoi{10.1051/0004-6361/202347459}

\bibitem[{{Kovlakas} {et~al.}(2021){Kovlakas}, {Zezas}, {Andrews}, {Basu-Zych}, {Fragos}, {Hornschemeier}, {Kouroumpatzakis}, {Lehmer}, \& {Ptak}}]{2021MNRAS.506.1896K}
{Kovlakas}, K., {Zezas}, A., {Andrews}, J.~J., {et~al.} 2021, \mnras, 506, 1896, \dodoi{10.1093/mnras/stab1799}

\bibitem[{{Kumar} {et~al.}(2017){Kumar}, {Lu}, \& {Bhattacharya}}]{2017MNRAS.468.2726K}
{Kumar}, P., {Lu}, W., \& {Bhattacharya}, M. 2017, \mnras, 468, 2726, \dodoi{10.1093/mnras/stx665}

\bibitem[{{Lacy} {et~al.}(2020){Lacy}, {Baum}, {Chandler}, {Chatterjee}, {Clarke}, {Deustua}, {English}, {Farnes}, {Gaensler}, {Gugliucci}, {Hallinan}, {Kent}, {Kimball}, {Law}, {Lazio}, {Marvil}, {Mao}, {Medlin}, {Mooley}, {Murphy}, {Myers}, {Osten}, {Richards}, {Rosolowsky}, {Rudnick}, {Schinzel}, {Sivakoff}, {Sjouwerman}, {Taylor}, {White}, {Wrobel}, {Andernach}, {Beasley}, {Berger}, {Bhatnager}, {Birkinshaw}, {Bower}, {Brandt}, {Brown}, {Burke-Spolaor}, {Butler}, {Comerford}, {Demorest}, {Fu}, {Giacintucci}, {Golap}, {G{\"u}th}, {Hales}, {Hiriart}, {Hodge}, {Horesh}, {Ivezi{\'c}}, {Jarvis}, {Kamble}, {Kassim}, {Liu}, {Loinard}, {Lyons}, {Masters}, {Mezcua}, {Moellenbrock}, {Mroczkowski}, {Nyland}, {O'Dea}, {O'Sullivan}, {Peters}, {Radford}, {Rao}, {Robnett}, {Salcido}, {Shen}, {Sobotka}, {Witz}, {Vaccari}, {van Weeren}, {Vargas}, {Williams}, \& {Yoon}}]{2020PASP..132c5001L}
{Lacy}, M., {Baum}, S.~A., {Chandler}, C.~J., {et~al.} 2020, \pasp, 132, 035001, \dodoi{10.1088/1538-3873/ab63eb}

\bibitem[{{Lan} {et~al.}(2024){Lan}, {Zhao}, {Wei}, \& {Wang}}]{2024ApJ...967L..44L}
{Lan}, H.-T., {Zhao}, Z.-Y., {Wei}, Y.-J., \& {Wang}, F.-Y. 2024, \apjl, 967, L44, \dodoi{10.3847/2041-8213/ad4ae8}

\bibitem[{{Lanman} {et~al.}(2022){Lanman}, {Andersen}, {Chawla}, {Josephy}, {Noble}, {Kaspi}, {Bandura}, {Bhardwaj}, {Boyle}, {Brar}, {Breitman}, {Cassanelli}, {Dong}, {Fonseca}, {Gaensler}, {Good}, {Kaczmarek}, {Leung}, {Masui}, {Meyers}, {Ng}, {Patel}, {Pearlman}, {Petroff}, {Pleunis}, {Rafiei-Ravandi}, {Rahman}, {Sanghavi}, {Scholz}, {Shin}, {Stairs}, {Tendulkar}, \& {Zwaniga}}]{2022ApJ...927...59L}
{Lanman}, A.~E., {Andersen}, B.~C., {Chawla}, P., {et~al.} 2022, \apj, 927, 59, \dodoi{10.3847/1538-4357/ac4bc7}

\bibitem[{{Leung} \& {CHIME/FRB Collaboration}(2025)}]{2025ATel17086....1L}
{Leung}, C., \& {CHIME/FRB Collaboration}. 2025, The Astronomer's Telegram, 17086, 1

\bibitem[{{Lin} {et~al.}(2022){Lin}, {Lin}, {Li}, {Tseng}, {Jiang}, {Wang}, {Cheng}, {Pen}, {Chen}, {Chen}, {Chen}, {Goto}, {Hashimoto}, {Hwang}, {King}, {Kubo}, {Kuo}, {Mills}, {Nam}, {Oshiro}, {Shen}, {Tseng}, {Wang}, {Wu}, {Bower}, {Chang}, {Chen}, {Chen}, {Chiang}, {Fedynitch}, {Gusinskaia}, {Ho}, {Hsiao}, {Hu}, {Huang}, {J{\'a}uregui Garc{\'\i}a}, {Kim}, {Kuo}, {Ling}, {On}, {Peterson}, {R. Raquel}, {Su}, {Uno}, {Wu}, {Yamasaki}, \& {Zhu}}]{2022PASP..134i4106L}
{Lin}, H.-H., {Lin}, K.-y., {Li}, C.-T., {et~al.} 2022, \pasp, 134, 094106, \dodoi{10.1088/1538-3873/ac8f71}

\bibitem[{{Lin} {et~al.}(2024){Lin}, {Scholz}, {Ng}, {Pen}, {Bhardwaj}, {Chawla}, {Curtin}, {Li}, {Newburgh}, {Reda}, {Sand}, {Tendulkar}, {Andersen}, {Bandura}, {Brar}, {Cassanelli}, {Cook}, {Dobbs}, {Dong}, {Eadie}, {Fonseca}, {Gaensler}, {Giri}, {Herrera-Martin}, {Hill}, {Kaczmarek}, {Kania}, {Kaspi}, {Khairy}, {Lanman}, {Leung}, {Masui}, {Mena-Parra}, {Meyers}, {Michilli}, {Milutinovic}, {Ordog}, {Pearlman}, {Pleunis}, {Rafiei-Ravandi}, {Rahman}, {Ransom}, {Sanghavi}, {Shin}, {Smith}, {Stairs}, {Stenning}, {Vanderlinde}, \& {Wulf}}]{2024ApJ...975...75L}
{Lin}, H.-H., {Scholz}, P., {Ng}, C., {et~al.} 2024, \apj, 975, 75, \dodoi{10.3847/1538-4357/ad779d}

\bibitem[{{Lorimer} {et~al.}(2007){Lorimer}, {Bailes}, {McLaughlin}, {Narkevic}, \& {Crawford}}]{Lorimer2007}
{Lorimer}, D.~R., {Bailes}, M., {McLaughlin}, M.~A., {Narkevic}, D.~J., \& {Crawford}, F. 2007, Science, 318, 777, \dodoi{10.1126/science.1147532}

\bibitem[{{Loudas} {et~al.}(2025){Loudas}, {Li}, {Strauss}, \& {Leja}}]{Loudas+25}
{Loudas}, N., {Li}, D., {Strauss}, M.~A., \& {Leja}, J. 2025, arXiv e-prints, arXiv:2502.15566, \dodoi{10.48550/arXiv.2502.15566}

\bibitem[{{Lyubarsky}(2014)}]{2014MNRAS.442L...9L}
{Lyubarsky}, Y. 2014, \mnras, 442, L9, \dodoi{10.1093/mnrasl/slu046}

\bibitem[{{Macquart} \& {Ekers}(2018)}]{2018MNRAS.480.4211M}
{Macquart}, J.~P., \& {Ekers}, R. 2018, \mnras, 480, 4211, \dodoi{10.1093/mnras/sty2083}

\bibitem[{Macquart {et~al.}(2020)Macquart, Prochaska, McQuinn, Bannister, Bhandari, \& et~al.}]{Macquart2020}
Macquart, J.-P., Prochaska, J.~X., McQuinn, M., {et~al.} 2020, Nature, 581, 391, \dodoi{10.1038/s41586-020-2288-8}

\bibitem[{{Macquart} {et~al.}(2019){Macquart}, {Shannon}, {Bannister}, {James}, {Ekers}, \& {Bunton}}]{2019ApJ...872L..19M}
{Macquart}, J.~P., {Shannon}, R.~M., {Bannister}, K.~W., {et~al.} 2019, \apjl, 872, L19, \dodoi{10.3847/2041-8213/ab03d6}

\bibitem[{{Macquart} {et~al.}(2015){Macquart}, {Keane}, {Grainge}, {McQuinn}, {Fender}, {Hessels}, {Deller}, {Bhat}, {Breton}, {Chatterjee}, {Law}, {Lorimer}, {Ofek}, {Pietka}, {Spitler}, {Stappers}, \& {Trott}}]{2015aska.confE..55M}
{Macquart}, J.~P., {Keane}, E., {Grainge}, K., {et~al.} 2015, in Advancing Astrophysics with the Square Kilometre Array (AASKA14), 55, \dodoi{10.22323/1.215.0055}

\bibitem[{{Madison} {et~al.}(2008){Madison}, {Li}, \& {Filippenko}}]{Madison2008}
{Madison}, D., {Li}, W., \& {Filippenko}, A.~V. 2008, Central Bureau Electronic Telegrams, 1239, 2

\bibitem[{{Marcote} {et~al.}(2017){Marcote}, {Paragi}, {Hessels}, {Keimpema}, {van Langevelde}, {Huang}, {Bassa}, {Bogdanov}, {Bower}, {Burke-Spolaor}, {Butler}, {Campbell}, {Chatterjee}, {Cordes}, {Demorest}, {Garrett}, {Ghosh}, {Kaspi}, {Law}, {Lazio}, {McLaughlin}, {Ransom}, {Salter}, {Scholz}, {Seymour}, {Siemion}, {Spitler}, {Tendulkar}, \& {Wharton}}]{2017ApJ...834L...8M}
{Marcote}, B., {Paragi}, Z., {Hessels}, J.~W.~T., {et~al.} 2017, \apjl, 834, L8, \dodoi{10.3847/2041-8213/834/2/L8}

\bibitem[{{Marcote} {et~al.}(2020){Marcote}, {Nimmo}, {Hessels}, {Tendulkar}, {Bassa}, {Paragi}, {Keimpema}, {Bhardwaj}, {Karuppusamy}, {Kaspi}, {Law}, {Michilli}, {Aggarwal}, {Andersen}, {Archibald}, {Bandura}, {Bower}, {Boyle}, {Brar}, {Burke-Spolaor}, {Butler}, {Cassanelli}, {Chawla}, {Demorest}, {Dobbs}, {Fonseca}, {Giri}, {Good}, {Gourdji}, {Josephy}, {Kirichenko}, {Kirsten}, {Landecker}, {Lang}, {Lazio}, {Li}, {Lin}, {Linford}, {Masui}, {Mena-Parra}, {Naidu}, {Ng}, {Patel}, {Pen}, {Pleunis}, {Rafiei-Ravandi}, {Rahman}, {Renard}, {Scholz}, {Siegel}, {Smith}, {Stairs}, {Vanderlinde}, \& {Zwaniga}}]{2020Natur.577..190M}
{Marcote}, B., {Nimmo}, K., {Hessels}, J.~W.~T., {et~al.} 2020, \nat, 577, 190, \dodoi{10.1038/s41586-019-1866-z}

\bibitem[{{Margalit} {et~al.}(2020){Margalit}, {Beniamini}, {Sridhar}, \& {Metzger}}]{Margalit+20_SGR1935}
{Margalit}, B., {Beniamini}, P., {Sridhar}, N., \& {Metzger}, B.~D. 2020, \apjl, 899, L27, \dodoi{10.3847/2041-8213/abac57}

\bibitem[{{Margalit} {et~al.}(2019){Margalit}, {Berger}, \& {Metzger}}]{2019ApJ...886..110M}
{Margalit}, B., {Berger}, E., \& {Metzger}, B.~D. 2019, \apj, 886, 110, \dodoi{10.3847/1538-4357/ab4c31}

\bibitem[{{Margalit} \& {Metzger}(2018)}]{2018ApJ...868L...4M}
{Margalit}, B., \& {Metzger}, B.~D. 2018, \apjl, 868, L4, \dodoi{10.3847/2041-8213/aaedad}

\bibitem[{{Margalit} {et~al.}(2018){Margalit}, {Metzger}, {Berger}, {Nicholl}, {Eftekhari}, \& {Margutti}}]{Margalit+18_CLOUDY}
{Margalit}, B., {Metzger}, B.~D., {Berger}, E., {et~al.} 2018, \mnras, 481, 2407, \dodoi{10.1093/mnras/sty2417}

\bibitem[{{McKinven} \& {CHIME/FRB Collaboration}(2022)}]{2022ATel15679....1M}
{McKinven}, R., \& {CHIME/FRB Collaboration}. 2022, The Astronomer's Telegram, 15679, 1

\bibitem[{{Metzger} {et~al.}(2017){Metzger}, {Berger}, \& {Margalit}}]{2017ApJ...841...14M}
{Metzger}, B.~D., {Berger}, E., \& {Margalit}, B. 2017, \apj, 841, 14, \dodoi{10.3847/1538-4357/aa633d}

\bibitem[{{Metzger} {et~al.}(2019){Metzger}, {Margalit}, \& {Sironi}}]{Metzger2019}
{Metzger}, B.~D., {Margalit}, B., \& {Sironi}, L. 2019, \mnras, 485, 4091, \dodoi{10.1093/mnras/stz700}

\bibitem[{{Michilli} {et~al.}(2023){Michilli}, {Bhardwaj}, {Brar}, {Gaensler}, {Kaspi}, {Kirichenko}, {Masui}, {Mckinven}, {Ng}, {Patel}, {Sand}, {Scholz}, {Shin}, {Siegel}, {Stairs}, {Cassanelli}, {Cook}, {Dobbs}, {Dong}, {Fonseca}, {Ibik}, {Kaczmarek}, {Leung}, {Pearlman}, {Petroff}, {Pleunis}, {Rafiei-Ravandi}, {Sanghavi}, {Shaw}, \& {Tendulkar}}]{2023ApJ...950..134M}
{Michilli}, D., {Bhardwaj}, M., {Brar}, C., {et~al.} 2023, \apj, 950, 134, \dodoi{10.3847/1538-4357/accf89}

\bibitem[{{Minhajur Rahaman} {et~al.}(2025){Minhajur Rahaman}, {Acharya}, {Beniamini}, \& {Granot}}]{2025arXiv250401125M}
{Minhajur Rahaman}, S., {Acharya}, S.~K., {Beniamini}, P., \& {Granot}, J. 2025, arXiv e-prints, arXiv:2504.01125, \dodoi{10.48550/arXiv.2504.01125}

\bibitem[{{Neustadt} {et~al.}(2021){Neustadt}, {Kochanek}, {Stanek}, {Basinger}, {Jayasinghe}, {Garling}, {Adams}, \& {Gerke}}]{2021MNRAS.508..516N}
{Neustadt}, J.~M.~M., {Kochanek}, C.~S., {Stanek}, K.~Z., {et~al.} 2021, \mnras, 508, 516, \dodoi{10.1093/mnras/stab2605}

\bibitem[{{Newburgh} {et~al.}(2016){Newburgh}, {Bandura}, {Bucher}, {Chang}, {Chiang}, {Cliche}, {Dav{\'e}}, {Dobbs}, {Clarkson}, {Ganga}, {Gogo}, {Gumba}, {Gupta}, {Hilton}, {Johnstone}, {Karastergiou}, {Kunz}, {Lokhorst}, {Maartens}, {Macpherson}, {Mdlalose}, {Moodley}, {Ngwenya}, {Parra}, {Peterson}, {Recnik}, {Saliwanchik}, {Santos}, {Sievers}, {Smirnov}, {Stronkhorst}, {Taylor}, {Vanderlinde}, {Van Vuuren}, {Weltman}, \& {Witzemann}}]{2016SPIE.9906E..5XN}
{Newburgh}, L.~B., {Bandura}, K., {Bucher}, M.~A., {et~al.} 2016, in Society of Photo-Optical Instrumentation Engineers (SPIE) Conference Series, Vol. 9906, Ground-based and Airborne Telescopes VI, ed. H.~J. {Hall}, R.~{Gilmozzi}, \& H.~K. {Marshall}, 99065X, \dodoi{10.1117/12.2234286}

\bibitem[{{Ng} \& {CHIME/FRB Collaboration}(2025)}]{2025ATel17081....1N}
{Ng}, M., \& {CHIME/FRB Collaboration}. 2025, The Astronomer's Telegram, 17081, 1

\bibitem[{{Nicholl} {et~al.}(2017){Nicholl}, {Williams}, {Berger}, {Villar}, {Alexander}, {Eftekhari}, \& {Metzger}}]{2017ApJ...843...84N}
{Nicholl}, M., {Williams}, P.~K.~G., {Berger}, E., {et~al.} 2017, \apj, 843, 84, \dodoi{10.3847/1538-4357/aa794d}

\bibitem[{{Niu} {et~al.}(2022){Niu}, {Aggarwal}, {Li}, {Zhang}, {Chatterjee}, {Tsai}, {Yu}, {Law}, {Burke-Spolaor}, {Cordes}, {Zhang}, {Ocker}, {Yao}, {Wang}, {Feng}, {Niino}, {Bochenek}, {Cruces}, {Connor}, {Jiang}, {Dai}, {Luo}, {Li}, {Miao}, {Niu}, {Anna-Thomas}, {Sydnor}, {Stern}, {Wang}, {Yuan}, {Yue}, {Zhou}, {Yan}, {Zhu}, \& {Zhang}}]{2022Natur.606..873N}
{Niu}, C.~H., {Aggarwal}, K., {Li}, D., {et~al.} 2022, \nat, 606, 873, \dodoi{10.1038/s41586-022-04755-5}

\bibitem[{{Ocker} {et~al.}(2022){Ocker}, {Cordes}, {Chatterjee}, \& {Gorsuch}}]{2022ApJ...934...71O}
{Ocker}, S.~K., {Cordes}, J.~M., {Chatterjee}, S., \& {Gorsuch}, M.~R. 2022, \apj, 934, 71, \dodoi{10.3847/1538-4357/ac75ba}

\bibitem[{{Pastorello} {et~al.}(2012){Pastorello}, {Pumo}, {Navasardyan}, {Zampieri}, {Turatto}, {Sollerman}, {Taddia}, {Kankare}, {Mattila}, {Nicolas}, {Prosperi}, {San Segundo Delgado}, {Taubenberger}, {Boles}, {Bachini}, {Benetti}, {Bufano}, {Cappellaro}, {Cason}, {Cetrulo}, {Ergon}, {Germany}, {Harutyunyan}, {Howerton}, {Hurst}, {Patat}, {Stritzinger}, {Strolger}, \& {Wells}}]{Pastorello2011}
{Pastorello}, A., {Pumo}, M.~L., {Navasardyan}, H., {et~al.} 2012, \aap, 537, A141, \dodoi{10.1051/0004-6361/201118112}

\bibitem[{{Piro}(2016)}]{Piro2016}
{Piro}, A.~L. 2016, \apjl, 824, L32, \dodoi{10.3847/2041-8205/824/2/L32}

\bibitem[{{Piro} \& {Gaensler}(2018)}]{Piro&Gaensler18}
{Piro}, A.~L., \& {Gaensler}, B.~M. 2018, \apj, 861, 150, \dodoi{10.3847/1538-4357/aac9bc}

\bibitem[{{Planck Collaboration} {et~al.}(2020){Planck Collaboration}, {Aghanim}, {Akrami}, {Ashdown}, {Aumont}, {Baccigalupi}, {Ballardini}, {Banday}, {Barreiro}, {Bartolo}, {Basak}, {Battye}, {Benabed}, {Bernard}, {Bersanelli}, {Bielewicz}, {Bock}, {Bond}, {Borrill}, {Bouchet}, {Boulanger}, {Bucher}, {Burigana}, {Butler}, {Calabrese}, {Cardoso}, {Carron}, {Challinor}, {Chiang}, {Chluba}, {Colombo}, {Combet}, {Contreras}, {Crill}, {Cuttaia}, {de Bernardis}, {de Zotti}, {Delabrouille}, {Delouis}, {Di Valentino}, {Diego}, {Dor{\'e}}, {Douspis}, {Ducout}, {Dupac}, {Dusini}, {Efstathiou}, {Elsner}, {En{\ss}lin}, {Eriksen}, {Fantaye}, {Farhang}, {Fergusson}, {Fernandez-Cobos}, {Finelli}, {Forastieri}, {Frailis}, {Fraisse}, {Franceschi}, {Frolov}, {Galeotta}, {Galli}, {Ganga}, {G{\'e}nova-Santos}, {Gerbino}, {Ghosh}, {Gonz{\'a}lez-Nuevo}, {G{\'o}rski}, {Gratton}, {Gruppuso}, {Gudmundsson}, {Hamann}, {Handley}, {Hansen}, {Herranz}, {Hildebrandt}, {Hivon}, {Huang}, {Jaffe}, {Jones}, {Karakci}, {Keih{\"a}nen},
  {Keskitalo}, {Kiiveri}, {Kim}, {Kisner}, {Knox}, {Krachmalnicoff}, {Kunz}, {Kurki-Suonio}, {Lagache}, {Lamarre}, {Lasenby}, {Lattanzi}, {Lawrence}, {Le Jeune}, {Lemos}, {Lesgourgues}, {Levrier}, {Lewis}, {Liguori}, {Lilje}, {Lilley}, {Lindholm}, {L{\'o}pez-Caniego}, {Lubin}, {Ma}, {Mac{\'\i}as-P{\'e}rez}, {Maggio}, {Maino}, {Mandolesi}, {Mangilli}, {Marcos-Caballero}, {Maris}, {Martin}, {Martinelli}, {Mart{\'\i}nez-Gonz{\'a}lez}, {Matarrese}, {Mauri}, {McEwen}, {Meinhold}, {Melchiorri}, {Mennella}, {Migliaccio}, {Millea}, {Mitra}, {Miville-Desch{\^e}nes}, {Molinari}, {Montier}, {Morgante}, {Moss}, {Natoli}, {N{\o}rgaard-Nielsen}, {Pagano}, {Paoletti}, {Partridge}, {Patanchon}, {Peiris}, {Perrotta}, {Pettorino}, {Piacentini}, {Polastri}, {Polenta}, {Puget}, {Rachen}, {Reinecke}, {Remazeilles}, {Renzi}, {Rocha}, {Rosset}, {Roudier}, {Rubi{\~n}o-Mart{\'\i}n}, {Ruiz-Granados}, {Salvati}, {Sandri}, {Savelainen}, {Scott}, {Shellard}, {Sirignano}, {Sirri}, {Spencer}, {Sunyaev}, {Suur-Uski}, {Tauber}, {Tavagnacco},
  {Tenti}, {Toffolatti}, {Tomasi}, {Trombetti}, {Valenziano}, {Valiviita}, {Van Tent}, {Vibert}, {Vielva}, {Villa}, {Vittorio}, {Wandelt}, {Wehus}, {White}, {White}, {Zacchei}, \& {Zonca}}]{2020A&A...641A...6P}
{Planck Collaboration}, {Aghanim}, N., {Akrami}, Y., {et~al.} 2020, \aap, 641, A6, \dodoi{10.1051/0004-6361/201833910}

\bibitem[{{Prentice} \& {Mazzali}(2017)}]{2017MNRAS.469.2672P}
{Prentice}, S.~J., \& {Mazzali}, P.~A. 2017, \mnras, 469, 2672, \dodoi{10.1093/mnras/stx980}

\bibitem[{{Ravi}(2019)}]{2019raviNat}
{Ravi}, V. 2019, Nature Astronomy, 3, 928, \dodoi{10.1038/s41550-019-0831-y}

\bibitem[{{Sarbadhicary} {et~al.}(2019){Sarbadhicary}, {Badenes}, {Chomiuk}, {Caprioli}, \& {Huizenga}}]{2019MNRAS.487.5813S}
{Sarbadhicary}, S.~K., {Badenes}, C., {Chomiuk}, L., {Caprioli}, D., \& {Huizenga}, D. 2019, \mnras, 487, 5813, \dodoi{10.1093/mnras/stz1490}

\bibitem[{{Sautron} {et~al.}(2025){Sautron}, {McEwen}, {Younes}, {P{\'e}tri}, {Beniamini}, \& {Huppenkothen}}]{2025arXiv250311875S}
{Sautron}, M., {McEwen}, A.~E., {Younes}, G., {et~al.} 2025, arXiv e-prints, arXiv:2503.11875, \dodoi{10.48550/arXiv.2503.11875}

\bibitem[{{Shah} {et~al.}(2025){Shah}, {Shin}, {Leung}, {Fong}, {Eftekhari}, {Amiri}, {Andersen}, {Andrew}, {Bhardwaj}, {Brar}, {Cassanelli}, {Chatterjee}, {Curtin}, {Dobbs}, {Dong}, {Dong}, {Fonseca}, {Gaensler}, {Halpern}, {Hessels}, {Ibik}, {Jain}, {Joseph}, {Kaczmarek}, {Kahinga}, {Kaspi}, {Kharel}, {Landecker}, {Lanman}, {Lazda}, {Main}, {Mas-Ribas}, {Masui}, {Mckinven}, {Mena-Parra}, {Meyers}, {Michilli}, {Nimmo}, {Pandhi}, {Patil}, {Pearlman}, {Pleunis}, {Prochaska}, {Rafiei-Ravandi}, {Sammons}, {Sand}, {Scholz}, {Smith}, \& {Stairs}}]{2025ApJ...979L..21S}
{Shah}, V., {Shin}, K., {Leung}, C., {et~al.} 2025, \apjl, 979, L21, \dodoi{10.3847/2041-8213/ad9ddc}

\bibitem[{{Shannon} {et~al.}(2018){Shannon}, {Macquart}, {Bannister}, {Ekers}, {James}, {Os{\l}owski}, {Qiu}, {Sammons}, {Hotan}, {Voronkov}, {Beresford}, {Brothers}, {Brown}, {Bunton}, {Chippendale}, {Haskins}, {Leach}, {Marquarding}, {McConnell}, {Pilawa}, {Sadler}, {Troup}, {Tuthill}, {Whiting}, {Allison}, {Anderson}, {Bell}, {Collier}, {G{\"u}rkan}, {Heald}, \& {Riseley}}]{2018Natur.562..386S}
{Shannon}, R.~M., {Macquart}, J.~P., {Bannister}, K.~W., {et~al.} 2018, \nat, 562, 386, \dodoi{10.1038/s41586-018-0588-y}

\bibitem[{{Sharma} {et~al.}(2023){Sharma}, {Somalwar}, {Law}, {Ravi}, {Catha}, {Chen}, {Connor}, {Faber}, {Hallinan}, {Harnach}, {Hellbourg}, {Hobbs}, {Hodge}, {Hodges}, {Lamb}, {Rasmussen}, {Sherman}, {Shi}, {Simard}, {Squillace}, {Weinreb}, {Woody}, {Yadlapalli}, \& {Deep Synoptic Array Team}}]{2023ApJ...950..175S}
{Sharma}, K., {Somalwar}, J., {Law}, C., {et~al.} 2023, \apj, 950, 175, \dodoi{10.3847/1538-4357/accf1d}

\bibitem[{{Sharma} {et~al.}(2024){Sharma}, {Ravi}, {Connor}, {Law}, {Ocker}, {Sherman}, {Kosogorov}, {Faber}, {Hallinan}, {Harnach}, {Hellbourg}, {Hobbs}, {Hodge}, {Hodges}, {Lamb}, {Rasmussen}, {Somalwar}, {Weinreb}, {Woody}, {Leja}, {Anand}, {Das}, {Qin}, {Rose}, {Dong}, {Miller}, \& {Yao}}]{2024Natur.635...61S}
{Sharma}, K., {Ravi}, V., {Connor}, L., {et~al.} 2024, \nat, 635, 61, \dodoi{10.1038/s41586-024-08074-9}

\bibitem[{{Shin} {et~al.}(2023){Shin}, {Masui}, {Bhardwaj}, {Cassanelli}, {Chawla}, {Dobbs}, {Dong}, {Fonseca}, {Gaensler}, {Herrera-Mart{\'\i}n}, {Kaczmarek}, {Kaspi}, {Leung}, {Merryfield}, {Michilli}, {M{\"u}nchmeyer}, {Pearlman}, {Rafiei-Ravandi}, {Smith}, {Stairs}, \& {Tendulkar}}]{2023ApJ...944..105S}
{Shin}, K., {Masui}, K.~W., {Bhardwaj}, M., {et~al.} 2023, \apj, 944, 105, \dodoi{10.3847/1538-4357/acaf06}

\bibitem[{{Shin} {et~al.}(2025){Shin}, {Curtin}, {Fine}, {Pandhi}, {Andrew}, {Bhardwaj}, {Chatterjee}, {Cook}, {Fonseca}, {Gaensler}, {Hessels}, {Jain}, {Kaspi}, {Kharel}, {Lanman}, {Lazda}, {Leung}, {Main}, {Masui}, {Michilli}, {Ng}, {Nimmo}, {Pearlman}, {Pen}, {Pleunis}, {Rafiei-Ravandi}, {Sammons}, {Sand}, {Scholz}, {Smith}, \& {Stairs}}]{2025arXiv250513297S}
{Shin}, K., {Curtin}, A., {Fine}, M., {et~al.} 2025, arXiv e-prints, arXiv:2505.13297, \dodoi{10.48550/arXiv.2505.13297}

\bibitem[{{Spitler} {et~al.}(2016){Spitler}, {Scholz}, {Hessels}, {Bogdanov}, {Brazier}, {Camilo}, {Chatterjee}, {Cordes}, {Crawford}, {Deneva}, {Ferdman}, {Freire}, {Kaspi}, {Lazarus}, {Lynch}, {Madsen}, {McLaughlin}, {Patel}, {Ransom}, {Seymour}, {Stairs}, {Stappers}, {van Leeuwen}, \& {Zhu}}]{2016Natur.531..202S}
{Spitler}, L.~G., {Scholz}, P., {Hessels}, J.~W.~T., {et~al.} 2016, \nat, 531, 202, \dodoi{10.1038/nature17168}

\bibitem[{{Sukhbold} {et~al.}(2016){Sukhbold}, {Ertl}, {Woosley}, {Brown}, \& {Janka}}]{2016ApJ...821...38S}
{Sukhbold}, T., {Ertl}, T., {Woosley}, S.~E., {Brown}, J.~M., \& {Janka}, H.~T. 2016, \apj, 821, 38, \dodoi{10.3847/0004-637X/821/1/38}

\bibitem[{{Taylor} {et~al.}(2014){Taylor}, {Cinabro}, {Dilday}, {Galbany}, {Gupta}, {Kessler}, {Marriner}, {Nichol}, {Richmond}, {Schneider}, \& {Sollerman}}]{2014ApJ...792..135T}
{Taylor}, M., {Cinabro}, D., {Dilday}, B., {et~al.} 2014, \apj, 792, 135, \dodoi{10.1088/0004-637X/792/2/135}

\bibitem[{{Thompson} {et~al.}(1980){Thompson}, {Clark}, {Wade}, \& {Napier}}]{1980ApJS...44..151T}
{Thompson}, A.~R., {Clark}, B.~G., {Wade}, C.~M., \& {Napier}, P.~J. 1980, \apjs, 44, 151, \dodoi{10.1086/190688}

\bibitem[{{Thornton} {et~al.}(2013){Thornton}, {Stappers}, {Bailes}, {Barsdell}, {Bates}, {Bhat}, {Burgay}, {Burke-Spolaor}, {Champion}, {Coster}, {D'Amico}, {Jameson}, {Johnston}, {Keith}, {Kramer}, {Levin}, {Milia}, {Ng}, {Possenti}, \& {van Straten}}]{Thornton2013}
{Thornton}, D., {Stappers}, B., {Bailes}, M., {et~al.} 2013, Science, 341, 53, \dodoi{10.1126/science.1236789}

\bibitem[{{Tsvetkov} {et~al.}(2004){Tsvetkov}, {Pavlyuk}, \& {Bartunov}}]{2004AstL...30..729T}
{Tsvetkov}, D.~Y., {Pavlyuk}, N.~N., \& {Bartunov}, O.~S. 2004, Astronomy Letters, 30, 729, \dodoi{10.1134/1.1819491}

\bibitem[{{Turatto} {et~al.}(2007){Turatto}, {Benetti}, \& {Pastorello}}]{2007AIPC..937..187T}
{Turatto}, M., {Benetti}, S., \& {Pastorello}, A. 2007, in American Institute of Physics Conference Series, Vol. 937, Supernova 1987A: 20 Years After: Supernovae and Gamma-Ray Bursters, ed. S.~{Immler}, K.~{Weiler}, \& R.~{McCray} (AIP), 187--197, \dodoi{10.1063/1.3682902}

\bibitem[{{Vanderlinde} {et~al.}(2019){Vanderlinde}, {Liu}, {Gaensler}, {Bond}, {Hinshaw}, {Ng}, {Chiang}, {Stairs}, {Brown}, {Sievers}, {Mena}, {Smith}, {Bandura}, {Masui}, {Spekkens}, {Belostotski}, {Dobbs}, {Turok}, {Boyle}, {Rupen}, {Landecker}, {Pen}, \& {Kaspi}}]{2019clrp.2020...28V}
{Vanderlinde}, K., {Liu}, A., {Gaensler}, B., {et~al.} 2019, in Canadian Long Range Plan for Astronomy and Astrophysics White Papers, Vol. 2020, 28, \dodoi{10.5281/zenodo.3765414}

\bibitem[{{Vink}(2012)}]{2012A&ARv..20...49V}
{Vink}, J. 2012, \aapr, 20, 49, \dodoi{10.1007/s00159-011-0049-1}

\bibitem[{{Vink}(2020)}]{2020pesr.book.....V}
---. 2020, {Physics and Evolution of Supernova Remnants} (Springer), \dodoi{10.1007/978-3-030-55231-2}

\bibitem[{{Wang} {et~al.}(2025){Wang}, {Bannister}, {Gupta}, {Deng}, {Pilawa}, {Tuthill}, {Bunton}, {Flynn}, {Glowacki}, {Jaini}, {Lee}, {Lenc}, {Lucero}, {Paek}, {Radhakrishnan}, {Thyagarajan}, {Uttarkar}, {Wang}, {Bhat}, {James}, {Moss}, {Murphy}, {Reynolds}, {Shannon}, {Spitler}, {Tzioumis}, {Caleb}, {Deller}, {Gordon}, {Marnoch}, {Ryder}, {Simha}, {Anderson}, {Ball}, {Brodrick}, {Cooray}, {Gupta}, {Hayman}, {Ng}, {Pearce}, {Phillips}, {Voronkov}, \& {Westmeier}}]{2025PASA...42....5W}
{Wang}, Z., {Bannister}, K.~W., {Gupta}, V., {et~al.} 2025, \pasa, 42, e005, \dodoi{10.1017/pasa.2024.107}

\bibitem[{{Xing} {et~al.}(2024){Xing}, {Yu}, {Yan}, {Zhang}, \& {Zhang}}]{2024arXiv241106996X}
{Xing}, Y., {Yu}, W., {Yan}, Z., {Zhang}, X., \& {Zhang}, B. 2024, arXiv e-prints, arXiv:2411.06996, \dodoi{10.48550/arXiv.2411.06996}

\bibitem[{Yao {et~al.}(2017)Yao, Manchester, \& Wang}]{Yao2017}
Yao, J.~M., Manchester, R.~N., \& Wang, N. 2017, The Astrophysical Journal, 835, 29, \dodoi{10.3847/1538-4357/835/1/29}

\bibitem[{{Yaron} {et~al.}(2020){Yaron}, {Ofek}, {Gal-Yam}, \& {Sass}}]{2020TNSAN..70....1Y}
{Yaron}, O., {Ofek}, E., {Gal-Yam}, A., \& {Sass}, A. 2020, Transient Name Server AstroNote, 70, 1

\bibitem[{{Zhang}(2023)}]{2023RvMP...95c5005Z}
{Zhang}, B. 2023, Reviews of Modern Physics, 95, 035005, \dodoi{10.1103/RevModPhys.95.035005}

\bibitem[{{Zonca} {et~al.}(2020){Zonca}, {Singer}, {Lenz}, {Reinecke}, {Rosset}, {Hivon}, \& {Gorski}}]{zonca2020healpy}
{Zonca}, A., {Singer}, L., {Lenz}, D., {et~al.} 2020, {healpy: Python wrapper for HEALPix}

\end{thebibliography}

\appendix

\section{Maximum DM-Excess constraint}
\label{app:max-dmexcess}

The observed DM of an FRB can be decomposed into contributions from distinct regions along the line of sight:
\begin{equation}
  \mathrm{DM}_{\mathrm{FRB}} = \mathrm{DM}_{\mathrm{ISM}} + \mathrm{DM}_{\mathrm{MW,halo}} + \mathrm{DM}_{\mathrm{IGM}} + \mathrm{DM}_{\mathrm{host}},
  \label{eq:DM}
\end{equation}
where $\mathrm{DM}_{\mathrm{ISM}}$ is the contribution from the Milky Way disk (estimated using the NE2001 \citep{Cordes2002} and YMW16 \citep{Yao2017} models), $\mathrm{DM}_{\mathrm{MW,halo}}$ is that from the Galactic halo, $\mathrm{DM}_{\mathrm{IGM}}$ arises from the intergalactic medium (IGM), and $\mathrm{DM}_{\mathrm{host}}$ is the contribution from the FRB’s host galaxy.

In the FRB Catalog-1, the excess DM is defined as
\begin{equation}
  \mathrm{DM}_{\mathrm{excess}} = \mathrm{DM}_{\mathrm{FRB}} - \mathrm{DM}_{\mathrm{ISM}},
\end{equation}
thereby isolating the combined contributions of the Milky Way halo, the IGM, and the host galaxy. We adopt a value of $\mathrm{DM}_{\mathrm{MW,halo}} = 50$~pc\,cm$^{-3}$ \citep{2021ApJ...910L..18B,2023ApJ...946...58C}. The intergalactic contribution, $\mathrm{DM}_{\mathrm{IGM}}$, is computed using the semi-analytical formulation of \citet{Macquart2020}.

The host galaxy contribution is modelled with a log-normal probability density function:
\begin{equation}
  f_{\mathrm{DM,host}}(x)=\frac{1}{\sqrt{2\pi}\,x\,\sigma}\exp\!\left[-\frac{(\ln x - \mu)^2}{2\sigma^2}\right],
\end{equation}
where $\mu$ and $\sigma$ denote, respectively, the mean and standard deviation of the natural logarithm of the rest-frame host DM, $\ln(\mathrm{DM}_{\mathrm{host,rf}})$. Following the analysis of \citet{Kovacs2024}, the full sample at $z\approx0$ is characterized by a median host DM of $82\pm4$~pc\,cm$^{-3}$ $-$ implies $\mu \simeq \ln(82) \simeq 4.41$ $-$ and the standard deviation $\sigma \simeq 0.8$.
Although subsamples defined by the projected offset of the FRB from the host center can produce higher median values, these parameters are adopted here to represent a conservative estimate of $\mathrm{DM}_{\mathrm{host}}$. 

\begin{figure}
    \centering
    \includegraphics[width=0.6\linewidth]{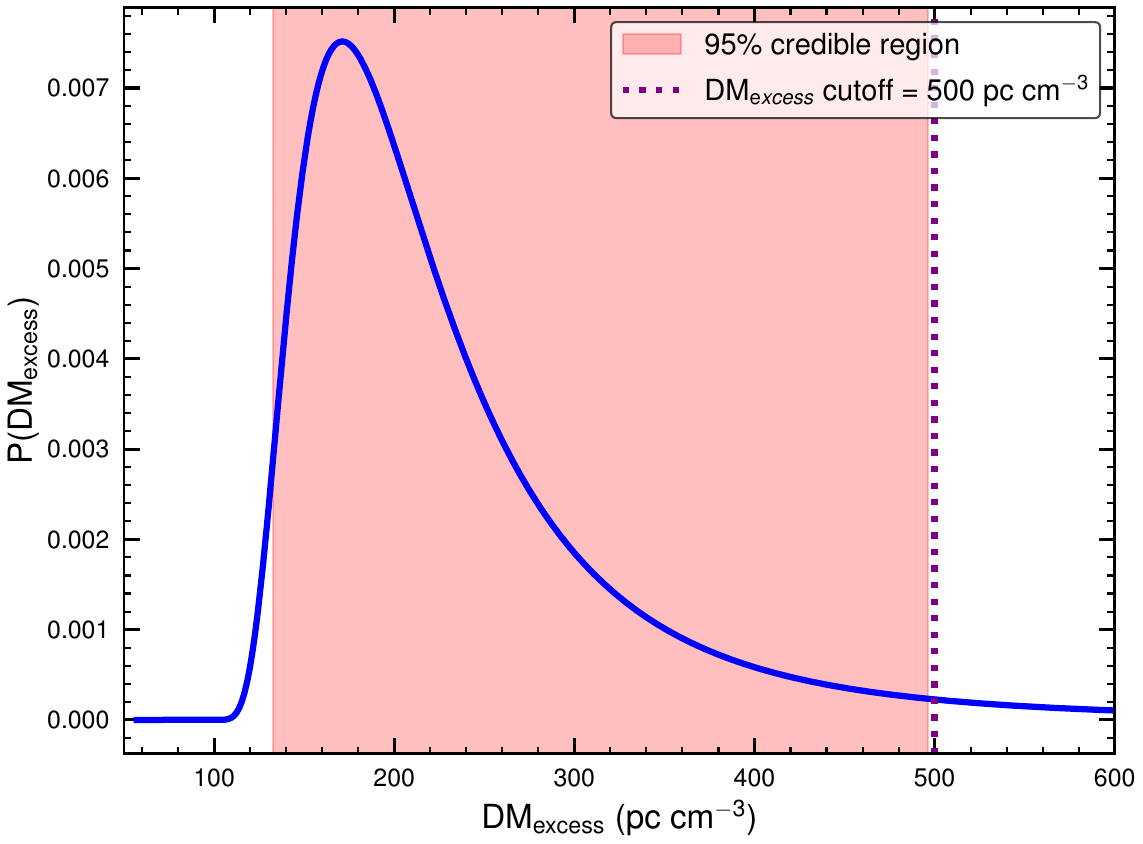}
    \caption{Posterior probability density of the extragalactic dispersion measure, \(\mathrm{DM}_{\mathrm{excess}}\), modelled as the sum of an IGM component, a log–normal host contribution, and a fixed Milky-Way halo term of 50 pc cm\(^{-3}\).  The red shaded region marks the 95\% credible interval; the adopted \(\mathrm{DM}_{\mathrm{excess}}\) cutoff is shown by the black dashed line.}
    \label{fig:dm_excess}
\end{figure}

For an FRB at $z=0.043$ (or 200~Mpc), the combination of these components leads to a posterior on the DM$_{\rm excess}$ which is shown in Figure \ref{fig:dm_excess}. From this posterior, we estimated 95\% credible region upper limit on the DM$_{\rm excess}$ $\approx$ 500~pc\,cm$^{-3}$.  We employ this DM$_{\rm excess}$ in \S\ref{sec:sample_frb} to select the FRBs which for the cross-match analysis. Post sample selection, we can also use the scattering time $\tau$ of the selected FRB to further constrain the selected sample for false positive as discussed in Appendix \ref{app:host_dm_scattering}.

\begin{figure}[htbp]
    \centering
    \renewcommand{\arraystretch}{0}  

    \begin{tabular}{c c}
        \includegraphics[width=0.5\linewidth]{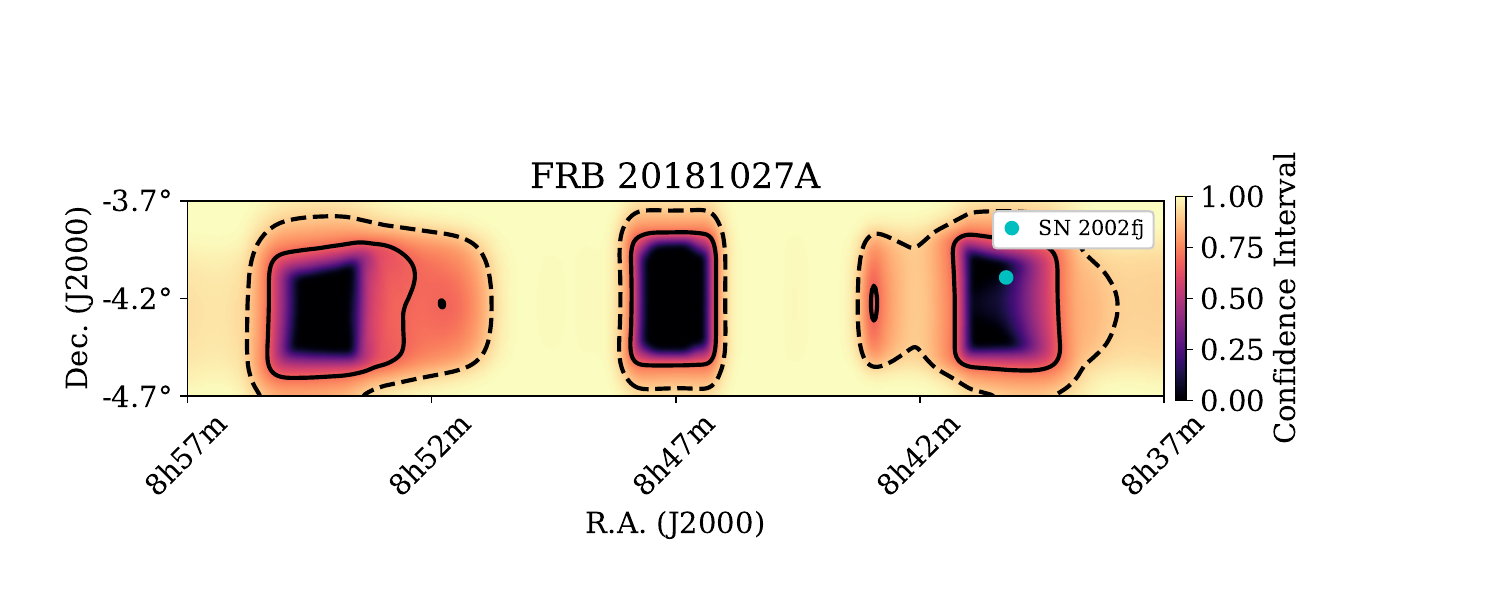} & 
        \includegraphics[width=0.5\linewidth]{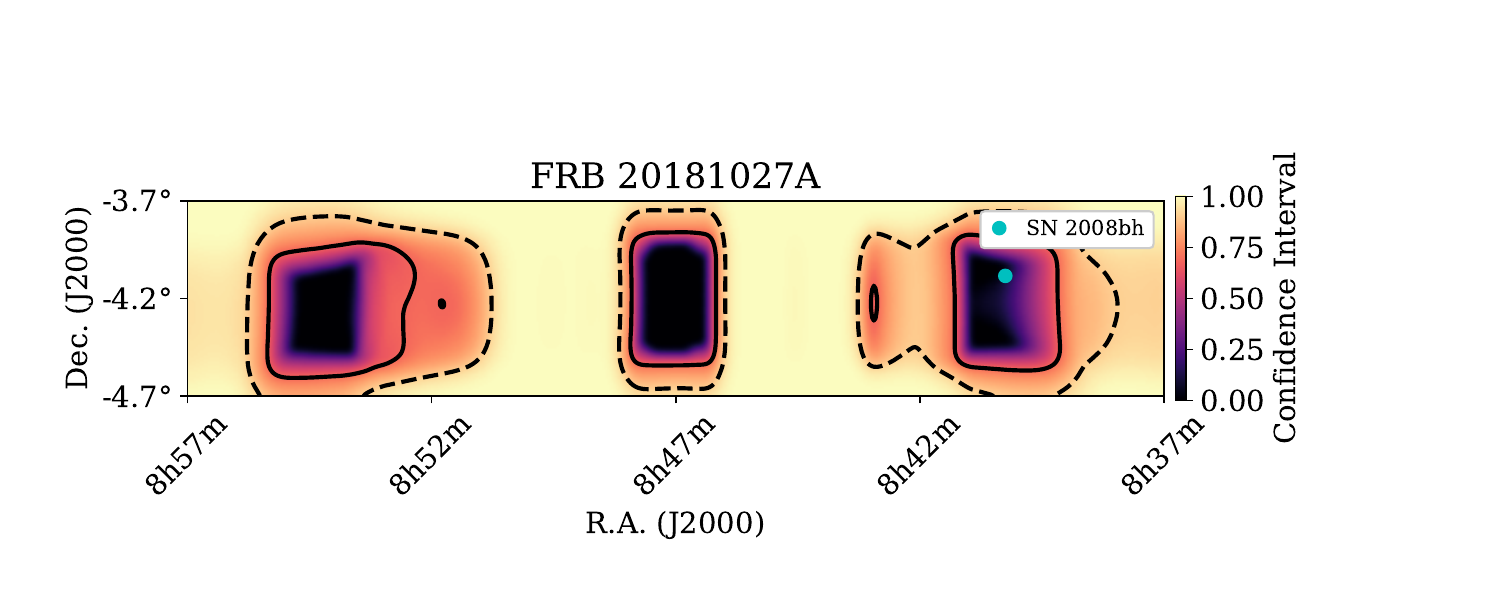} \\

        \includegraphics[width=0.5\linewidth]{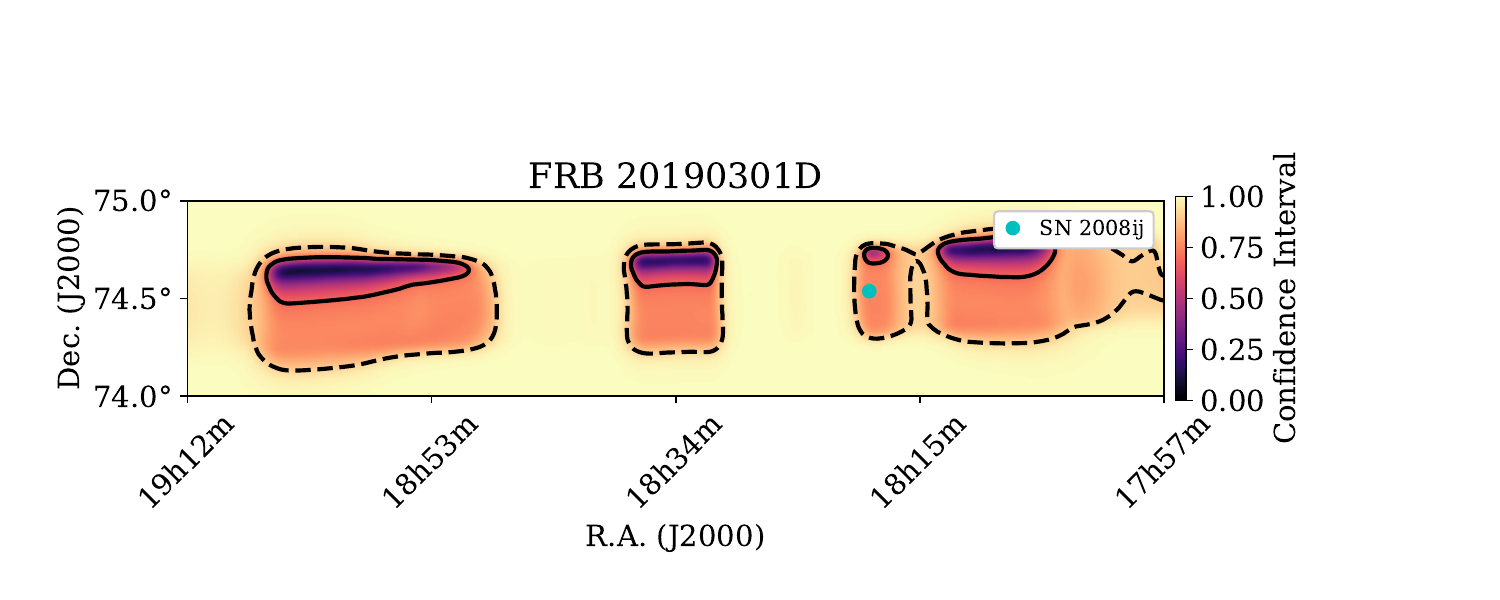} & 
        \includegraphics[width=0.5\linewidth]{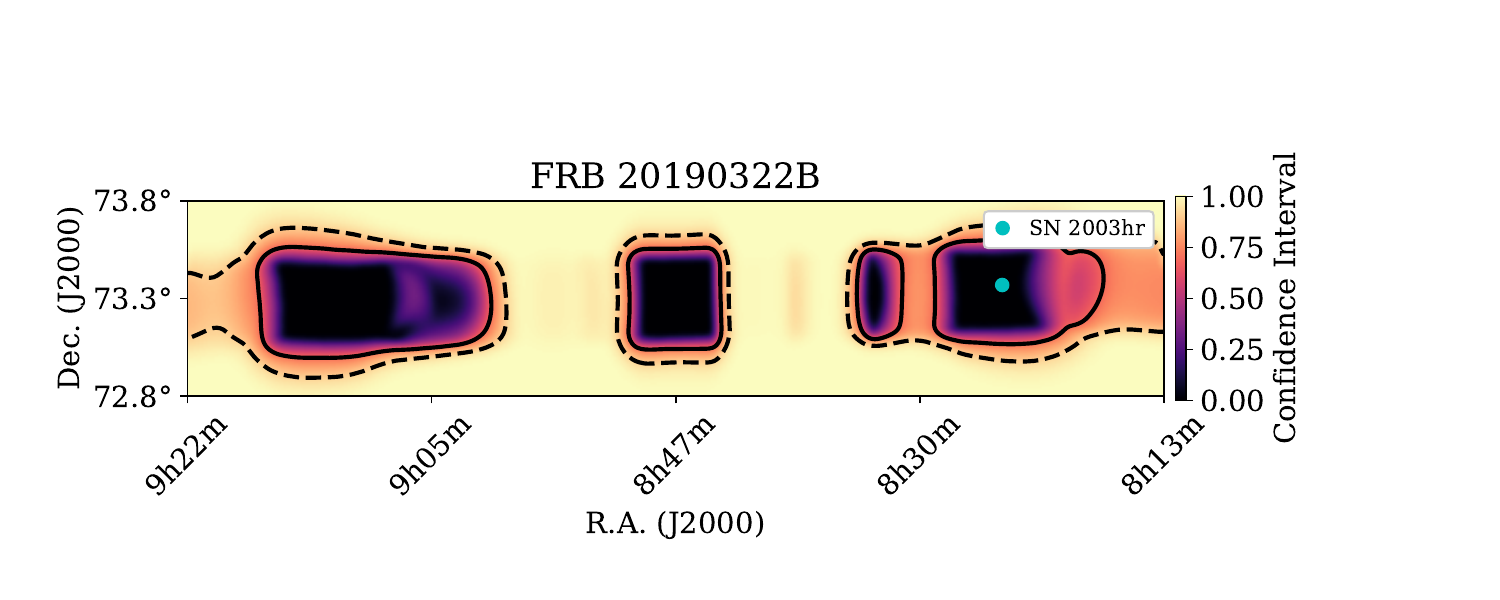} \\

        \includegraphics[width=0.5\linewidth]{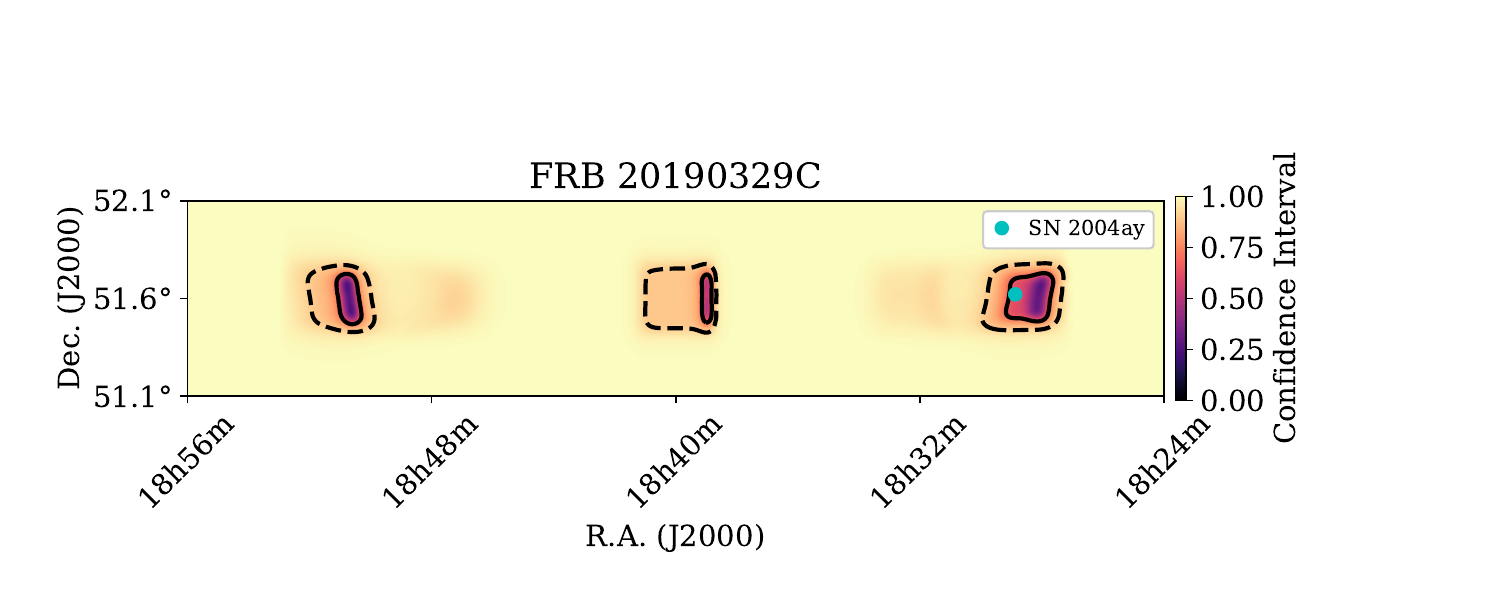} & 
        \includegraphics[width=0.5\linewidth]{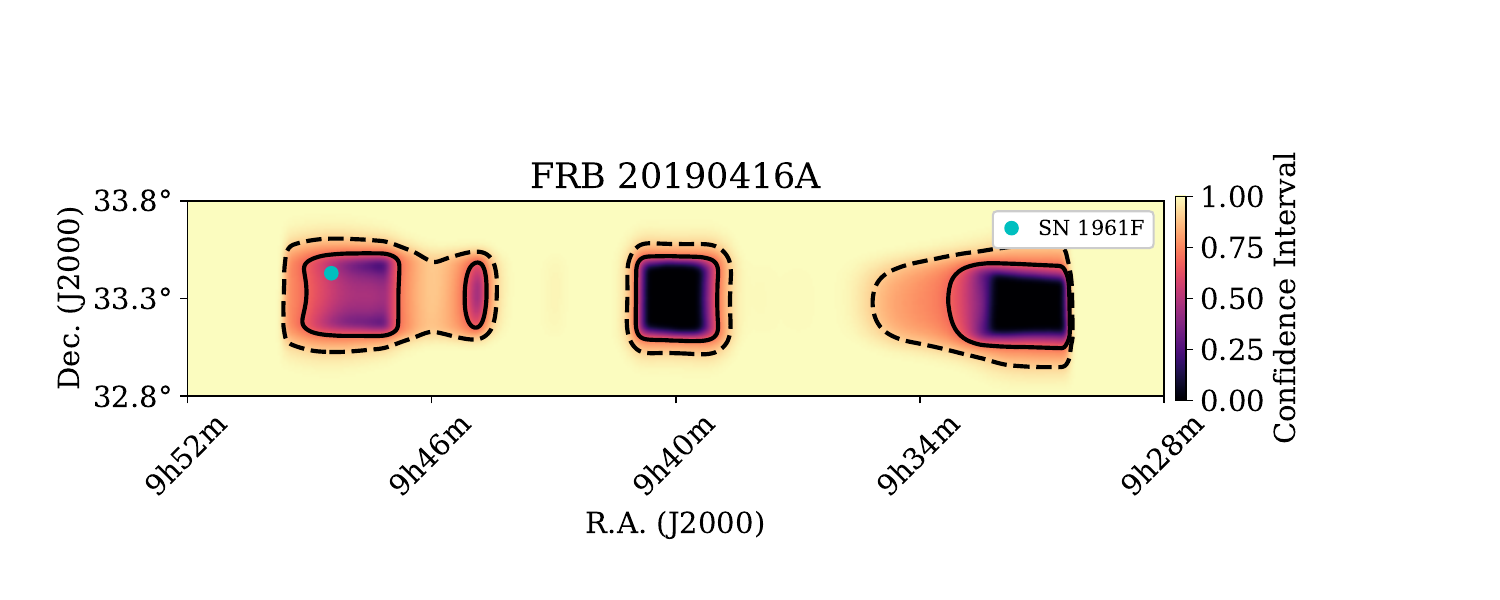} \\
    \end{tabular}
    \includegraphics[width=0.5\linewidth]{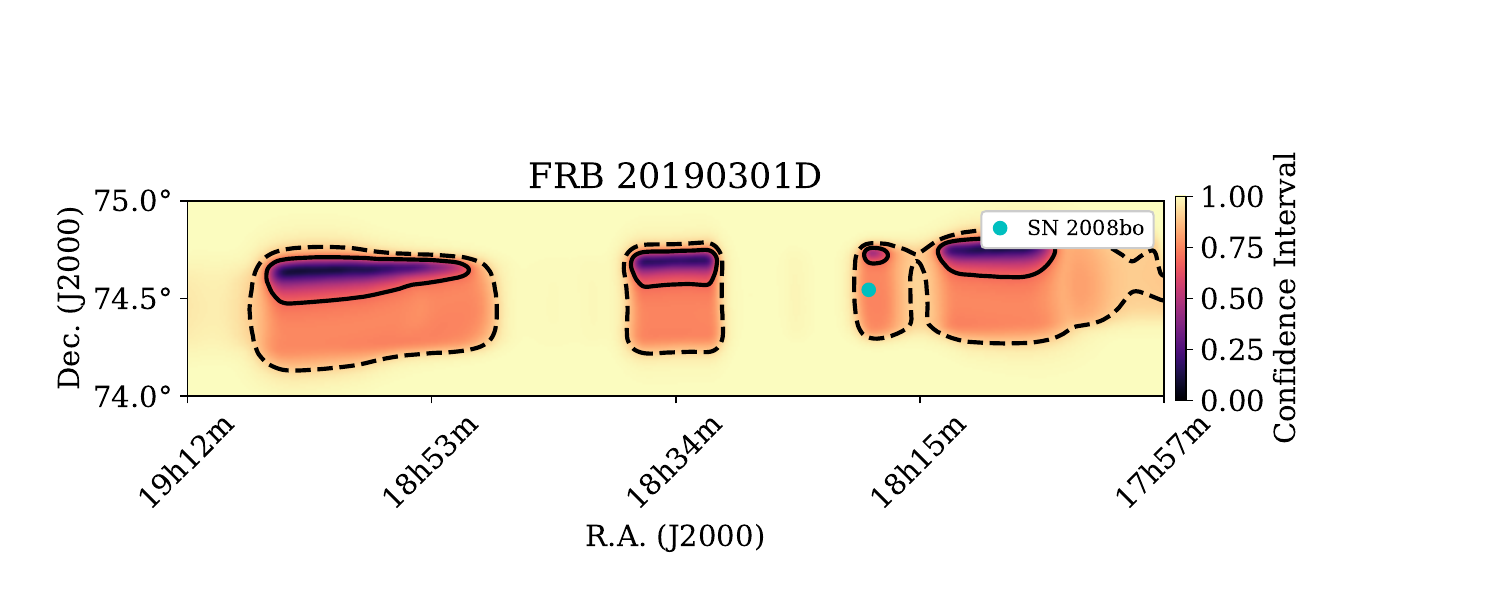} \\

    \caption{Header localization confidence regions for additional seven CHIME/FRB sources that coincide with a supernova if the DM$_{\rm excess}$ cut-off constraint discussed in Appendix \ref{app:max-dmexcess} is relaxed. The description of the plots is same as in Figure \ref{fig:frb_loc}.}
    \label{fig:frb_loc_seven}
\end{figure}

Finally, if we relax the DM cut-off and hence, consider all 474 non-repeating FRBs for the cross-matching analysis, we find additional seven FRB-SN pairs. Important properties and localization region plots of the seven FRB-SN pairs are shown in Table \ref{app_tab:1}, \ref{app_tab:2} and Figure \ref{fig:frb_loc_seven}, respectively. However, using the DM$_\mathrm{host,\tau}$ values estimated using the formalism described in Appendix \ref{app:host_dm_scattering}, none of the seven FRBs satisfy the Equation \ref{eq:DM} and are therefore unlikely to be associated with their corresponding supernova host galaxies. Furthermore, no compact VLASS source (persistent or transient) is detected at the positions of the seven CCSNe that overlap CHIME localizations (see \S\ref{sec:frb_prs}).

\begin{table}[ht]
\centering
\caption{Additional Detection Pairs: SN Properties (same as Table \ref{tab:sn_properties})}
\begin{tabular}{ccccccccc}
\toprule
\shortstack{Pair \\ Number} &
\shortstack{SN \\ $ $} &
\shortstack{Host Galaxy \\ $ $} &
\shortstack{$z$ \\ $ $} &
\shortstack{Type \\ $ $} &
\shortstack{RA \\ (deg)} &
\shortstack{Dec \\ (deg)} &
\shortstack{CHIME Exposure \\ T$_{\mathrm{exp}}$ (hrs)} &
\shortstack{Proj. SN host offset \\ (kpc)} \\
\midrule
5 & 2002fj & NGC2642 & 0.014 & IIn & 130.1879 & -4.1274 & 17.2 & 6.8 \\
6 & 2008bh & NGC2642 & 0.014 & II & 130.1923 & -4.1193 & 16.5 & 8.2 \\
7 & 2008ij & NGC6643 & 0.005 & II & 274.9659 & 74.5653 & 68.7 & 2.4 \\
8 & 2003hr & NGC2551 & 0.008 & II & 126.1628 & 73.4065 & 58.2 & 8.3 \\
9 & 2004ay & UGC11255 & 0.032 & IIn & 277.2399 & 51.6488 & 29.9 & 5.1 \\
10 & 1961F & NGC3003 & 0.005 & IIPec & 147.1596 & 33.4264 & 12.4 & 3.8 \\
11 & 2008bo & NGC6643 & 0.005 & Ib/c & 274.9765 & 74.5726 & 68.7 & 3.5 \\
\bottomrule
\end{tabular}
\label{app_tab:1}
\end{table}

\begin{table}[ht]
\centering
\caption{Additional Detection Pairs: FRB Properties (same as Table \ref{tab:frb_properties})}
\begin{tabular}{cccccccccc}
\toprule
\shortstack{Pair \\ Number} &
\shortstack{FRB \\ Name } &
\shortstack{RA \\ (deg)$^{a}$} &
\shortstack{Dec \\ (deg)$^{a}$} &
\shortstack{DM \\ (pc cm$^{-3}$)} &
\shortstack{DM$_\mathrm{excess\_NE2001}$ \\ (pc cm$^{-3}$)} &
\shortstack{DM$_\mathrm{excess\_YMW16}$ \\ (pc cm$^{-3}$)} &
\shortstack{DM$_\mathrm{Host,\tau,max}$ \\ (pc cm$^{-3}$)} &
\shortstack{$\tau$ \\ (ms)} &
\shortstack{Temp.\ offset$^{\,b}$ \\ (months)} \\
\midrule
5 & 20181027A & 131.9 & -4.240 & 727.7 & 664 & 670 & 115 & 0.0047(7) & 151 \\
6 & 20181027A & 131.9 & -4.240 & 727.7 & 664 & 670 & 112 & 0.0047(7) & 114 \\
7 & 20190301D & 278.7 & 74.68 & 1160.7 & 1108 & 1112 & 39 & 0.00053(8) & 84 \\
8 & 20190322B & 132.0 & 73.34 & 577.0 & 530 & 536 & 70 & $<$0.0017 & 155 \\
9 & 20190329C & 279.8 & 51.63 & 1256.4 & 1196 & 1202 & 48 & 0.0008(1) & 155 \\
10 & 20190416A & 145.0 & 33.30 & 2287.3 & 2248 & 2261 & 70 & 0.0017(4) & 156 \\
11 & 20190301D & 278.7 & 74.68 & 1161.7 & 1108 & 1112 & 37 & 0.00053(8) & 132 \\

\bottomrule
\end{tabular}

\label{app_tab:2}
\end{table}

\section{Scattering time constraint}\label{app:host_dm_scattering}

In Appendix~\ref{app:max-dmexcess}, we adopted literature priors on the
host contribution to the dispersion measure (DM\(_{\rm host}\)) as part
of a pre-planned hypothesis test \citep{2024ApJ...971L..51B}.  After
identifying the four candidate FRB–SN pairs, we refine the host
contribution using the observed FRB pulse broadening.
CHIME/FRB Catalog-1 provides scattering time \(\tau\) for all
four bursts.  Following the formalism of \cite{2022ApJ...931...88C},
we convert \(\tau\) into an upper limit on the host contribution
DM\(_{\rm host,\tau}\); combined with the IGM term at the supernova
redshift, this yields an expected DM\(_{\rm excess}\) that can be
compared with the measured value.

Contributions to the observed scattering are additive, i.e.,
\begin{equation}
  \tau(\nu) = \tau_{\rm MW} + \tau_{\rm halo} + \tau_{\rm IGM}(z)
              + \frac{\tau_{\rm IGH}}{(1+z_{\rm IGH})^{3}}
              + \frac{\tau_{\rm host}}{(1+z_{\rm host})^{3}}
              > \frac{\tau_{\rm host}}{(1+z_{\rm host})^{3}},
\label{eq:tau_budget}
\end{equation}
where \(\tau_{\rm IGH}\) denotes an optional intervening galaxy or
halo.  Assuming the host dominates the scattering, we constrain
DM\(_{\rm host}\) with the Equation 1 from \cite{2024Natur.634.1065B}
\begin{equation}
  \tau(\nu) \simeq 48.0 \,{\rm ns}\,
      A_\tau\,\widetilde{F}\,G_{\rm scatt}\,
      (1+z)^{-3}\,\nu^{-4}\,
      {\rm DM}_{\rm host}^{2},
\label{eq:tau_dm}
\end{equation}
where \(\nu\) is the observed frequency in GHz.  As discussed in \cite{2024Natur.634.1065B}, we adopt \(A_\tau=1\) and
\(G_{\rm scatt}=1\), appropriate when the scattering layer is
co-spatial with the burst source and geometrically thin relative to the
line-of-sight distance. Moreover, we assume that the host galaxies of the candidate SNe are late-type
(i.e.\ spiral or irregular) systems, consistent with the host morphology reported in the HECATE catalog (see \S\ref{sec:cross-match}).  In these galaxies, the thin disc contains the highest average free-electron density and dominates both the dispersion and scattering of Galactic pulsars
(e.g.\ \cite{Cordes2002}).  We therefore parameterize the turbulence
coefficient \(\widetilde{F}\) with the log-normal distribution for
spiral thin discs from \cite{2022ApJ...934...71O}.  Adopting this
distribution also implicitly accounts for variations with stellar mass
and star-formation rate across the HECATE sample.
The turbulence parameter
\(\widetilde{F}\) is drawn from the log-normal distribution for spiral
galaxy thin discs tabulated by \cite{2022ApJ...934...71O},
\(\widetilde{F}=1\pm0.5\;{\rm(pc^2\,km)^{-1/3}}\), which brackets the
range found in galaxies of similar stellar mass and morphology that of the four SNe.  With
\(\nu\) fixed at 0.6 GHz, the observed \(\tau\) values yield
90\% credible region upper limits on DM\(_{\rm host}\); these are reported in
Table~\ref{tab:frb_properties}.

If $\mathrm{DM}_{\rm excess}
  - \mathrm{DM}_{\rm MW,halo}
  - \mathrm{DM}_{\rm IGM}(z_{\rm SN})
  - \mathrm{DM}_{\rm host,\tau}
  \lesssim 0,$
the scattering constraint is satisfied and the FRB–SN association is
strengthened.  Only the FRB20190412B–SN2009gi pair meets this
criterion, making it the most plausible association among the
candidates listed in Tables~\ref{tab:sn_properties},
\ref{tab:frb_properties}, \ref{app_tab:1}, and \ref{app_tab:2}.

Note that the conversion in Equation \ref{eq:tau_dm} is valid when the plasma that produces the bulk of DM\(_{\rm host}\) also dominates the multi-path
scattering, as observed for young pulsars located within a few hundred
parsecs of the Galactic plane \citep{2022ApJ...931...88C}.  For the
FRB–SN sample this assumption is justified if (i) the burst originates
inside the ionized ISM of the host and (ii) any foreground screens
(e.g. intervening halos) make a negligible contribution to \(\tau\).
If scattering instead occurs in a foreground galaxy, the DM\(_{\rm
host,\tau}\) overestimates the true host dispersion; conversely, if the
screen is highly anisotropic, the same DM produces less temporal
smearing, weakening the constraint.  High-resolution polarization or
scintillation measurements could discriminate between these scenarios
and refine DM\(_{\rm host}\) estimates in future work.

\section{Assessing the Chance-Association Probability of FRB-CCSN Coincidences}
\label{app:pcc}

To quantify the false‐alarm probability that a random CCSN lies inside an FRB’s localization region purely by chance, we first estimate the sky density of young magnetars within our volume‐limited sample (\(z\le0.043\)).  Analytically, the CCSN volumetric rate is taken as $ \mathcal R_{\rm CCSN}
  = 1.06\times10^{-4}\,(h/0.7)^{3}\;\mathrm{yr^{-1}\,Mpc^{-3}}$
\citep{2014ApJ...792..135T}. Assuming that a fraction of CCSN \(f_{\rm NS}\approx0.8\) of CCSNe leave neutron‐star remnants \citep{2016ApJ...821...38S,2021Natur.589...29B}, and of these roughly \(f_{\rm mag}\approx0.2\) are born as magnetars \citep{2019MNRAS.487.1426B}.  Integrating to \(z=0.043\) gives  
\[
\dot N_{\rm mag}
  = f_{\rm NS}\,f_{\rm mag}
    \int_{0}^{0.043}
      \mathcal R_{\rm CCSN}(z)\,\frac{dV}{dz}\,dz
  \simeq 4.3\times10^{2}\;\mathrm{yr^{-1}}.
\]
Over the 129‐yr span of recorded CCSNe used in \S\ref{sec:sample_sn} (1885–2014) this implies \(\sim5.6\times10^{4}\) magnetars within z = 0.043.  Spreading these uniformly over \(4\pi\)sr (or 41,253 deg$^{-2}$) yields
\[
\Sigma_{\rm mag}^{\rm(an)}
  = \frac{5.6\times10^{4}}{4\pi}\times\frac{1}{3282.8}
  \simeq1.3\;\mathrm{deg^{-2}}.
\]

By contrast, the SIA catalog of historic SNe lists \(\Sigma_{\rm CCSN}=7.7\) deg\(^{-2}\) over the same period and sky area \citep{2004AstL...30..729T}.  Multiplying by \(f_{\rm NS}f_{\rm mag}=0.16\) gives an empirical magnetar surface density $\Sigma_{\rm mag}^{\rm(emp)}
  = 0.16\times7.7
  \simeq1.2\;\mathrm{deg^{-2}}.$
The \(\sim10\%\) agreement between the analytic and empirical estimates suggests that the SIA catalog is \(\gtrsim90\)\% complete to \(z=0.043\) ensuring volumetric completeness of the SN sample used in this study.  We therefore adopt $\Sigma_{\rm mag}\simeq1.3\;\mathrm{deg^{-2}}$ in the chance‐association calculation.

Next, we compute the daily probability \(P_{\rm cc}\) of at least one random overlap between an FRB and a young extragalactic magnetar (within 129 yr). CHIME/FRB Catalog‐1 (DM‐excess cutoff \(\le500\)\,pc\,cm\(^{-3}\)) contains 241 bursts over roughly 1 year (\(N_{\rm day}\approx0.5\)\,d\(^{-1}\)) \citep{2021ApJS..257...59C}. For a Gaussian localization uncertainty with \(1\sigma\) radius \(r_{1\sigma}\), the 90\% containment radius is $ r_{90}=2.146\,r_{1\sigma},$ enclosing an area  $
A_{90}=\pi\,r_{90}^{2}
\quad(\text{in deg}^2).$  
Each FRB therefore overlaps \(\Sigma_{\rm mag}\,A_{90}\) magnetars on average.  Over one day the Poisson expectation is  
\[
\lambda = N_{\rm day}\,\Sigma_{\rm mag}\,A_{90},
\qquad
P_{\rm cc}(\text{day}) = 1 - e^{-\lambda}.
\]

\begin{figure}
  \centering
  \includegraphics[width=0.6\linewidth]{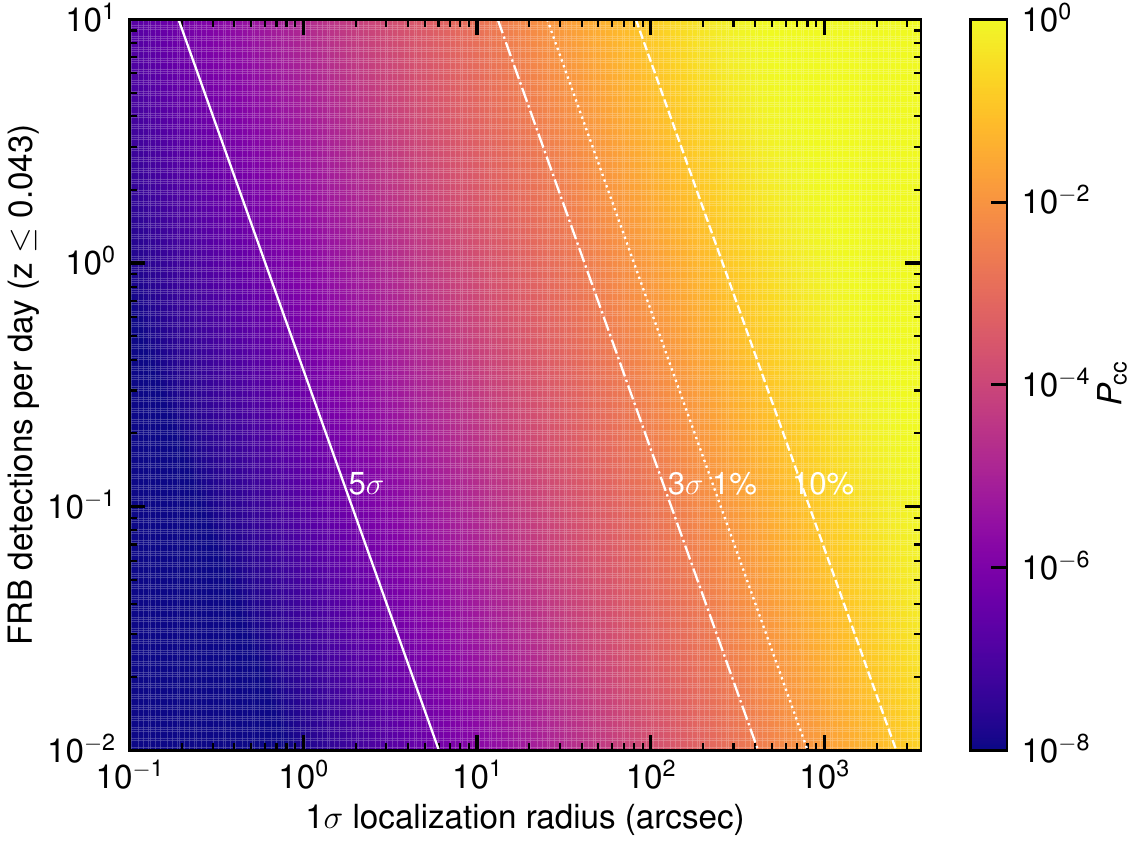}
  \caption{Probability \(P_{\mathrm{cc}}\) that at least one unrelated CCSN lies within an FRB 90 per cent localization region, for \(z\le0.043\).  The color scale shows \(\log_{10}P_{\mathrm{cc}}\) and white curves mark the 10\%, 1\%, 3$\sigma$, and 5$\sigma$ thresholds, which can separately by different line-styles.}
  \label{fig:Pcc_200mpc}
\end{figure}

Figure~\ref{fig:Pcc_200mpc} shows contours of \(P_{\rm cc}(\text{day})\) as a function of \(r_{1\sigma}=0.01''\)–\(3600''\) and \(N_{\rm day}=0.01\)–10\,d\(^{-1}\). At the CHIME/FRB header localisation scale (\(r_{1\sigma}\approx30'\)), the daily chance probability exceeds \(\sim30\%\) for \(N_{\rm day}\approx0.5\)\,d\(^{-1}\). In contrast, sub‐arcsecond precision (\(\lesssim0.1''\)) yields \(P_{\rm cc}\ll10^{-4}\), even for the highest anticipated FRB rates.

Extending to \(z>0.043\) would raise both the FRB detection rate and the magnetar surface density, shifting all contours upward and further increasing \(P_{\rm cc}\) at coarse localisation.  However, the historic SN catalog becomes progressively incomplete at greater distances, since SN remnants fade below detection limits (optical, radio, X-ray) on \(\sim30\)–50 yr timescales beyond \(\sim20\) Mpc \citep{2012A&ARv..20...49V,2015A&ARv..23....3D,2019MNRAS.487.5813S,2020pesr.book.....V}.  Consequently, the empirical \(\Sigma_{\rm mag}\) used here should be regarded as a lower bound, and the plotted \(P_{\rm cc}\) curves represent conservative estimates of the false‐alarm rate.

Looking forward, next‐generation wide‐field surveys—Rubin Observatory \citep{2023PASP..135j5002H}, Nancy Grace Roman Space Telescope \citep{2019arXiv190205569A}, NewAthena and AXIS \citep{2019ApJ...873..111I,2024Univ...10..316A}, SKA1-Mid and ngVLA \citep{2009IEEEP..97.1482D,2019clrp.2020...32D,2025NatAs...9...36C}—will extend the searchable SN remnant horizon by an order of magnitude in both depth and age.  At that point, false‐coincidence statistics must be revisited to account for the newly revealed population of older magnetar remnants.

\section{Derivation of the Frequency-Dependent Galaxy-Integrated FRB Rate Ratio}
\label{sec:rate_ratio_derivation}

In this appendix, we derive the expression used in \S\ref{sec:magnetar_constraints} for the ratio of the galaxy‐integrated FRB rates at two observing frequencies, \(\nu_a\) and \(\nu_b\).  We assume that FRBs originate from a population of magnetars whose formation rate tracks the CCSN rate and whose burst activity decays in time.  Free–free absorption in the expanding SN ejecta suppresses low‐frequency emission at early times.  Under these hypotheses, the rate ratio can be written entirely in terms of the intrinsic spectral index \(\Gamma\), the temporal decay index \(\beta\), and the differential transparency of the ejecta at the two frequencies.

We begin by considering a single magnetar of age \(t\).  At frequency \(\nu\) its instantaneous FRB burst rate above some energy threshold \(E_{\min}(\nu)\) is given by
\[
r_{\rm FRB}(t;\nu)
\;=\;
\nu^{-\Gamma}
\;\exp\bigl[-\tau_{\rm ff}(t,\nu)\bigr]
\;\int_{E_{\min}(\nu)}^\infty
r_{0}\,
\Bigl(\tfrac{t}{t_{0}}\Bigr)^{-\beta}\,
\frac{1}{E_*}\,
\Bigl(\tfrac{E}{E_*}\Bigr)^{-\gamma}
\exp\!\bigl(-E/E_*\bigr)
\,dE.
\]
Here:
\begin{itemize}
  \item \(\nu^{-\Gamma}\) accounts for the intrinsic FRB spectrum (\(F_\nu\propto\nu^{-\Gamma}\));
  \item \(\bigl(t/t_0\bigr)^{-\beta}\) captures the secular decline of burst activity with age;
  \item the Schechter–like energy function has characteristic energy \(E_*\) and power‐law index \(\gamma\);
  \item \(\exp[-\tau_{\rm ff}(t,\nu)]\) is the suppression due to free–free absorption in the SN ejecta.
\end{itemize}
Carrying out the change of variable \(x=E/E_*\) and using the incomplete gamma function \(\Gamma(s,x)=\int_x^\infty t^{\,s-1}e^{-t}dt\) with \(s=1-\gamma\), one finds
\[
r_{\rm FRB}(t;\nu)
=
\nu^{-\Gamma}
\;\exp\bigl[-\tau_{\rm ff}(t,\nu)\bigr]\;
r_0\,\Bigl(\tfrac{t}{t_0}\Bigr)^{-\beta}\;
E_*^{\,1-\gamma}\;
\Gamma\!\Bigl(1-\gamma,\tfrac{E_{\min}(\nu)}{E_*}\Bigr).
\]

Next, we sum over the entire magnetar population in a galaxy.  If magnetars form at a steady rate \(R_m\) (tracked by the CCSN rate and a magnetar‐formation fraction), then the total FRB rate at frequency \(\nu\) is
\[
R(\nu)
= R_m
  \int_{0}^{T_{\rm active}}
    r_{\rm FRB}(t;\nu)\,dt
= R_m\,r_0\,E_*^{1-\gamma}
  \int_{0}^{T_{\rm active}}
    \nu^{-\Gamma}
    \Bigl(\tfrac{t}{t_0}\Bigr)^{-\beta}
    \Gamma\!\Bigl(1-\gamma,\tfrac{E_{\min}(\nu)}{E_*}\Bigr)
    \exp\!\bigl[-\tau_{\rm ff}(t,\nu)\bigr]
  \,dt.
\]
Because \(E_*\), \(r_0\), and \(R_m\) do not depend on \(\nu\), they cancel when forming a ratio at two frequencies.  Furthermore, as discussed in \S\ref{sec:magnetar_constraints}, in our comparison both CHIME (600 MHz) and CRAFT ASKAP (1.4 GHz) rates are evaluated at the same fluence threshold (26 Jy ms) over effectively identical bandwidths \citep{2023ApJS..264...53C,2018Natur.562..386S}.  Hence the implied energy cut \(E_{\min}=4\pi D_L^2F_{\rm th}\Delta\nu\) is identical at both \(\nu_a\) and \(\nu_b\), so that
\[
\Gamma\!\Bigl(1-\gamma,\tfrac{E_{\min}(\nu_a)}{E_*}\Bigr)
\;\approx\;
\Gamma\!\Bigl(1-\gamma,\tfrac{E_{\min}(\nu_b)}{E_*}\Bigr)
\]
and their ratio is $\approx~1$ even after accounting for the redshift evolution of E$_{\rm min}$ \citep{2018MNRAS.480.4211M}.

Thus all normalization factors and the incomplete gamma term drop out, leaving

\begin{equation}
\label{equation:ratio}
\frac{R(\nu_a)}{R(\nu_b)}
=
\left(\frac{\nu_a}{\nu_b}\right)^{-\Gamma}
\;\times\;
\frac{
  \displaystyle
  \int_{0}^{T_{\rm active}}
    t^{-\beta}\,
    \exp\!\left[-\tau_{\rm ff}(t,\nu_a)\right]
  \,dt
}{
  \displaystyle
  \int_{0}^{T_{\rm active}}
    t^{-\beta}\,
    \exp\!\left[-\tau_{\rm ff}(t,\nu_b)\right]
  \,dt
}\,.
\end{equation}

In this final form:
\begin{itemize}
  \item The factor \(\bigl(\nu_a/\nu_b\bigr)^{-\Gamma}\) encodes the intrinsic spectral scaling.
  \item The two integrals quantify the different epochs at which the ejecta become transparent at \(\nu_a\) versus \(\nu_b\), weighted by the age decay \(t^{-\beta}\).
  \item \(T_{\rm active}\) is the maximum magnetar active lifetime (we adopt \(10^5\) yr), based on theoretical expectations for field decay \citep{2017ApJ...843L..26B,2022ASSL..465..245D}.
  \item \(\tau_{\rm ff}(t,\nu)\) may be taken from Equation~\eqref{eq:t_ff}, e.g.\ \(\propto\nu^{-2.1}t^{-5}\) for homogeneous ejecta.
\end{itemize}
Equation \ref{equation:ratio} is solved ratio numerically in \S\ref{sec:magnetar_constraints} to constrain \(\Gamma\) and \(\beta\) given the observed all‐sky rate ratio between 600 MHz and 1.4 GHz.


\end{document}